\documentclass{ptephy_v1}%%%%%% to generate preprint number with ptep logo

\usepackage{natbib}
\usepackage{wrapfig}
\usepackage{float}
\usepackage[caption = false]{subfig}

\usepackage{color}

\newcommand{\xmax}{X$_{\mathrm{max}}$ }

\begin{document}

\title{Radiowave Detection of Ultra-High Energy Neutrinos and Cosmic Rays}

\author{Tim Huege}
\author{Dave Besson}

\affil{Karlsruhe Institute of Technology, Karlsruhe 76021, Germany\email{tim.huege@kit.edu}} 
\affil{KU Physics Dept., 1082 Malott Hall, Lawrence KS, 66045, USA; National Research Nuclear University, Moscow Engineering Physics Institute, 31 Kashirskoye Shosse, Moscow 115409, Russia\email{zedlam@ku.edu}}

\begin{abstract}
Radio waves, perhaps because they are uniquely transparent in our terrestrial atmosphere, as well as the cosmos beyond, or perhaps because they are macroscopic, so the basic instruments of detection (antennas) are easily constructable, arguably occupy a privileged position within the electromagnetic spectrum, and, correspondingly, receive disproportionate attention experimentally. Detection of radio-frequency radiation, at macroscopic wavelengths, has blossomed within the last decade as a
competitive method for measurement of cosmic particles, particularly charged cosmic rays and neutrinos. 
Cosmic-ray detection via radio emission from extensive air showers has been demonstrated to be a reliable technique that has reached a reconstruction quality of the cosmic-ray parameters competitive with more traditional approaches.
Radio detection of neutrinos in dense media seems to be the most promising technique to achieve the gigantic detection volumes required to measure neutrinos at energies beyond the PeV-scale flux established by IceCube.
In this article, we review radio detection both of cosmic rays in the atmosphere, as well as neutrinos in dense media.
\end{abstract}

%\subjectindex{xxxx, xxx}

\maketitle

\section{Introduction}

Detection of ultra-high energy cosmic rays requires sensitivity to large volumes of target material. While established techniques for the detection of cosmic rays and high-energy neutrinos have been extremely successful, as outlined throughout the remainder of this volume, many alternative techniques for probing the rare incident fluxes have been proposed. One technique that has received ever-growing attention is the use of radio antennas to detect particle showers in the air and dense media, by their coherent emission. In air, such showers are initiated by cosmic rays, while in dense media, experiments seek detection of neutrinos. In this article, we review the state of radio detection of particle showers in both air and dense media.

Air-shower radio detection has progressed tremendously in the past decade, and radio detectors have by now become a valuable addition to several cosmic-ray experiments. As discussed in Chapter 2, all relevant cosmic-ray parameters -- the arrival direction, the energy of the primary particle and the mass-sensitive depth of shower maximum -- can be determined with competitive resolution using radio detection techniques. This has, in particular, become possible because the physics of radio emission is by now very well understood, and signal calculations have been able, within errors, to reproduce all experimental measurements made so far. We discuss the successes achieved to date as well as the potential for future applications.

For dense media, in particular ice, the hope is to be able to lower the energy thresholds for radio detection such that the high-energy extrapolation of the detected IceCube neutrino flux at energies of tens to hundreds of PeV as well as the cosmogenic neutrino flux expected at yet higher energies can be measured with large-scale radio detectors. In Chapter 3 we describe the challenges of in-ice radio detection of neutrinos, in particular the challenges presented by a target medium with a non-uniform complex dielectric permittivity, and discuss current projects and future plans.

\section{Radio detection of extensive air showers}

\begin{figure}[!htb]
  \centering
  \includegraphics[width=0.7\textwidth]{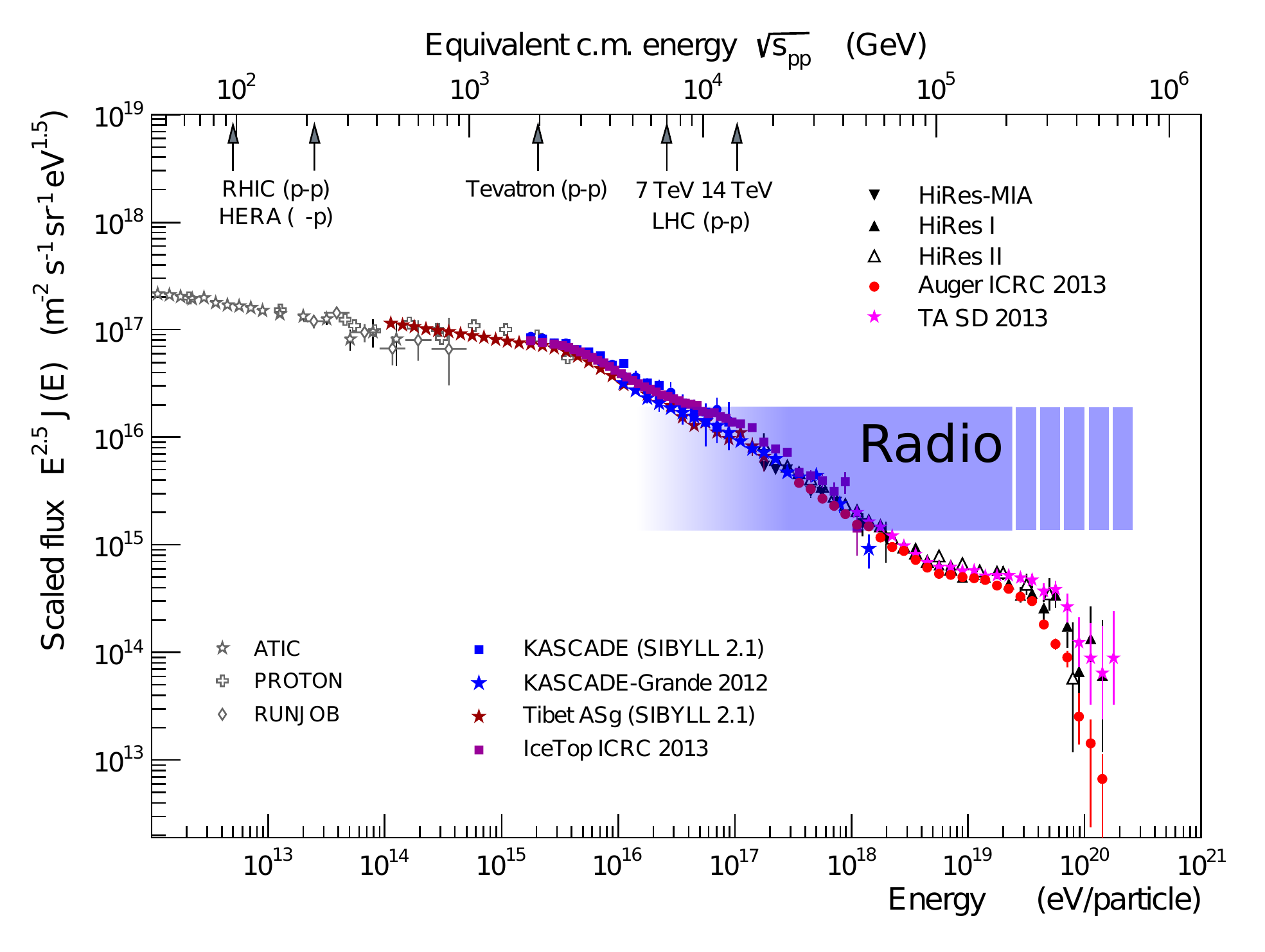}
  \caption{Energy spectrum of the highest energy cosmic rays 
  as measured by various air-shower experiments. The energy range accessible to 
  radio measurements using extant techniques is indicated.
  Diagram updated and adapted from \citep{Engel:2011zzb}, reprinted from \citep{HuegePLREP}.}
  \label{fig:crspectrum}
 \end{figure}

Over the past decade, radio detection of cosmic-ray air showers has matured from pioneering work using small prototype setups to full-fledged application in hybrid cosmic-ray observatories \citep{HuegePLREP,SchroederReview}.
The energy range to which today's concepts for radio detection can be readily applied is indicated in Figure \ref{fig:crspectrum}.
The limitations at low and high energies arise from different factors:
At energies below a few times 10$^{16}$~eV, detection is very difficult as the signals are overwhelmed by the always-present Galactic background. With the exception of near-horizontal air showers,
at energies beyond 10$^{19}$~eV, challenges have yet to be tackled to equip vast detection areas with the requisite number of radio antennas in an economical fashion.

In the following, we will first give a concise summary of the early history of cosmic-ray radio detection.
We will then discuss the current knowledge of the radio emission physics, which is now understood at a level of detail 
that allows a consistent description of all experimental data acquired so far.
This theoretical understanding constitutes a solid foundation for devising future radio-detection experiments, and for developing strategies for sophisticated data analyses.
Afterwards, we present a summary of currently existing radio detection experiments and showcase the currently achieved experimental results, in particular with respect to the reconstruction of cosmic-ray energies and the mass-dependent depth of air-shower maximum.
Finally, we offer an outlook on possible future directions in air-shower radio detection.

%----------------------------------------------------------------
\subsection{Early history}

Radio detection of cosmic-ray showers was first demonstrated in 1965 by Jelley et al. \citep{JelleyFruinPorter1965},
leading to concerted activity over the ensuing decade \citep{Allan1971}.
In spite of the limitations of the pre-digital, analog detection techniques, 
several groups succeeded in detecting radio emission from air showers, and in studying a number of emission properties,
leading to consensus on the following aspects:

\begin{itemize}
\item Below 100~MHz, the radio emission from air showers is generally coherent, i.e., the radiated power scales quadratically with the number of particles and thus the energy of the primary cosmic ray.
\item The emission is dominated by a geomagnetic effect as there is a clear correlation of the emission strength with the angle between the air-shower axis and the geomagnetic field (the so-called ``geomagnetic angle'').
\item The signal falls off quickly with lateral distance to the shower axis, roughly following an exponential decay with a scale radius of order 100~m.
\end{itemize}

Many other points, however, were not universally agreed upon within the community:

\begin{itemize}
\item Many secondary emission mechanisms were discussed, among them the Askaryan charge-excess emission \citep{Askaryan1962a,Askaryan1965}, but it was not at all clear whether any signal component was present in addition to the well-established geomagnetic one.
\item Effects of the refractive index gradient in the air as well as the sensitivity of the lateral signal distribution on the depth-of-shower maximum were discussed, however only on a purely theoretical basis \citep{AllanRefractive1971,HoughXmaxLDF}.
\item There was vast disagreement, by orders of magnitude, in the radio-emission amplitudes measured at a given energy of a cosmic-ray primary \citep{AtrashkevichVedeneevAllan1978}, most likely explained by difficulties in the absolute calibration of the radio detectors.
\item It was unclear how critical the influence of strong atmospheric electric fields might be, e.g.\ during thunderstorms \citep{Mandolesi19741431}, and it was feared that they would hamper a reliable cosmic-ray energy determination using radio detectors.
\end{itemize}

In the mid-1970s activities more or less ceased completely, mostly because of difficulties in a consistent interpretation of the data.
Other techniques, in particular fluorescence detection, also seemed more promising at the time.
It was only with the availability of powerful digital signal processing techniques that the field was revived at the beginning of the new millenium.

%----------------------------------------------------------------

\subsection{Emission physics}\label{sec:emissionphysics}

\begin{figure}[!htb]
\centering
\includegraphics[width=0.35\textwidth]{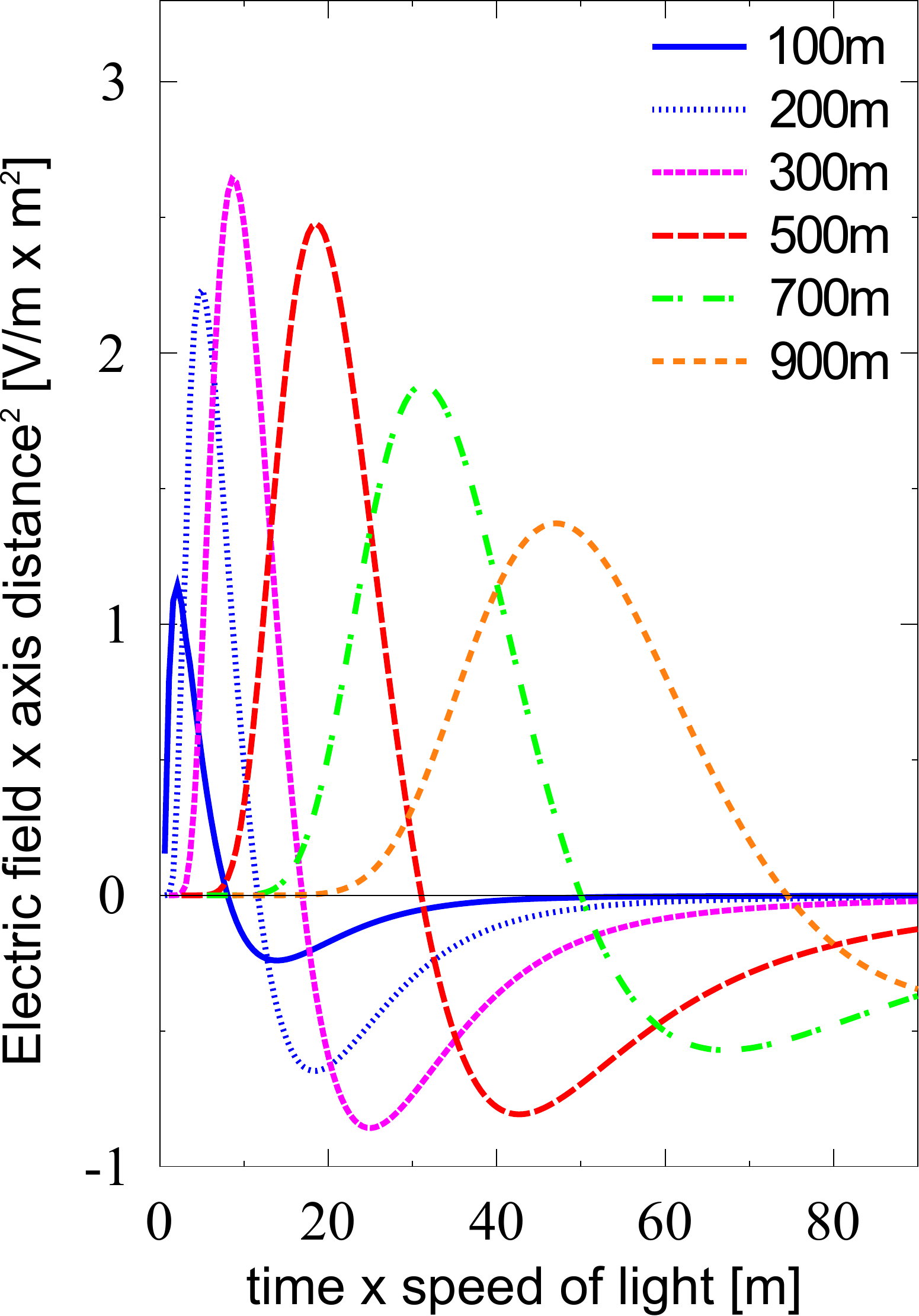}
\includegraphics[width=0.47\textwidth]{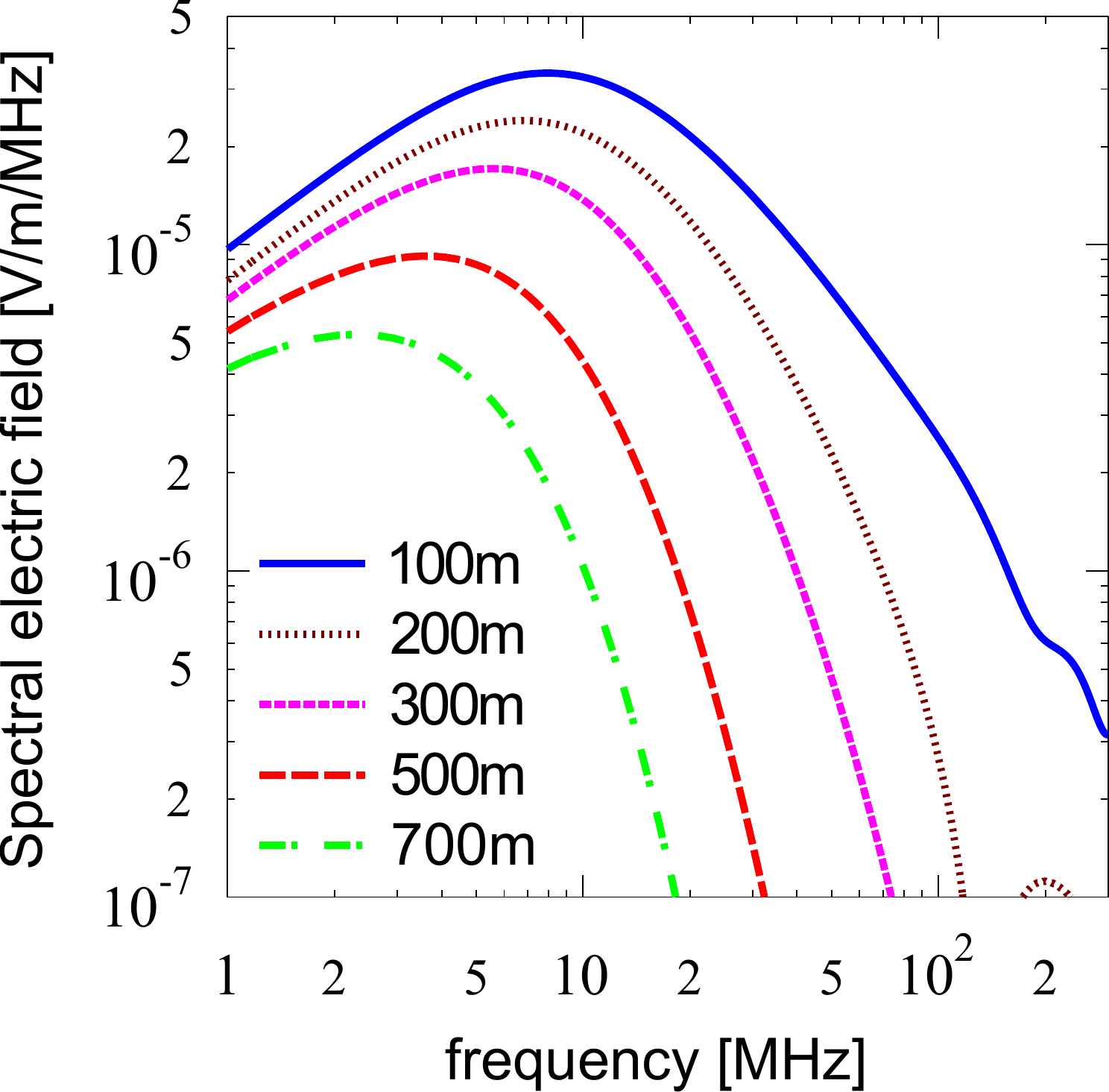}
\caption{Modeled radio pulses (left) due to geomagnetic effect in a $10^{17}$~eV air 
shower as observed at various observer distances from the shower 
axis as well as corresponding frequency spectra (right). Effects due to the refractive 
index of the atmosphere are not taken into account. Adapted from \citep{ScholtenWernerRusydi2008}, reprinted from \citep{HuegePLREP}.\label{fig:radiopulses}}
\end{figure}

The most important breakthrough with respect to radio detection of cosmic rays has been achieved with the detailed understanding of the radio emission physics.
The main effect causing the pulsed radio emission is due to the geomagnetic field: secondary electrons and positrons in the electromagnetic cascade of the air shower are continuously accelerated by the Lorentz force.
However, these particles also continuoulsy interact with the 
molecules and atoms of the atmosphere, which counteracts the geomagnetic acceleration.
The situation is akin to that of electrons in a conductor to which an electric potential is applied.
As a result, electrons and positrons undergo a net drift in the direction of the Lorentz force, perpendicular to the air-shower axis, leading to ``transverse currents'' \citep{KahnLerche1966,ScholtenWernerRusydi2008}.
During the evolution of the air shower, the currents first grow, then reach a maximum, and then decline again as the electromagnetic component dies out due to ionization losses. 
It is this time-variation of the transverse currents that leads to the radio emission, which has a characteristic linear polarization along the direction of the Lorentz force.
Furthermore, both the radiating particles and the radio emission propagate with approximately the speed of light, i.e., the emission becomes relativistically beamed in a narrow cone along the forward direction of the air shower, and becomes compressed in time to a pulse with a typical length of order ten nanoseconds.
This timescale also governs the typical coherence limit of the pulses: at frequencies below approximately 100~MHz, the emission is coherent, and its power scales with the square of the number of radiating particles, which in turn is roughly proportional to the energy of the primary particle.
The pulse length, however, depends on geometry, for observers far away from the shower axis geometrical time-delays broaden the pulses, and coherence is only obtained for low frequencies.
Typical pulses and corresponding frequency spectra for the geomagnetic radiation component are shown in Fig.\ \ref{fig:radiopulses}.

In addition to this main emission mechanism, there is a secondary component, typically contributing at about the 10\%-level of the electric field amplitudes, i.e., 1\%-level of the radiated power, as follows.
During the air shower evolution, the atmosphere is continuously ionized.
The freed electrons propagate with the air shower disk, whereas the much heavier positive ions stay behind.
This leads to a charge imbalance in the air shower: there are more electrons than positrons, at a 
level of roughly 10-20\%.
As the shower evolves, the net charge grows with the total number of particles, 
reaches a maximum, and then declines again.
One can interpret this as a ``longitudinal current'', which again is time-varying and 
thus yields electromagnetic radiation.
In combination with refractive-index effects (see below), this is equivalent to the ``Askaryan effect'' \citep{Askaryan1962a,Askaryan1965}, which is the sole relevant source of radio emission in dense media, cf.\ section \ref{sec:askaryanDM}.
As is the case for the geomagnetic emission, this radiation is relativistically forward-beamed because emitters and emission both travel with approximately the speed of light.
The emission is again linearly polarized, but the electric field vectors point radially towards the air shower axis, i.e., the orientation of the electric field vector depends on the relative position of the observer with respect to the shower axis.

\begin{figure}[!htb]
  \centering
  \includegraphics[width=0.23\textwidth,clip=true,trim=0cm 0cm 0cm 12cm]{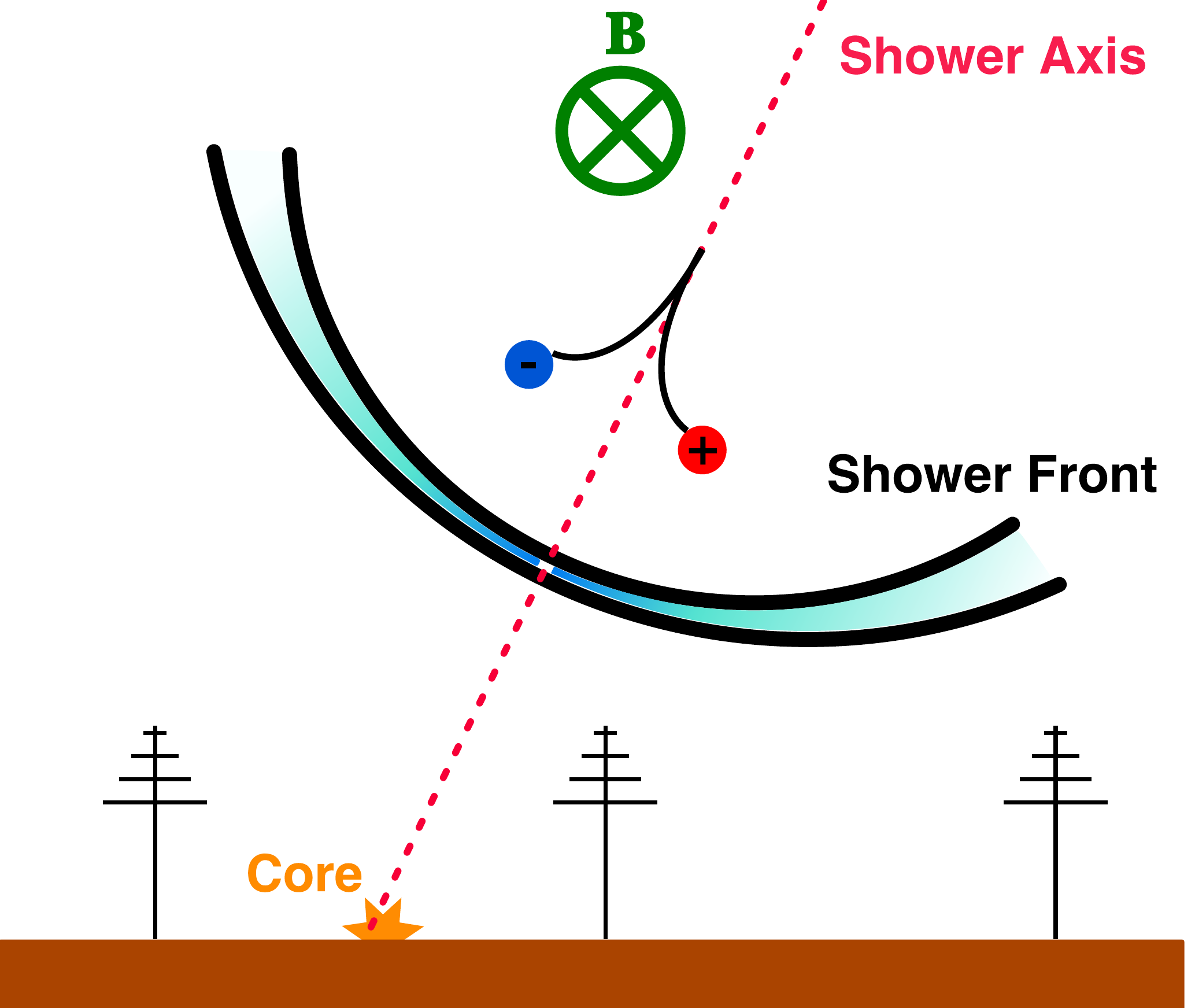}
  \includegraphics[width=0.21\textwidth]{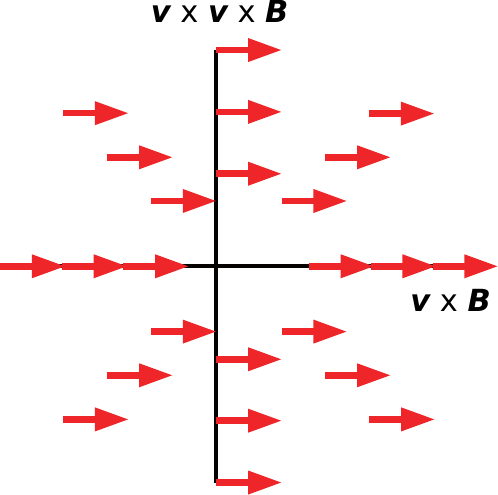}
  \hspace{0.06\textwidth}
  \includegraphics[width=0.23\textwidth,clip=true,trim=0cm 0cm 0cm 12cm]{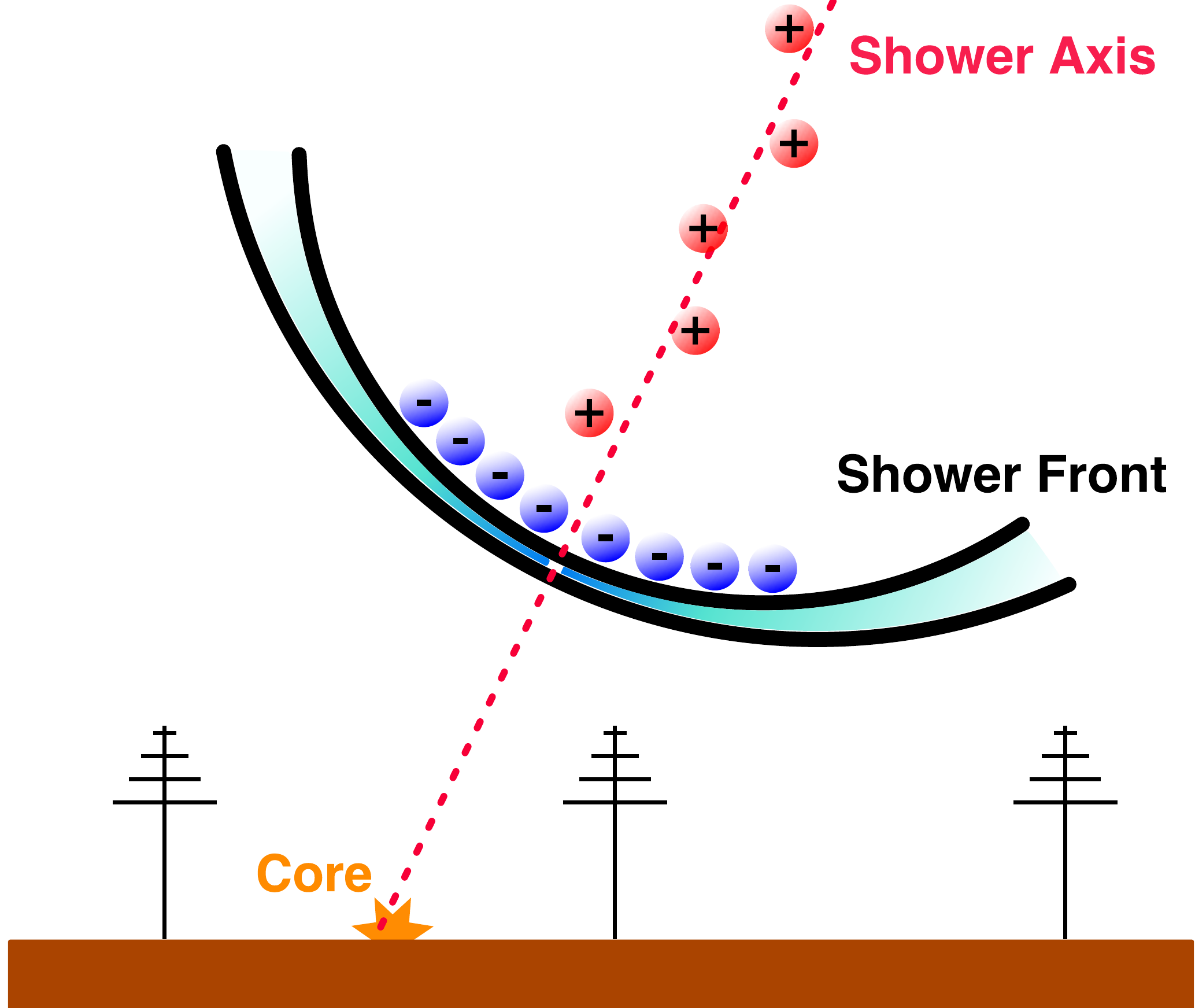}
  \includegraphics[width=0.21\textwidth]{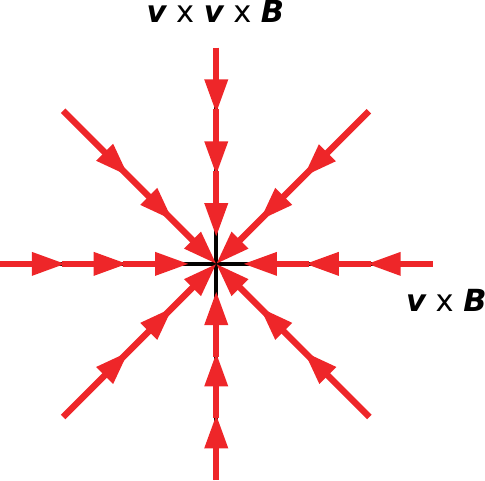}
  \caption{Left: Characterisation of the geomagnetic radiation mechanism; 
  the arrows denote the directions of the electric field vector in the plane 
  perpendicular to the air shower axis. The emission is uniformly and linearly
  polarized along the direction given by the Lorentz force, $\vec{v} \times \vec{B}$ (east-west for 
  vertical air showers).
  Right: 
  Characterisation of the charge-excess (Askaryan) emission. The arrows 
  denote the direction of the electric field vectors which are oriented
  radially with respect to the shower axis. Diagrams have been adapted from 
  \citep{SchoorlemmerThesis2012} and \citep{deVries2012S175}, reprinted from \citep{HuegePLREP}.}
  \label{fig:mechanisms}
 \end{figure}

\begin{figure}[!htb]
  \vspace{2mm}
  \centering
  \includegraphics[width=0.49\textwidth]{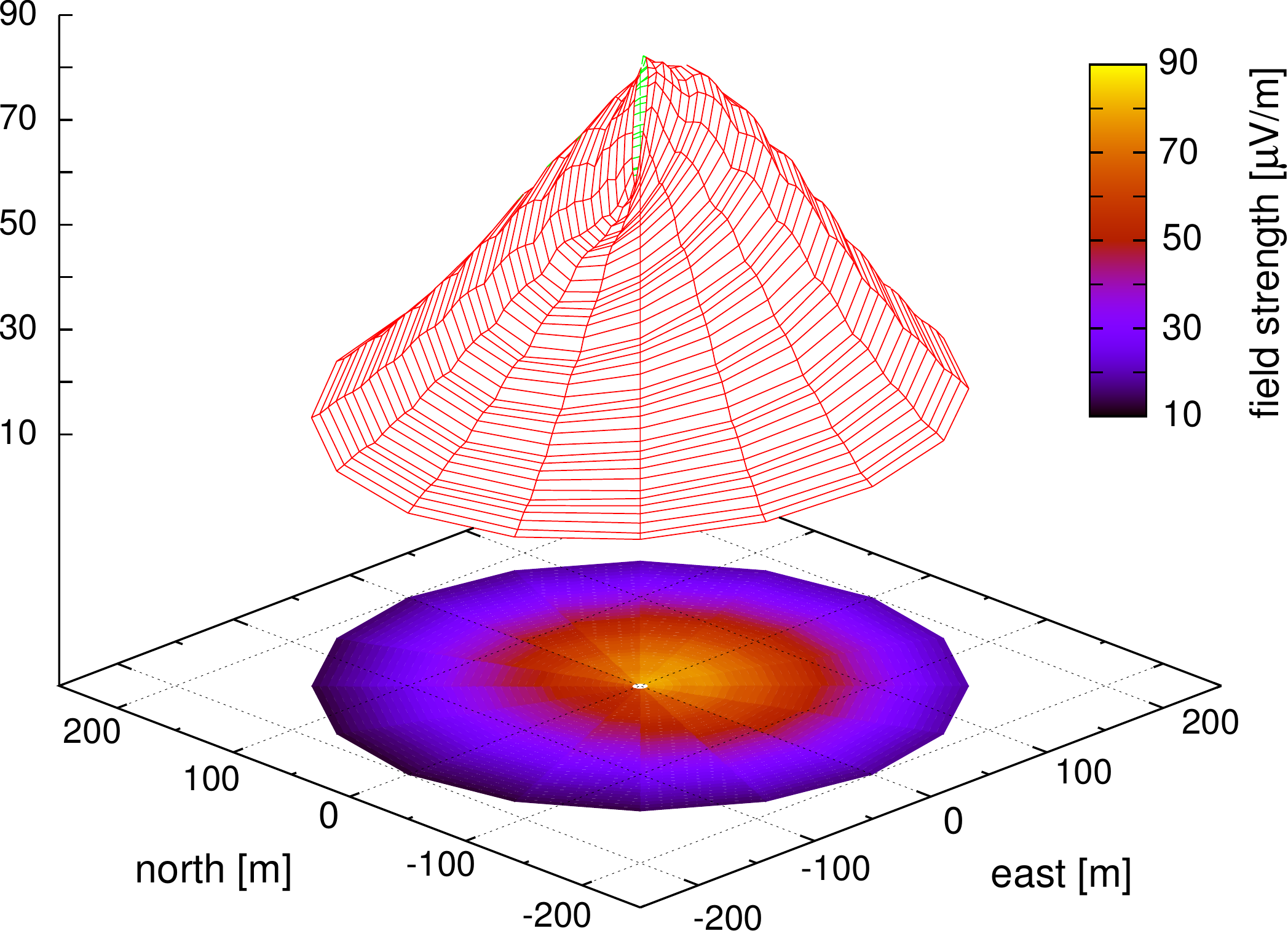}
  \caption{Simulation of the total electric field amplitude for a vertical cosmic-ray air shower at the site of 
  the LOPES experiment after filtering to the 40-80~MHz band. The pattern is asymmetric due to 
  the superposition of the geomagnetic and charge-excess emission 
  contributions with their charateristic polarizations. Refractive index effects have been taken into account. Adapted from 
  \citep{HuegeARENA2012a}, reprinted from \citep{HuegePLREP}.}
  \label{fig:footprint}
 \end{figure}

Both mechanisms and their characteristic polarization characteristics are shown in Fig.\ \ref{fig:mechanisms}.
Superposition of these two contributions with their characteristic polarizations leads to an asymmetry in the received amplitude as a function of observer position, as shown in Fig.\ \ref{fig:footprint}. 

\begin{figure}[!htb]
\centering
\includegraphics[width=0.48\textwidth]{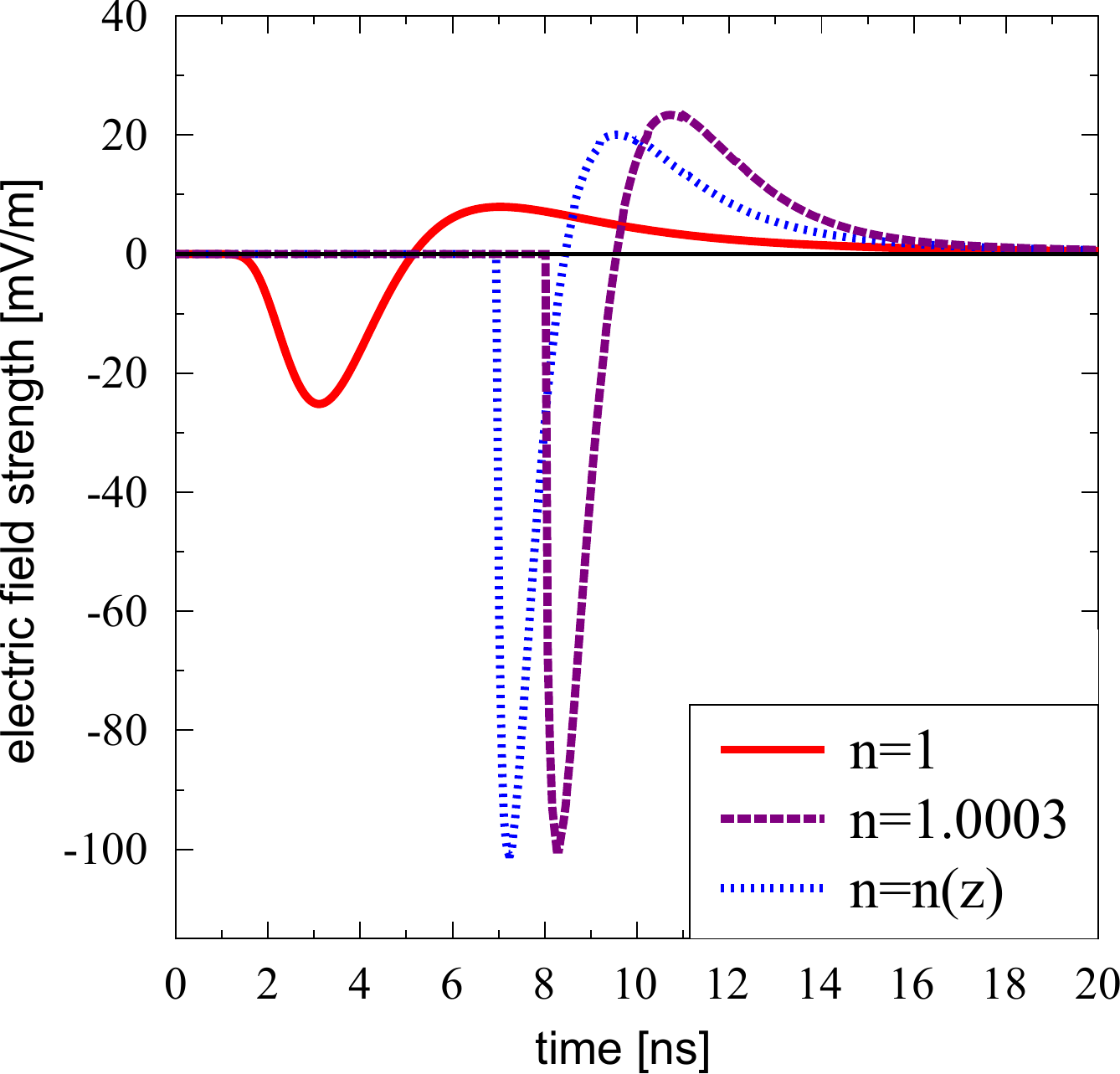}
\caption{Simulated radio pulses of a $5 \times 10^{17}$~eV air shower as
received by an observer at an axis distance of 100~m. The refractive index $n$ has been set to unity (vacuum),
1.0003 (sea level) and a realistic gradient in the 
atmosphere $n(z)$, demonstrating the resulting time compression of the radio 
pulses. The particle distribution is approximated to have no lateral extent.
Adapted from \citep{DeVriesBergScholten2011}, reprinted from \citep{HuegePLREP}.\label{fig:cherenkovcompression}}
\end{figure}

Another important ingredient is the fact that the atmosphere has a refractive index of $n > 1.0$ with a typical value at sea level of $n \approx 1.0003$, and decreasing with atmospheric density to higher altitudes.
Consequently, the emission propagates slightly more slowly than the cloud of emitting particles, modifying the coherence conditions.
In particular, at the Cherenkov angle\footnote{As the refractive index of the atmosphere follows the density gradient of the atmosphere, the Cherenkov angle varies over the course of the air-shower evolution.} the radio emission from the complete longitudinal air shower evolution arrives simultaneously, i.e., the already-short radio pulses are compressed further in time leading to significant and measurable pulse amplitudes up to GHz frequencies \citep{DeVriesBergScholten2011,AlvarezMunizCarvalhoZas2012}.
An example of this temporal pulse compression is shown in Fig.\ \ref{fig:cherenkovcompression}.
We stress that the emission from air showers should not be confused with regular ``Cherenkov radiation'', rather the geomagnetic and charge-excess radiation is time-compressed by Cherenkov-like effects.

\begin{figure}[tbh]
\centering
\includegraphics[width=0.65\textwidth]{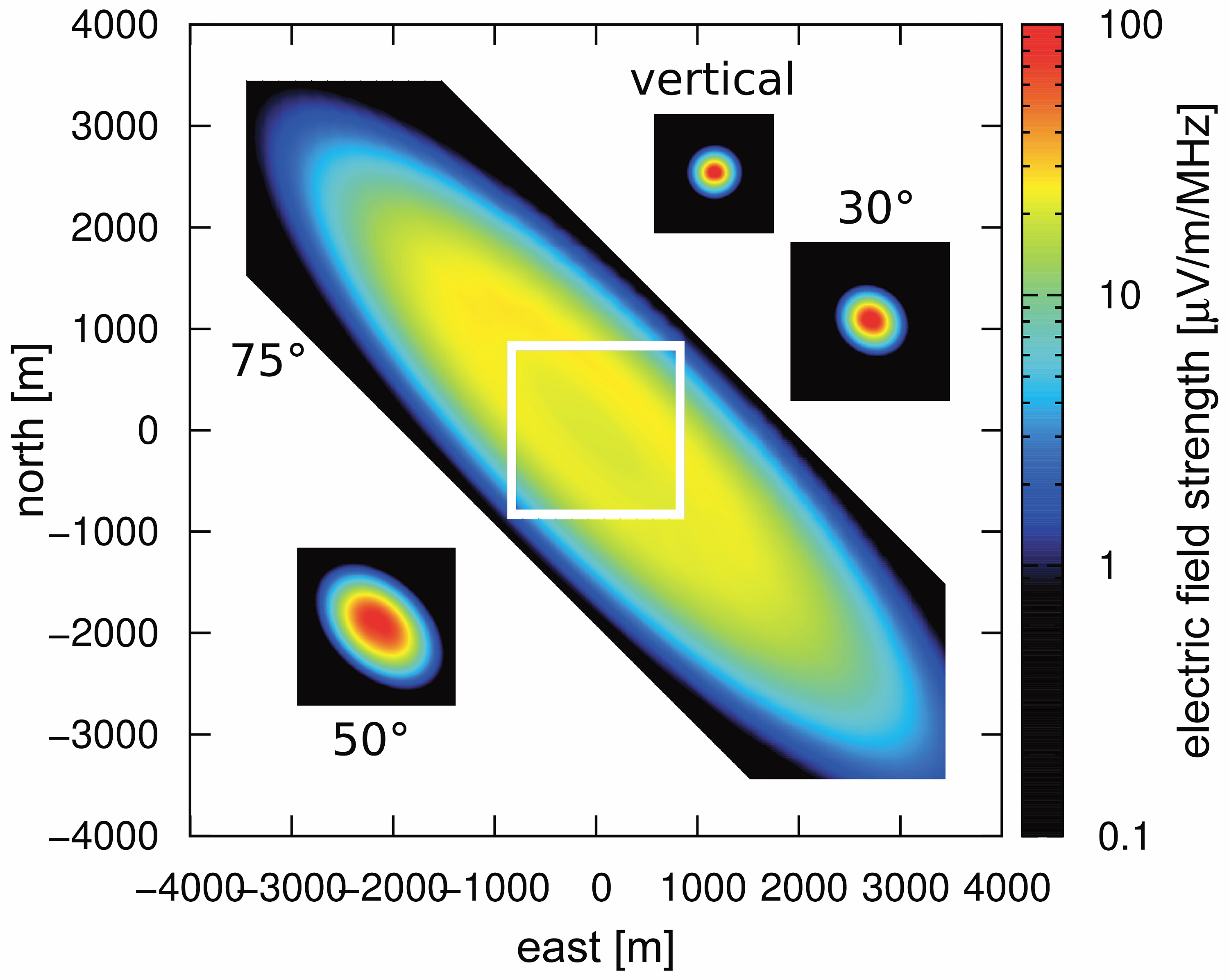}
\caption{Simulated radio emission footprints for air 
showers with various zenith angles in the 30-80~MHz frequency band.
The primary energy is $5 \times 10^{18}$~eV. The detection threshold 
governed by Galactic noise typically lies at 
$\approx$~1-2~$\mu$V/m/MHz. The footprint is small for air showers with zenith angles below $\approx 
50^{\circ}$, but becomes very extended for inclined showers with zenith 
angles of $70^{\circ}$ or larger. The white rectangle denotes 
the size of the $50^{\circ}$ inset.
Adapted from \citep{HuegeUHECR2014}, reprinted from \citep{HuegePLREP}.}
\label{fig:inclined}
\end{figure}

The strong forward-beaming of the radio emission has an important consequence: the area illuminated by the radio signals is typically rather small.
This is demonstrated in Fig. \ref{fig:inclined}, where one can see that the radio emission from air showers with zenith angles up to 50$^{\circ}$ is only detectable up to distances of a few hundred meters from the shower axis.
The effect is governed by the beaming angle in combination with the source distance and thus is more or less independent of the particle energy.
Consequently, air-shower detection typically (for zenith angles up to 60$^\circ$) requires rather dense arrays with grid spacings of at most 300~m to ensure coincident measurements in at least three radio detectors.
For inclined air showers with zenith angles exceeding 60$^{\circ}$,
the air shower develops to its maximum much higher in the atmosphere and thus the radio-emission source is far away, 
resulting in measurable radio signal distributed over a much larger area \citep{KambeitzARENA2016}.
The lateral distribution of the radio signal also `flattens', with a
correspondingly lower amplitude. Nevertheless,
as long as the received amplitudes are observable above the Galactic noise background, an instrument with a sparse antenna array (grid spacings of a km or more) is sufficient in this case.

The paradigm that has been laid out here is the commonly accepted interpretation of the radio emission physics in extensive air showers.
However, it should be noted that the only way to simulate radio signals in complete agreement with all observations is currently given by microscopic Monte Carlo simulations in which the acceleration of individual particles is used to calculate the associated electromagnetic radiation using discretized formalisms of classical electrodynamics \citep{ZHS,JamesFalckeHuege2012}.
Macroscopic models that try to superpose the emission along the above-described ``mechanisms'' exist \citep{ScholtenWernerRusydi2008,WernerScholten2008,KonstantinovMacroscopic2011} and provide semi-analytic results which give a good qualitative description of the emission.
However, there are significant deviations between predictions and measurements, in particular for the emission close to the shower axis, that lead to limitations when using macroscopic calculations in analyses.
An additional problem lies in the presence of free parameters in macroscopic approaches that need to be set to obtain agreement with measurements. In contrast, the two existing microscopic simulation codes CoREAS \citep{HuegeARENA2012a} and ZHAireS \citep{AlvarezMunizCarvalhoZas2012} have no free parameters and make no assumptions about the emission mechanisms -- they calculate the radio signal from first principles (in the form of classical electrodynamics) and have so far been able to describe all experimental data within the systematic uncertainties of the measurements.

%----------------------------------------------------------------

\subsection{Experiments and results}

A detailed discussion of the experiments involved in radio detection of extensive air showers is beyond the scope of this article, and we kindly refer the reader to more extensive review articles \citep{HuegePLREP,SchroederReview}.
Here, we will only briefly introduce the first- and second-generation digital experiments.
When radio detection was revived at the beginning of the new millenium, it was in particular the LOPES \citep{FalckeNature2005} and CODALEMA \citep{ArdouinBelletoileCharrier2005} efforts which performed the pioneering work.
LOPES extended the well-established KASCADE-Grande \citep{ApelArteagaBadea2010} experiment with radio antennas based on LOFAR \citep{LOFARCosmicRays} prototype designs.
CODALEMA followed the opposite approach and equipped an existing array of radio antennas with particle detectors; later, custom-made radio antennas for cosmic-ray detection were deployed.
Both projects started with relatively small-scale installations, but grew and evolved over the coming years.
In spite of their pioneering character, many important results could be achieved with these first-generation experiments: the dominance of the geomagnetic emission was confirmed early on \citep{FalckeNature2005,ArdouinBelletoileCharrier2006,CodalemaGeoMag}, the signal coherence and thus linear scaling of radio amplitudes with energy of the primary particle was demonstrated \citep{FalckeNature2005}, quantitative comparisons with signal simulations were made \citep{SchroederLOPESCoREAS2013,HuegeLOPESIcrc2015,LOPESrecalibration}, and even the first experimental evidence for the sensitivity of the radio lateral distribution on the depth of shower maximum was obtained \citep{ApelArteagaBaehren2012c}.
(We will discuss these results in more detail below.)

\begin{figure}
\centering
\includegraphics[width=1.0\textwidth]{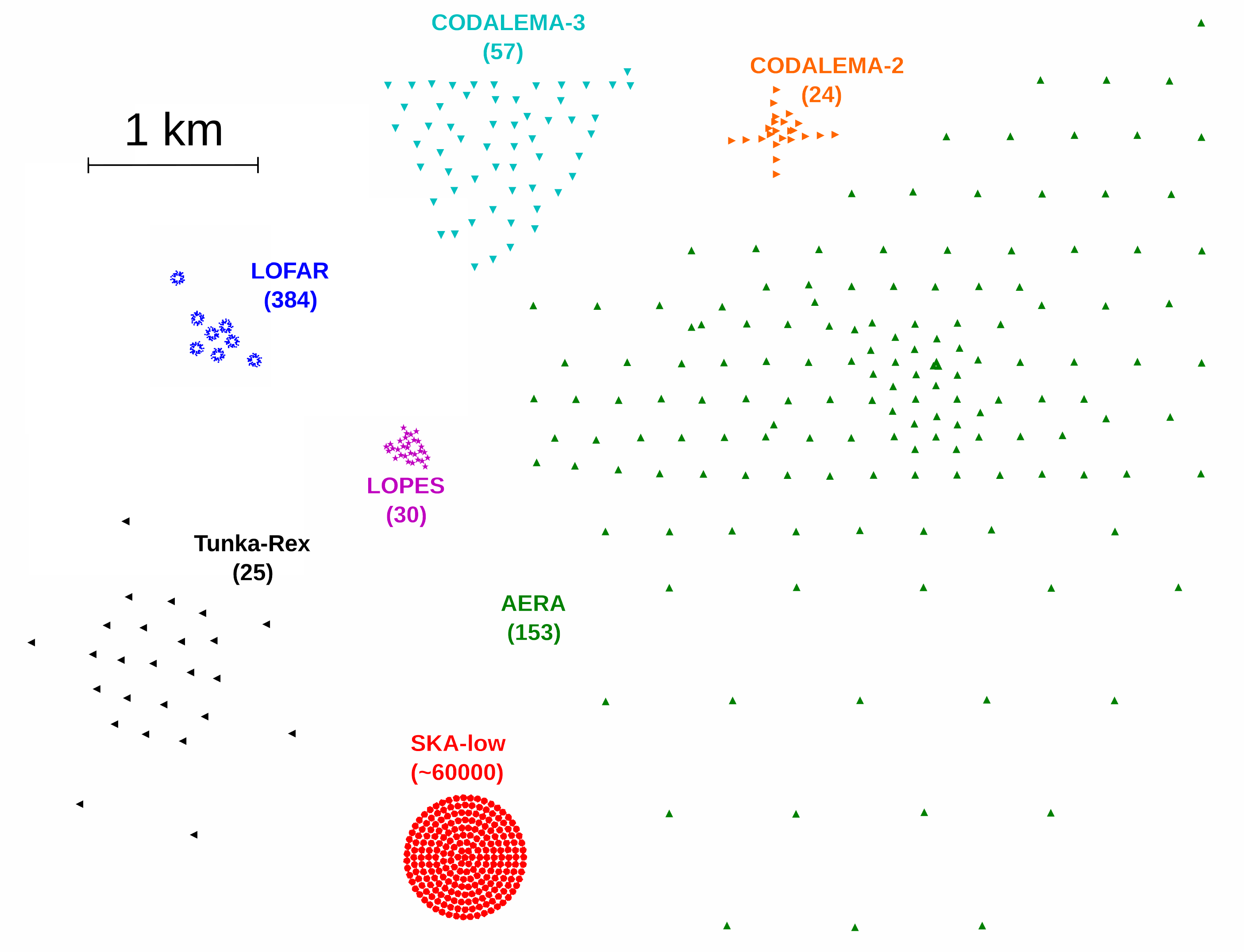}
\caption{Compilation of modern cosmic-ray radio detection experiments.
Each symbol represents one radio detector (typically a 
dual-polarised antenna), except for the SKA where individual detectors 
are not discernible because of their very high density. The number in 
brackets lists the total number of antennas used in each of the experiments. Reprinted from \citep{HuegePLREP}.\label{fig:experiments}}
\end{figure}

Based on the successes of these first-generation cosmic-ray radio detectors, a second generation of experiments was envisaged and implemented.
This includes in particular the Auger Engineering Radio Array (AERA) \citep{SchulzIcrc2015} at the Pierre Auger Observatory \citep{AugerNIM2014}, the Tunka-Rex radio extension of Tunka-133 \citep{TunkaRexInstrument}, and the use of the radio-astronomy observatory LOFAR as a cosmic-ray detector \citep{LOFARCosmicRays}.
AERA and Tunka-Rex can be considered ``sparse'' arrays in which the antennas are spaced several hundred metres apart to reach instrumented areas of approximately 17 km$^2$ and 1 km$^2$, respectively.
Both radio detectors have access to independent information on the depth of shower maximum (\xmax), in the case of AERA with the Auger fluorescence telescopes, and in the case of Tunka-Rex with the optical Tunka Cherenkov light detectors.
They can thus independently validate the precision and accuracy of the \xmax reconstruction from radio data.
In contrast to the sparse arrays, LOFAR's cosmic ray detection capability currently covers a comparatively small sensitive area of only $\sim 0.1$~km$^{2}$, with hundreds of antennas.
It is therefore more limited in event statistics and energy range, but can study individual radio events with an exceptionally high level of detail.
This is very useful, in particular, to gauge our understanding of the radio emission physics.
An overview of the most important experiments and their array layouts is shown on the same scale in Fig.\ \ref{fig:experiments}.

In the following, we will first discuss the quality of the radio emission modelling achieved today, as verified by experimental data, and then present results on the reconstruction of the arrival direction, the particle energy and the depth of shower maximum.

\subsubsection{Understanding of the radio-emission physics}

A solid understanding of the radio-emission physics in extensive air showers is of paramount importance for the planning of experiments as well as the interpretation of experimental data.
In the past few years, especially the microscopic full Monte Carlo simulations based on first-principle calculations have been validated extensively with measurements, and as of today, they have been able to reproduce all measurements within their systematic uncertainties.
This involves all aspects of the radio emission such as the absolute amplitudes of the signals, the polarization characteristics, and the asymmetric lateral distribution function with the presence of a ``Cherenkov bump'' \citep{HuegeLOPESIcrc2015,LOPESrecalibration,SchroederAERAIcrc2013,TunkaRexInstrument,LOFARXmaxMethod2014}.
In the following, we present the most relevant experimental results that illustrate the current level of understanding of the radio emission physics.

\begin{figure}[!htb]
  \centering
  \includegraphics[width=0.33\textwidth]{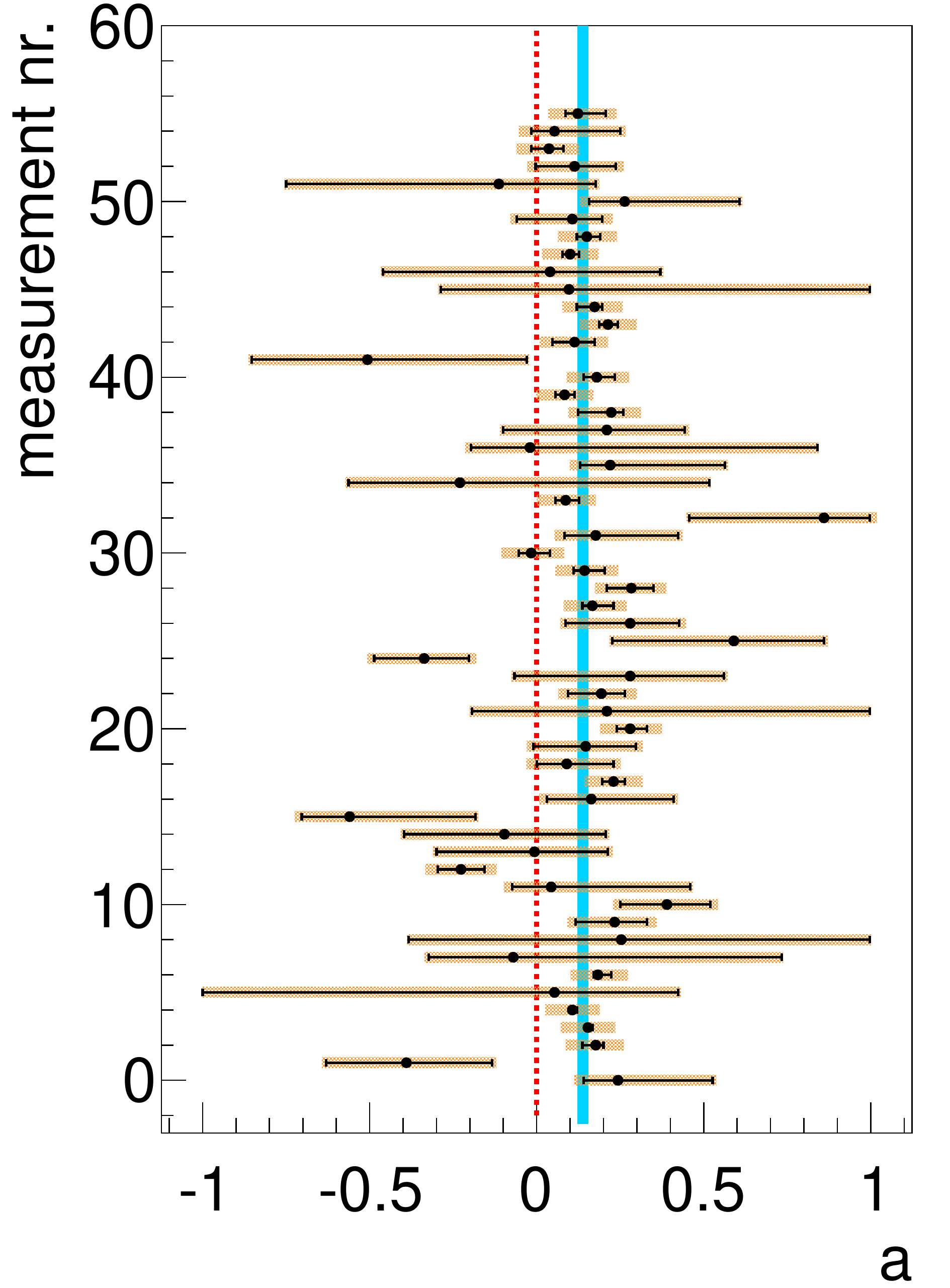}
  \includegraphics[width=0.57\textwidth]{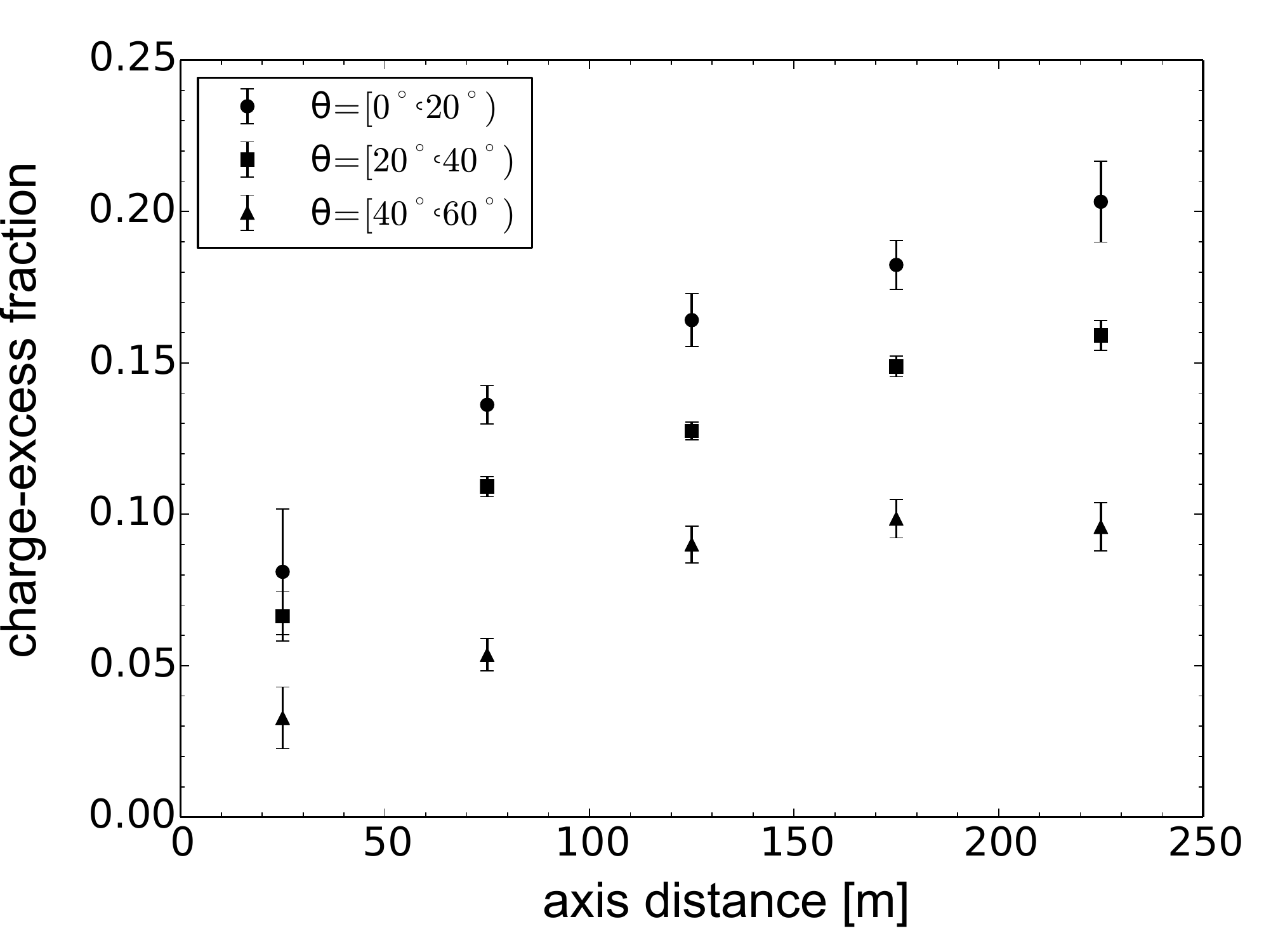}
  \caption{Left: Fraction of radially polarized emission
  $a$ relative to the contribution from geomagnetic emission as measured by AERA. Positive values 
  characterize the radial polarisation expected for charge-excess emission. The average value of $a$
  for the studied data set amounts to 14\%. Adapted from \citep{AERAPolarization2014}, reprinted from \citep{HuegePLREP}.
  Right: Detailed measurement of the charge-excess fraction as a function of air-shower zenith angle 
  and observer lateral distance from the shower axis performed with LOFAR. Adapted from 
  \citep{LOFARChargeExcess2014}, reprinted from \citep{HuegePLREP}.\label{fig:chargeexcess}}
\end{figure}

The paradigm described in section \ref{sec:emissionphysics} involves the presence of a sub-dominant contribution from charge-excess emission which leads to an asymmetry in the radio-emission footprint.
This prediction has been clearly confirmed by measurements.
The first evidence was found by CODALEMA, which showed that the core position estimated from radio signals on the basis of a rotationally symmetric lateral distribution function had a systematic offset to the true core position, a clear sign for an asymmetry in the radio lateral distribution \citep{CODALEMACoreShift,Belletoile201550}.
This was followed-up with more quantitative measurements both by AERA \citep{AERAPolarization2014} and LOFAR \citep{LOFARChargeExcess2014}.
In Fig.\ \ref{fig:chargeexcess}, the fraction of emission with radial linear polarization (as expected for the Askaryan charge-excess emission) is quantified.
AERA first measured this fraction to be at an average level of 14\% for a specific set of measurements.
LOFAR later provided a detailed measurement of the scaling of this percentage with shower zenith angle and distance from the shower axis.
This scaling can in fact be understood as a dependence of the density of the medium into which the particle shower radiates \citep{Glaser:2016qso}.

\begin{figure}[t]
    \centering
    \includegraphics[width=0.48\textwidth]{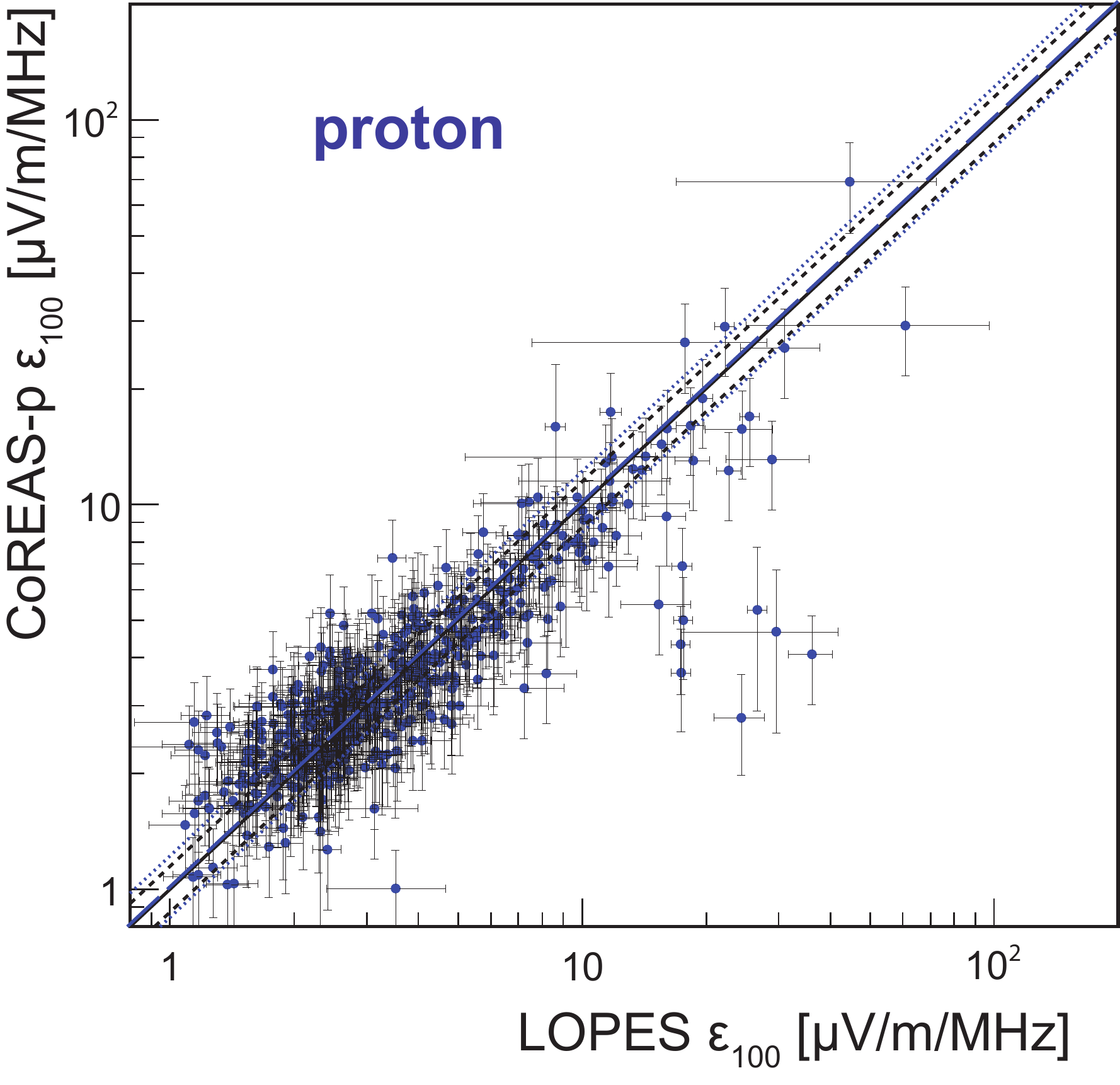}
    \hspace{3mm}
    \includegraphics[width=0.48\textwidth]{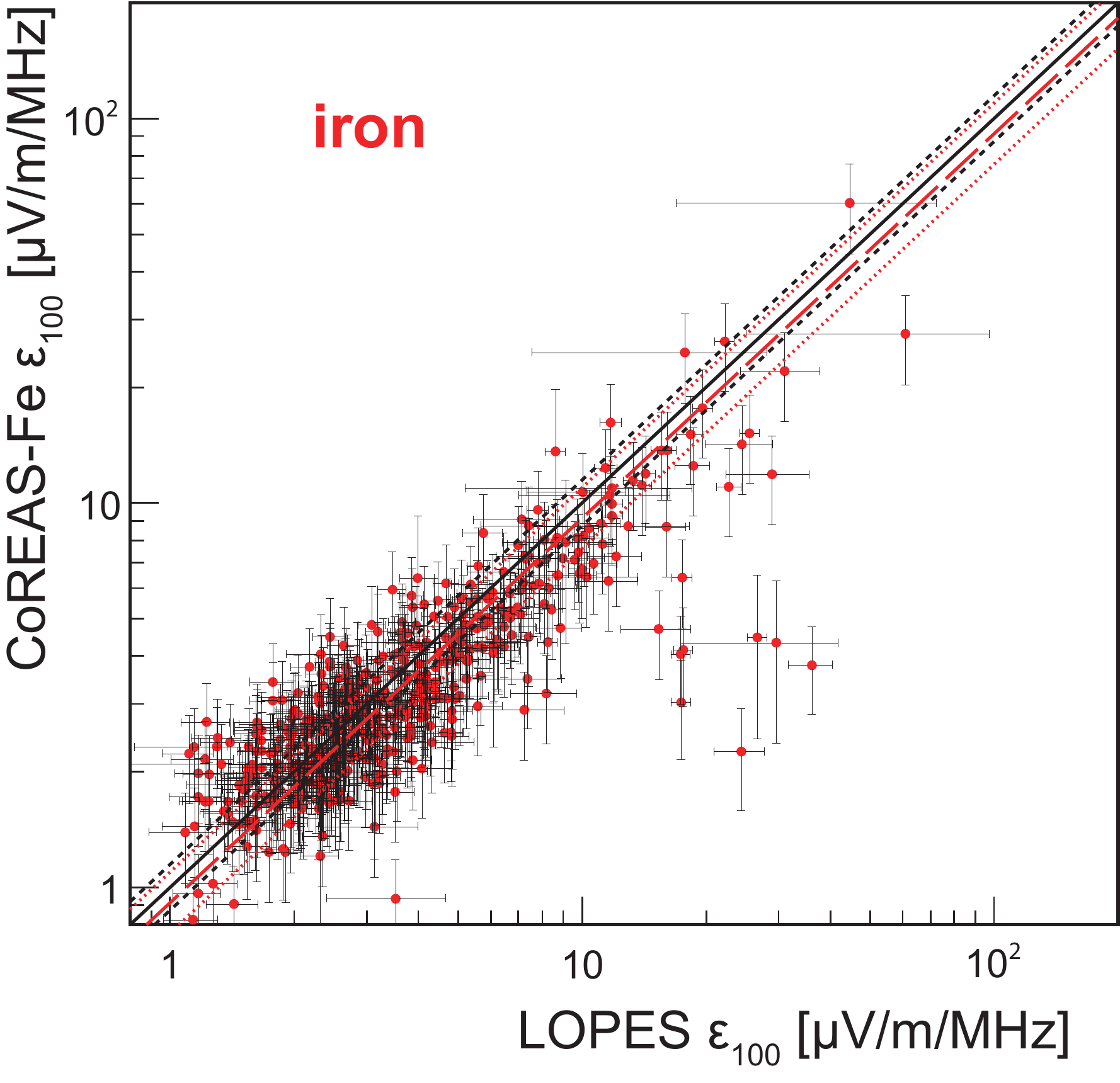}
    \caption{Comparison of the electric-field amplitudes of individual air showers at a lateral 
    distance of 100~m derived from LOPES measurements and from CoREAS predictions for  
    proton-induced showers (left) and iron-induced showers (right). 
    The black lines denote the 1:1 expectation (solid) and the 
    16\% systematic scale uncertainty of the LOPES amplitude 
    calibration (dashed). The colored lines illustrate the actual 
    correlation between simulations and data (long-dashed) and the 
    systematic uncertainty of the simulated amplitudes due to the 
    20\% systematic uncertainty of the energy 
    reconstruction of KASCADE-Grande (dotted). Adapted from 
    \citep{HuegeLOPESIcrc2015}, reprinted from \citep{HuegePLREP}.}
    \label{fig:epscomparison}
\end{figure}

\begin{figure}[!htb]
\centering
\includegraphics[width=0.49\textwidth]{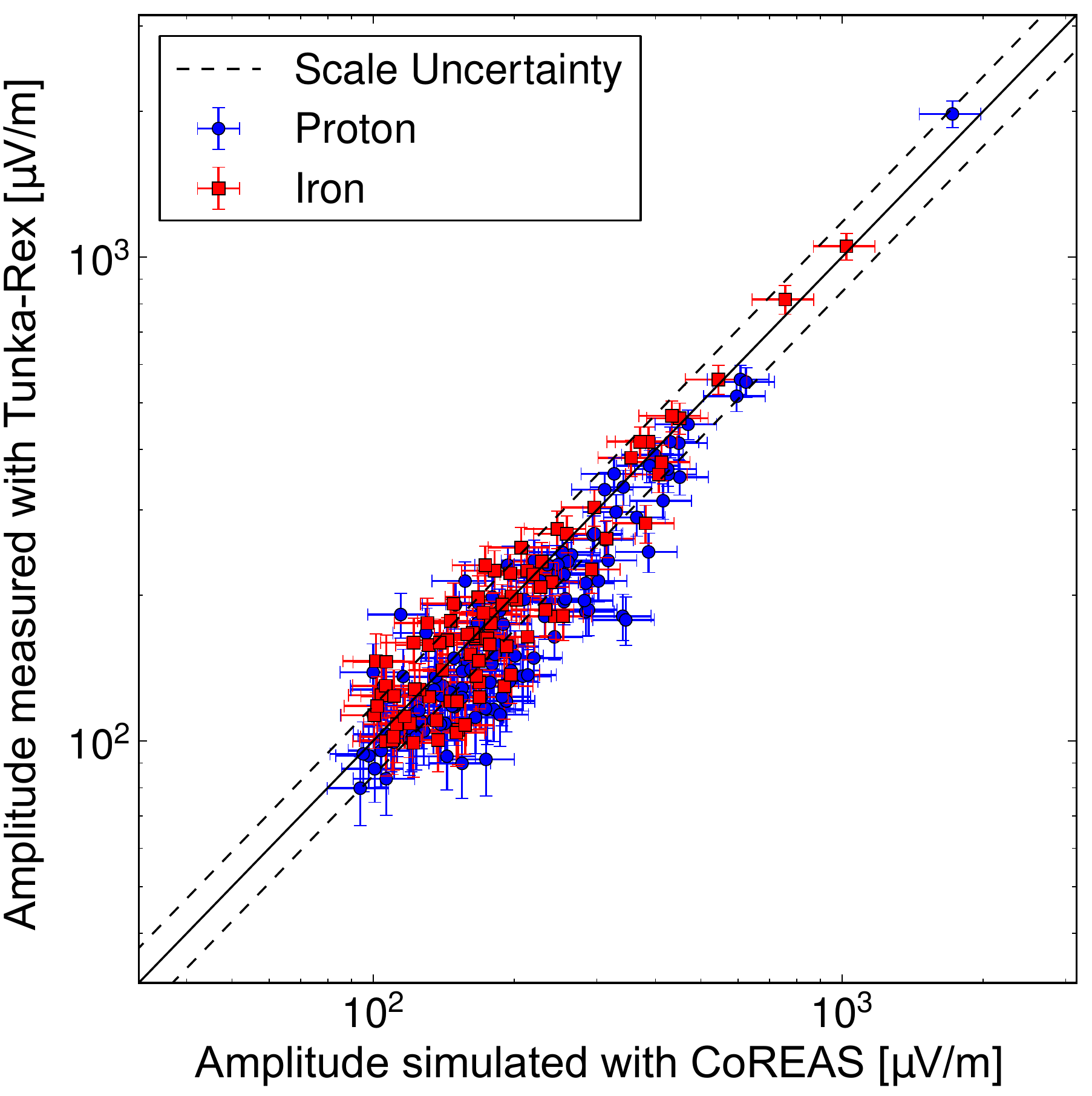}
\caption{Comparison of electric-field amplitudes measured in 
individual Tunka-Rex antennas with 
corresponding CoREAS simulations. Each event is simulated once for a proton primary and once for an iron primary. Adapted from 
\citep{TunkaRexInstrument}, reprinted from \citep{HuegePLREP}.\label{fig:tunkarexvscoreas}}
\end{figure}

An accurate prediction of the absolute electric field amplitudes by radio-emission simulations is of particular importance, as these can be used to set the absolute energy scale of a cosmic-ray detector.
In the 1970s, electric field amplitudes measured by different groups deviated by orders of magnitude, and quantitative comparisons with emission models were not available.
Today, measurements are well-calibrated with typical systematic uncertainties on the order of 15\%.
Simulations with full Monte Carlo codes, devoid of any free parameters, can indeed reproduce the measured signal amplitudes within this level of uncertainty. 
This is demonstrated in Fig.\ \ref{fig:epscomparison} for measurements made with LOPES and for Tunka-Rex measurements in Fig.\ \ref{fig:tunkarexvscoreas}, both of which are compared with CoREAS simulations.
Both experiments share the same calibration, which is also used by LOFAR \citep{LOFARCalibration}, and both are consistent within systematic uncertainties with the full Monte Carlo simulations.

LOFAR, with its hundreds of measurements for each individual air-shower detection, can test the shape of the lateral signal distribution, including its asymmetry and the presence of a Cherenkov bump, with high accuracy. (The LOFAR absolute amplitude scale has so far not been finalized; the energy resolution of the associated particle detector array is also worse than those of KASCADE, Tunka and AERA.) One such example comparison is shown in the middle panel of Fig.\ \ref{fig:xmaxlofar}.
LOFAR has detected hundreds of such events, and there is very good agreement between the measurements and CoREAS simulations.
We can thus be very confident that the radio-emission physics is well-understood by now, at least on the level of 10-15\%.

\begin{figure}[!htb]
  \centering
  \includegraphics[width=0.35\textwidth]{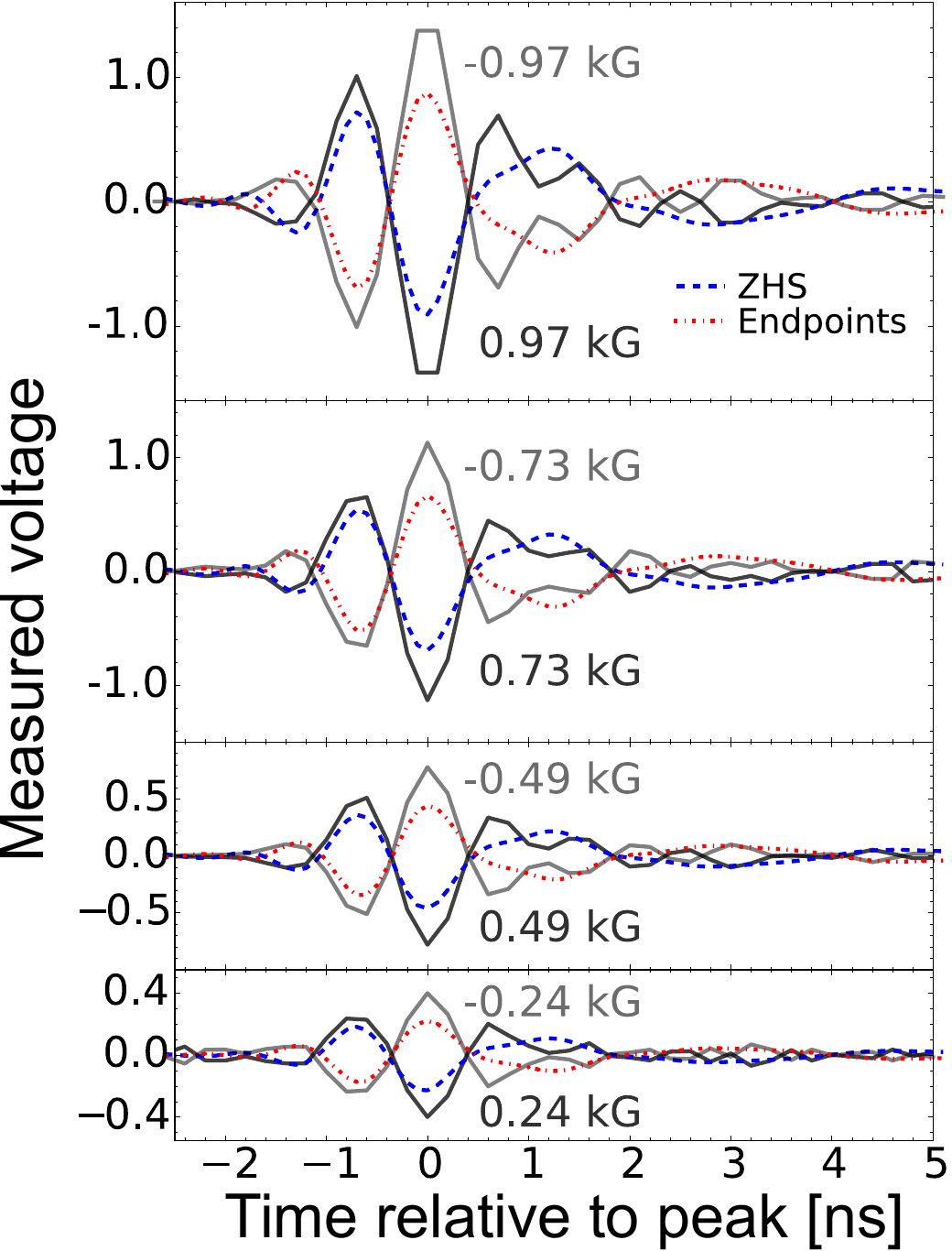}
  \caption{Radio pulses (Voltage at oscilloscope) resulting from the magnetic emission contribution of an electromagnetic particle shower in a high-density polyethylene target as measured by the SLAC T-510 experiment for various magnetic field configurations. The comparison with predictions from microscopic simulations shows good qualitative agreement. The deviations of the signal amplitudes are within  the systematic uncertainties of the measurement. Adapted from \citep{SLACT510-PRL}, reprinted from \citep{HuegePLREP}.
  \label{fig:t-510result}}
 \end{figure}

Further confidence in our understanding of the radio-emission physics is inspired by measurements under laboratory conditions.
In the SLAC T-510 experiment \citep{SLACT510-PRL}, a well-defined electron beam was shot in a target of high-density polyethylene, which was embedded in a strong, tunable magnetic field.
Detailed microscopic Monte Carlo simulations \citep{AnneT510ARENA2016} could reproduce the measurements well.
This includes details such as the frequency-dependent position and width of the Cherenkov cone, the decoupling of the Askaryan and magnetic emission contributions via the signal polarization and the linear scaling and polarity of the magnetic emission with the applied magnetic field, cf.\ Fig.\ \ref{fig:t-510result}.
The absolute amplitudes predicted by two different microscopic formalisms (the endpoint formalism \citep{JamesFalckeHuege2012} and the Zas-Halzen-Stanev formalism \citep{ZHS}) were found to be in agreement to within $\approx 5\%$.
The simulated pulses underpredict the measured ones by $\approx 35$\%, but this effect is explainable by reflections on the bottom surface of the target which were not fully controlled in the experimental setup.
Future measurements have the potential to lower the systematic uncertainty to a level of 10\% or lower.

\subsubsection{Reconstruction of cosmic-ray parameters}

There are three main reconstruction parameters relevant for cosmic-ray detection.
The first and easiest is the arrival direction, which can be reconstructed from the arrival times of the radio pulses in individual radio antennas.
Under a plane-wave assumption, the arrival direction can be reconstructed with a precision of 1-2$^{\circ}$.
Using a more refined wavefront model (e.g., a hyperbolical wavefront), precisions of well below 0.5$^{\circ}$ have been achieved \citep{HuegePLREP,SchroederReview}

\begin{figure}[htb]
  \centering
  \includegraphics[width=0.55\textwidth]{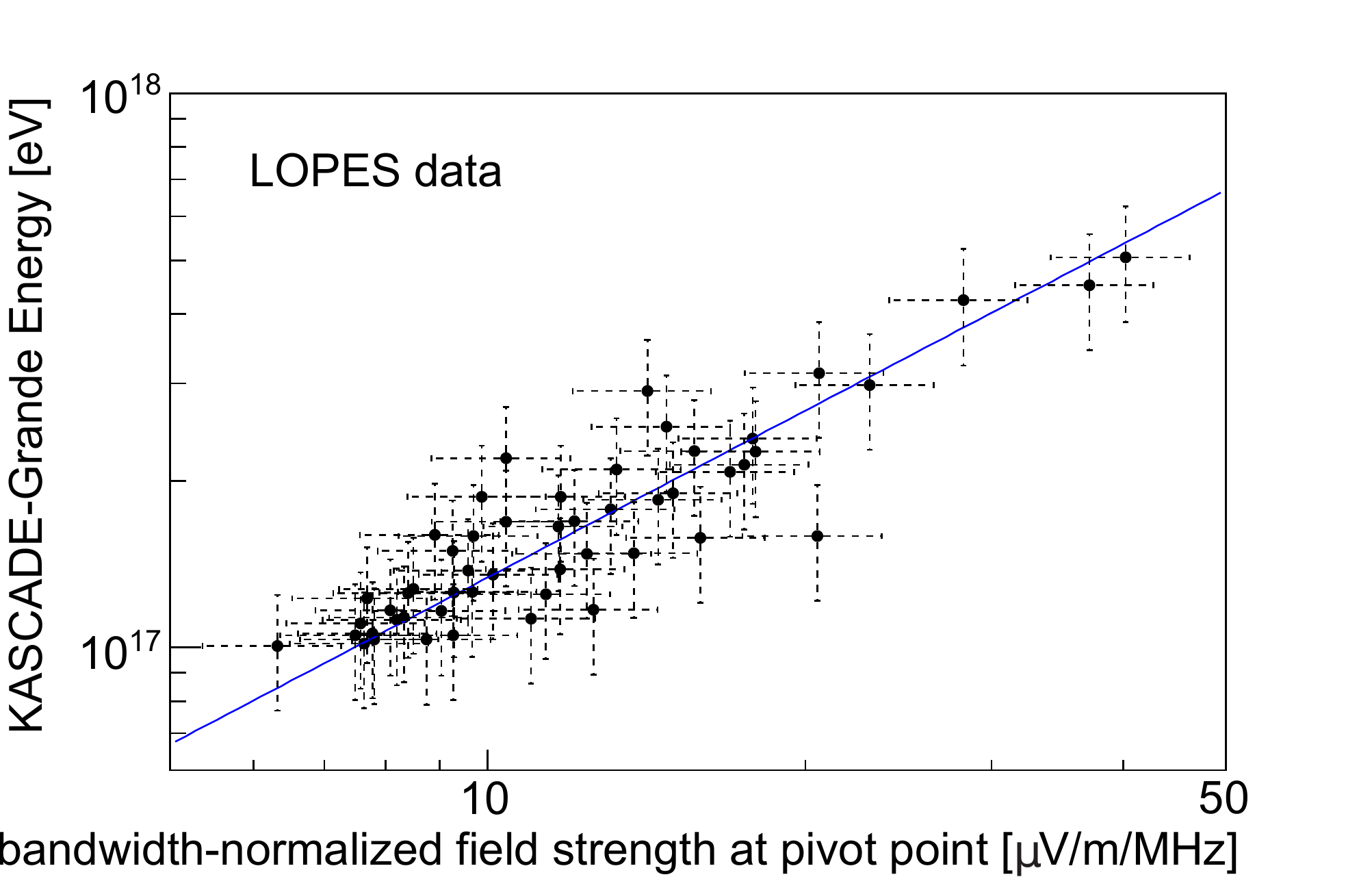}
  \includegraphics[width=0.44\textwidth]{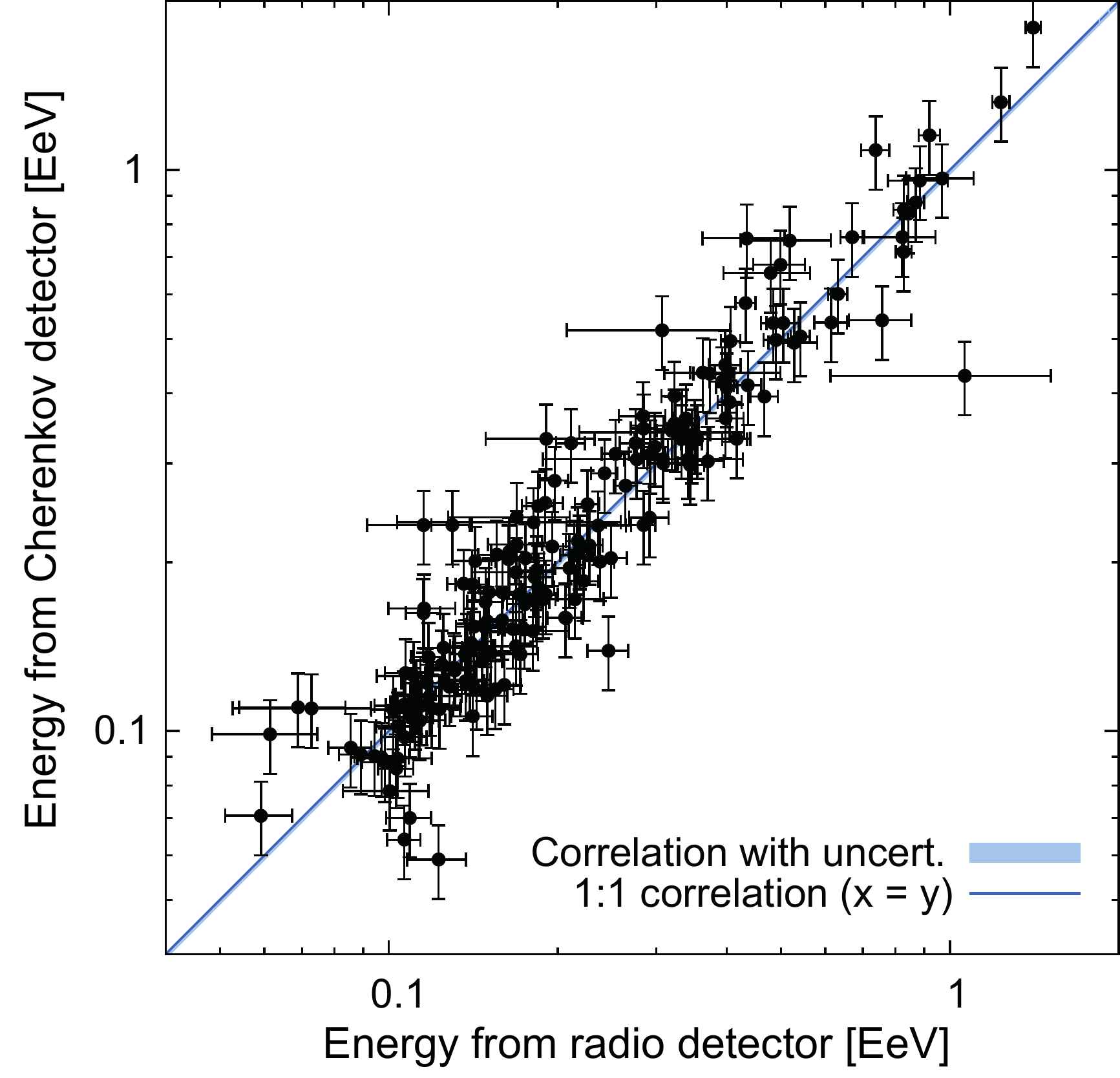}
  \caption{Left: Correlation between the radio amplitude measured by LOPES at the characteristic
  lateral distance and the energy reconstructed by KASCADE-Grande.
  Amplitudes have been normalized appropriately for the angle
  to the geomagnetic field. These amplitudes still have to be scaled down
  by a factor of 2.6 due to the later-revised calibration of the experiment.
  Adapted from \citep{ApelArteagaBaehren2014}, reprinted from \citep{HuegePLREP}.
  Right: Correlation of the radio-based energy estimator of Tunka-Rex, 
  in comparison with the energy reconstructed with the Tunka-133 optical Cherenkov detectors. Adapted from 
\citep{TunkaRexCrossCalibration}, reprinted from \citep{HuegePLREP}.\label{fig:lopestunkaenergy}
  }
\end{figure}

Reconstruction of the energy of the primary cosmic ray was achieved with good accuracy and high precision fairly early on in the development of the field.
This exploits the coherence of the radio emission at frequencies below $\approx 100$~MHz.
As long as coherence is fulfilled, the amplitudes of the measured electric fields scale linearly with the number of electrons and positrons, which in turn scales approximately linearly with the energy of the cosmic-ray primary. (The power radiated in the form of radio signals correspondingly scales quadratically with the energy of the cosmic-ray.)
Measured amplitudes need to be corrected for $\sin \alpha$, the sine of the geomagnetic angle, to take into account the direction dependence of the dominant geomagnetic emission component.
To reach maximum precision, one would like to measure the amplitude in a way that is least sensitive to fluctuations of $X_{\mathrm{max}}$.
In fact, this can be achieved with an amplitude measurement at a characteristic lateral distance (which is dependent on observation altitude) from the shower axis  and typically is of order 100~m.
This characteristic behavior was shown early on in simulation studies \citep{HuegeUlrichEngel2008}, and was later confirmed both theoretically and experimentally.
The earliest quantitative analysis following this approach was presented by LOPES and is reproduced here in the left panel of Fig.\ \ref{fig:lopestunkaenergy}.
The energy resolution achieved with this analysis was of order 20-30\%.
This includes the combined energy resolution of the LOPES radio measurements and the KASCADE-Grande particle measurements; the latter individually amounts to approximately 20\%.
The intrinsic fluctuations in the radio amplitude expected from simulations are very low, well below 10\% and possibly even below 5\%, which means that there is still room for improvement, if the 
experimental uncertainties can be reduced.
A similar analysis was performed by Tunka-Rex, with the added improvement that asymmetries due to the Askaryan charge-excess component are corrected for explicitly in Tunka-Rex while they were only averaged out implicitly in the LOPES analysis.
The achieved resolution, combined for the Tunka Cherenkov and Tunka-Rex radio detectors, is illustrated in the right panel of Fig.\ \ref{fig:lopestunkaenergy}, corresponding to
approximately 20\% and again largely dominated by the 15\% uncertainty of the Tunka Cherenkov-light detectors.

\begin{figure}[!htb]
\centering
\includegraphics[width=0.54\textwidth]{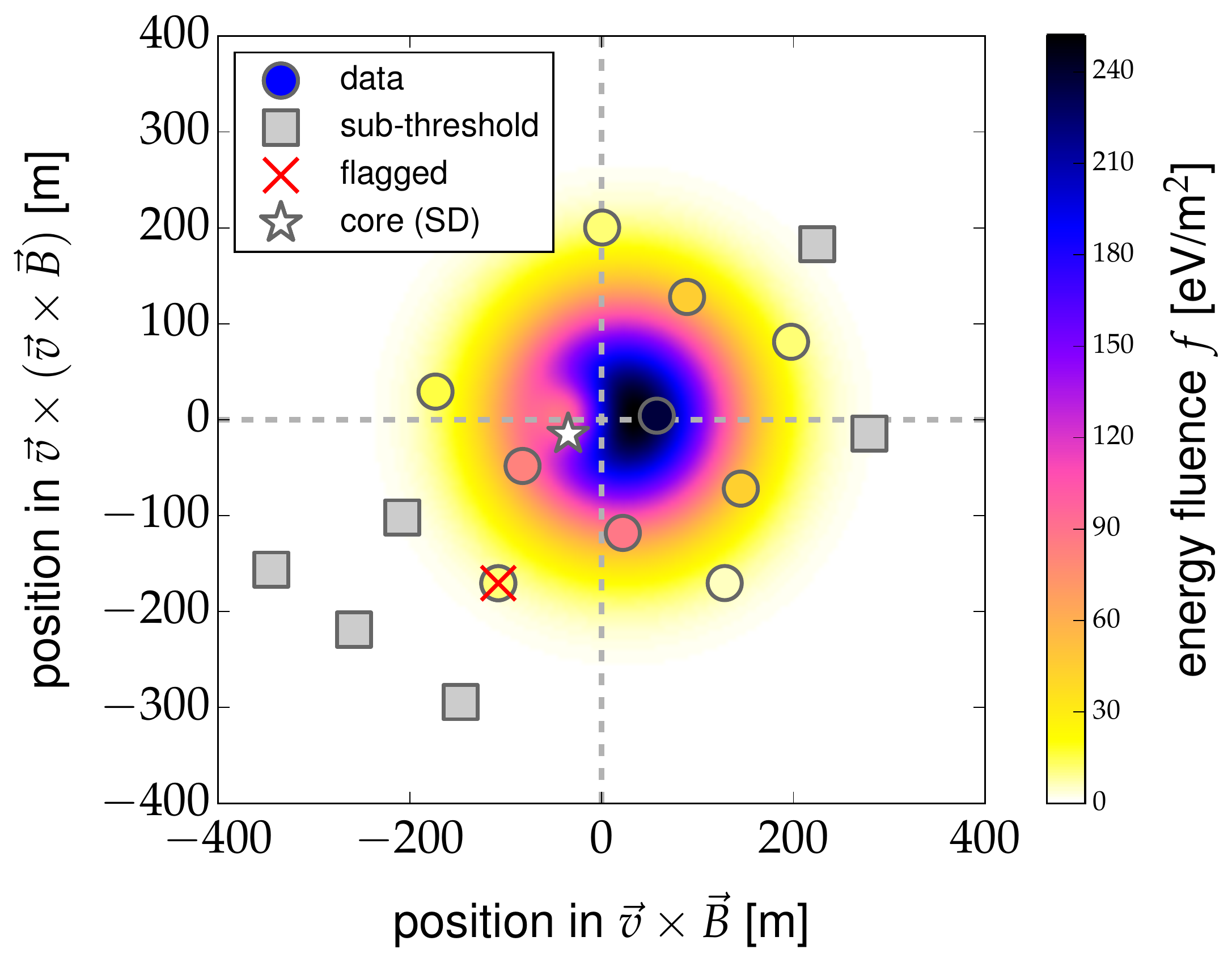}
\includegraphics[width=0.45\textwidth]{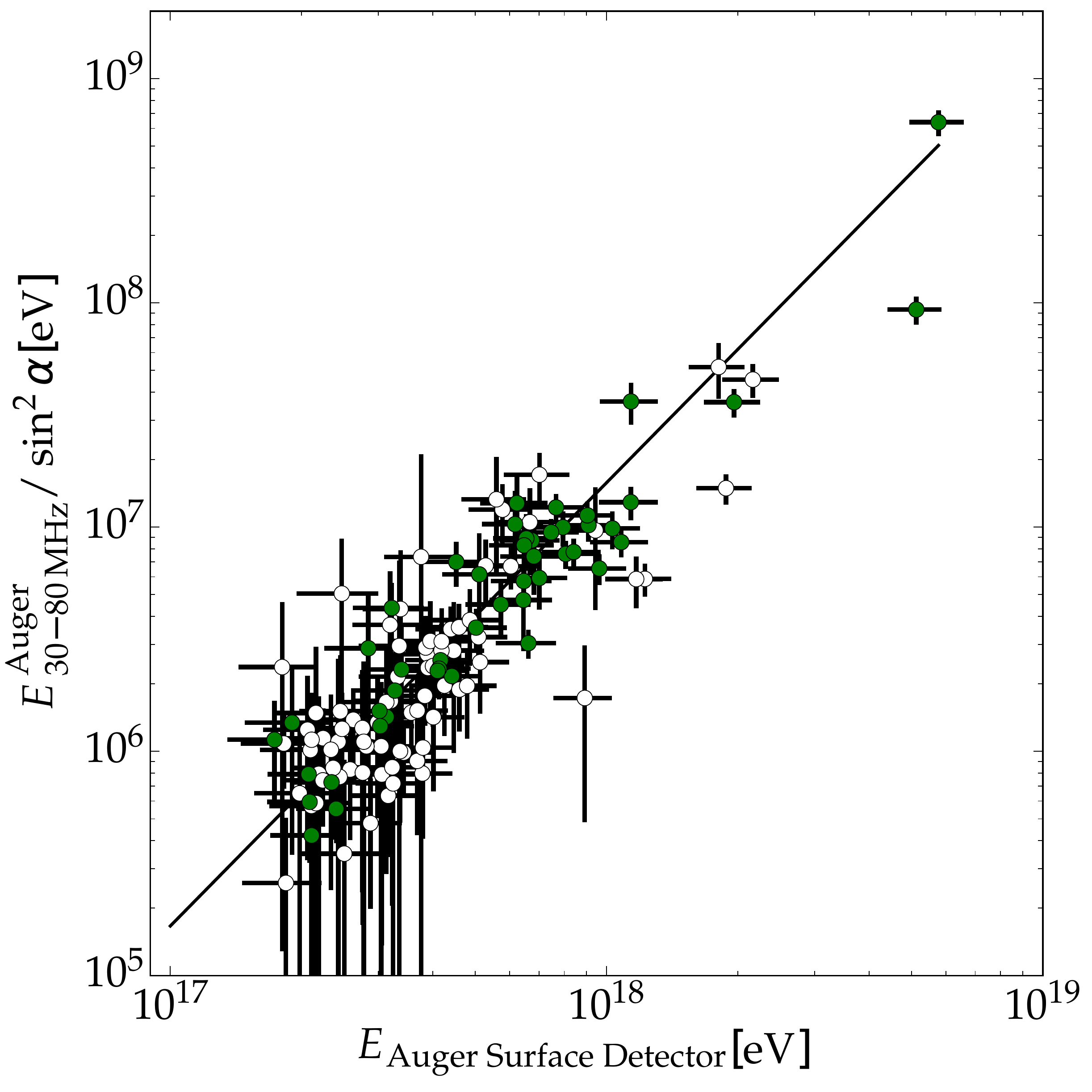}
\caption{Left: Energy fluence measured with individual AERA antennas and
their fit with a two-dimensional lateral signal distribution
function. Both antennas with a detected signal (\emph{data}) as
well as antennas with a signal below detection threshold 
(\emph{sub-threshold}) are included in the fit. Measurements with
polarisation characteristics deviating from expectations are excluded (\emph{flagged}) 
to suppress transient radio-frequency interference.
The best-fitting core position of the air shower is at the origin of the plot, 
slightly offset from that reconstructed with the Auger surface 
detector (\emph{core (SD)}). The background-color map denotes the two-dimensional 
signal distribution fit. For an ideal fit, the color of the data points blends in with the background-color map.
Adapted from \citep{AERAEnergyPRD}, reprinted from \citep{HuegePLREP}.
Right: Correlation of the radiation energy, normalized for 
perpendicular incidence to the geomagnetic field, with the cosmic-ray energy
measured by the Auger surface detector. Open circles denote air showers with radio signals 
detected in three or four AERA antennas. Filled circles 
denote showers with five or more detected radio signals. Adapted 
from \citep{AERAEnergyPRD}, reprinted from \citep{HuegePLREP}.\label{fig:aeraenergy}}
\end{figure}

A more universal concept for the determination of the particle energy was recently introduced by AERA.
Here, it is not the electric-field amplitude at a characteristic lateral distance that is used as an energy estimator.
Instead, the total energy deposited on the ground in the form of radio emission is reconstructed from the measured radio signals.
This means that the electric field measured at a particular location is converted to the Poynting flux and then integrated in time to determine the energy fluence (eV/m$^{2}$) at this location.
Afterwards, a two-dimensional signal distribution model \citep{NellesLDF,AERAEnergyPRD} which takes into account the asymmetry and the Cherenkov bump is fit to the energy fluence measurements, and then the total signal distribution is integrated over area to determine the total energy in the radio signal, the \emph{radiation energy}.
This procedure is illustrated for a sample event in the left panel of Fig.\ \ref{fig:aeraenergy}.
The radiation energy needs to be corrected by $\sin^{2} \alpha$, following which it 
scales quadratically with the cosmic-ray energy, as is shown in the right panel of Fig.\ \ref{fig:aeraenergy}.
The achieved resolution for the radio measurement alone was quantified to be 17\% in the AERA measurement, and is thus of roughly equal precision as that of the amplitude-based methods.
However, the radiation energy has a strong conceptual advantage: in contrast to the amplitude measurement, where both the absolute amplitudes and the optimal characteristic distance at which to measure
are dependent on observation altitude, the radiation energy is independent of observation altitude, since
the radio emission in air showers undergoes no significant absorption in the atmosphere.
The radiation energy thus gives a direct measurement of the energy in the electromagnetic component of the air shower, very similar to the integral of the longitudinal air shower evolution profile achieved with fluorescence detectors.
It is thus very well-suited to cross-calibrate detectors -- either radio detectors against each other or radio detectors against other detectors \citep{AERAEnergyPRL}.
According to the latest AERA results \citep{AERAEnergyPRL}, the radiation energy amounts to
\begin{eqnarray}
 E_{30-80\,{\mathrm{MHz}}} &=& \left(15.8~ \pm 0.7\,\mathrm{(stat)} \pm 6.7 
\,\mathrm{(sys)}\right)\,\mathrm{MeV} \nonumber \\ &\times& \left(\sin\alpha 
\,\frac{E_\mathrm{CR}}{10^{18}\,\mathrm{eV}} \, \frac{B_\mathrm{Earth}}{0.24\,\mathrm{G}} \right)^2.
\end{eqnarray}
An extensive air shower with a primary energy of $10^{18}$~eV arriving perpendicular to a geomagnetic field with a strength of 0.24~G thus radiates 15.8~MeV in the form of radio signals in the frequency range from 30 to 80~MHz.
This is a minute fraction of the energy of the primary cosmic ray, yet it can be measured reliably and without any limitations or fluctuations by photon statistics (the average photon energy is of order 10$^{-7}$~eV).
The systematic uncertainty on this measurement stems in equal parts from the uncertainty on the absolute energy scale of the Pierre Auger Observatory (currently set by the Auger fluorescence detector) and the absolute calibration of the logarithmic-periodic dipole antennas used in AERA, the latter of which still have room for significant improvement.

An important aspect that is closely related to the cosmic-ray energy measurement using radio measurements is the potential to independently determine the absolute energy scale of a cosmic-ray detector using radio antennas.
This is possible because the radio emission can be predicted on an absolute scale purely on the basis of the electromagnetic component of air showers, which is the component that is best-understood and least sensitive to hadronic interaction physics.
Activities in this direction have already begun \citep{AERAEnergyPRL} and will certainly be pushed further in the near future.

\begin{figure}[!htb]
\centering
\includegraphics[width=0.51\textwidth]{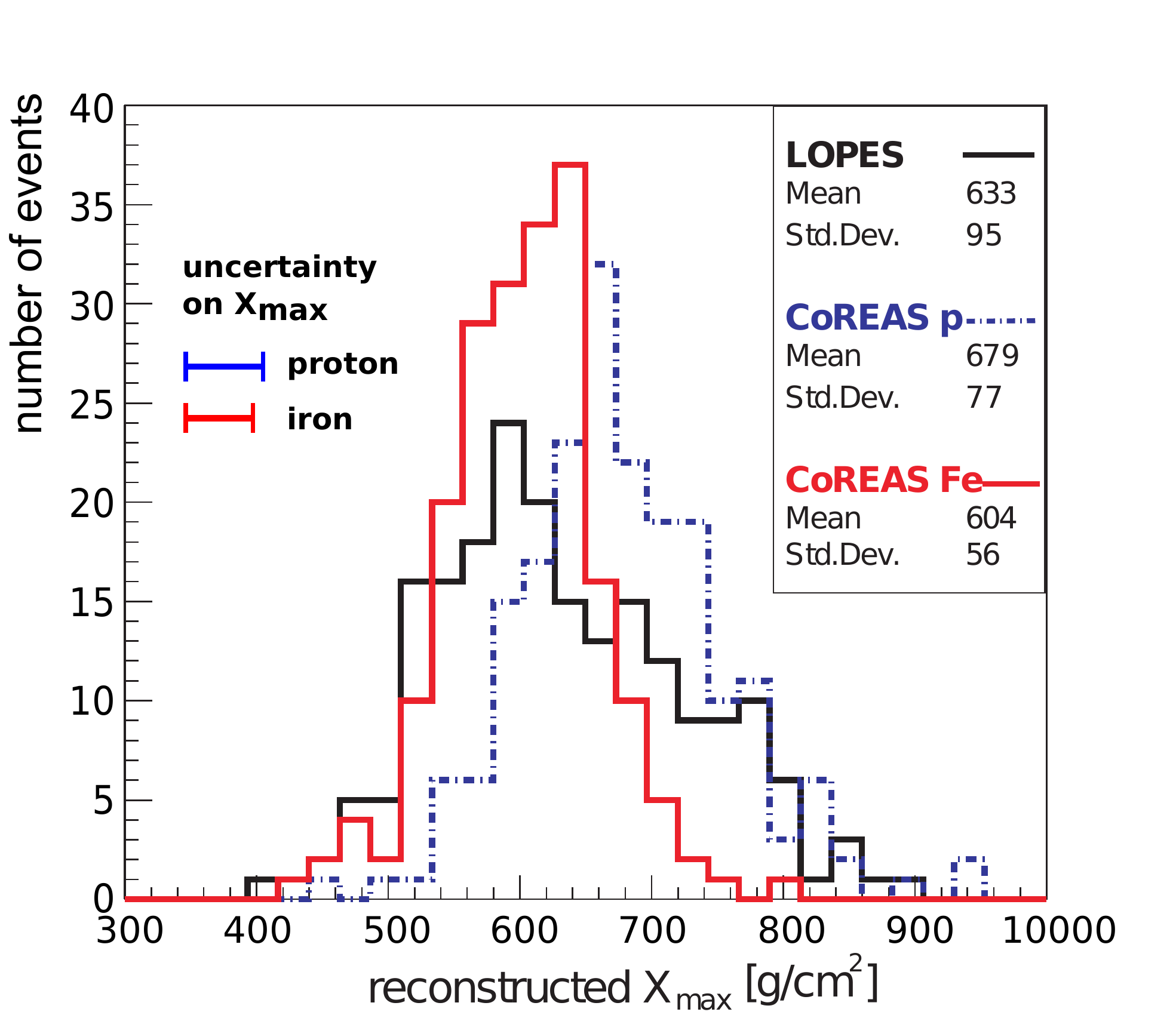}
\includegraphics[width=0.48\textwidth]{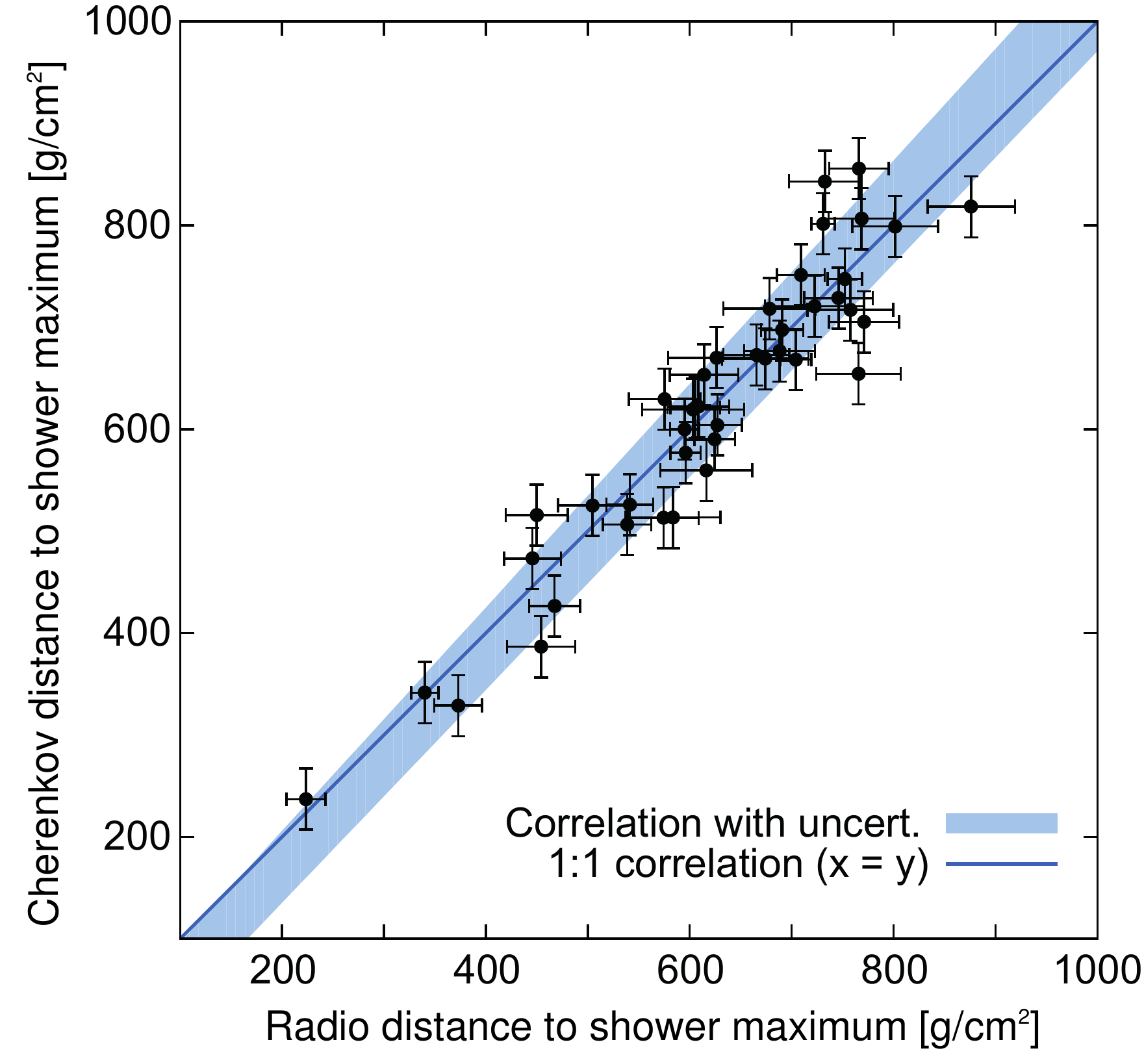}
\caption{Left: Distribution of air-shower \xmax values
  reconstructed from the slopes of the radio lateral distribution 
  functions measured with LOPES compared to the distributions
  of CoREAS simulations for proton and iron primaries. Adapted from
  \citep{ApelArteagaBaehren2014}, reprinted from \citep{HuegePLREP}.
  Right: Atmospheric depth in g/cm$^{2}$ between observer location and \xmax as determined from Tunka-Rex radio measurements and 
the Tunka-133 Cherenkov detectors. Adapted from \citep{TunkaRexCrossCalibration}, reprinted from \citep{HuegePLREP}.\label{fig:tunkarexxmax}}
\end{figure}

It was also evident early on that the radio signal of extensive air showers harbors sensitivity to the depth of shower maximum on an event-by-event basis \citep{HuegeUlrichEngel2008}.
The most important signature is the slope of the lateral signal distribution function: deeper showers have steeper lateral distributions than less-penetrating showers.
Other signatures exist in the opening angle of the radio-signal wavefront \citep{LOPESWavefront2014}, which can be determined with precise arrival-time measurements, and possibly in the spectral index of the measured frequency spectrum of the radio pulses \citep{GrebeARENA2012}.

The sensitivity of the lateral signal distribution to the depth of shower maximum has already been successfully 
exploited experimentally.
The proof of principle has been made by the LOPES experiment and is shown in the left panel of Fig.\ \ref{fig:tunkarexxmax}.
While the method was shown to work in principle, the achieved resolution was not competitive, and an independent cross-check with a direct \xmax measurement was not available.
This has now been achieved with Tunka-Rex, as is shown in the right panel of Fig.\ \ref{fig:tunkarexxmax}, where the slope of the radio signal distribution has been converted into an \xmax measurement and has then been compared with the independent measurement provided by the Cherenkov-light detectors.
The combined resolution of the two detectors amounts to $\sim 50$~g/cm$^{2}$, while the uncertainty of the Tunka-133 \xmax reconstruction alone is specified as 28~g/cm$^{2}$.
The resolution of the radio reconstruction can thus be estimated of order $\sim 40$~g/cm$^{2}$, although there is certainly still room for improvement in the analysis strategies.
Recent results from AERA show a similar agreement between the \xmax values reconstructed from radio data and the measurement with the Auger fluorescence detectors \citep{GateARENA2016}.

\begin{figure}[htb]
  \includegraphics[width=0.32\textwidth]{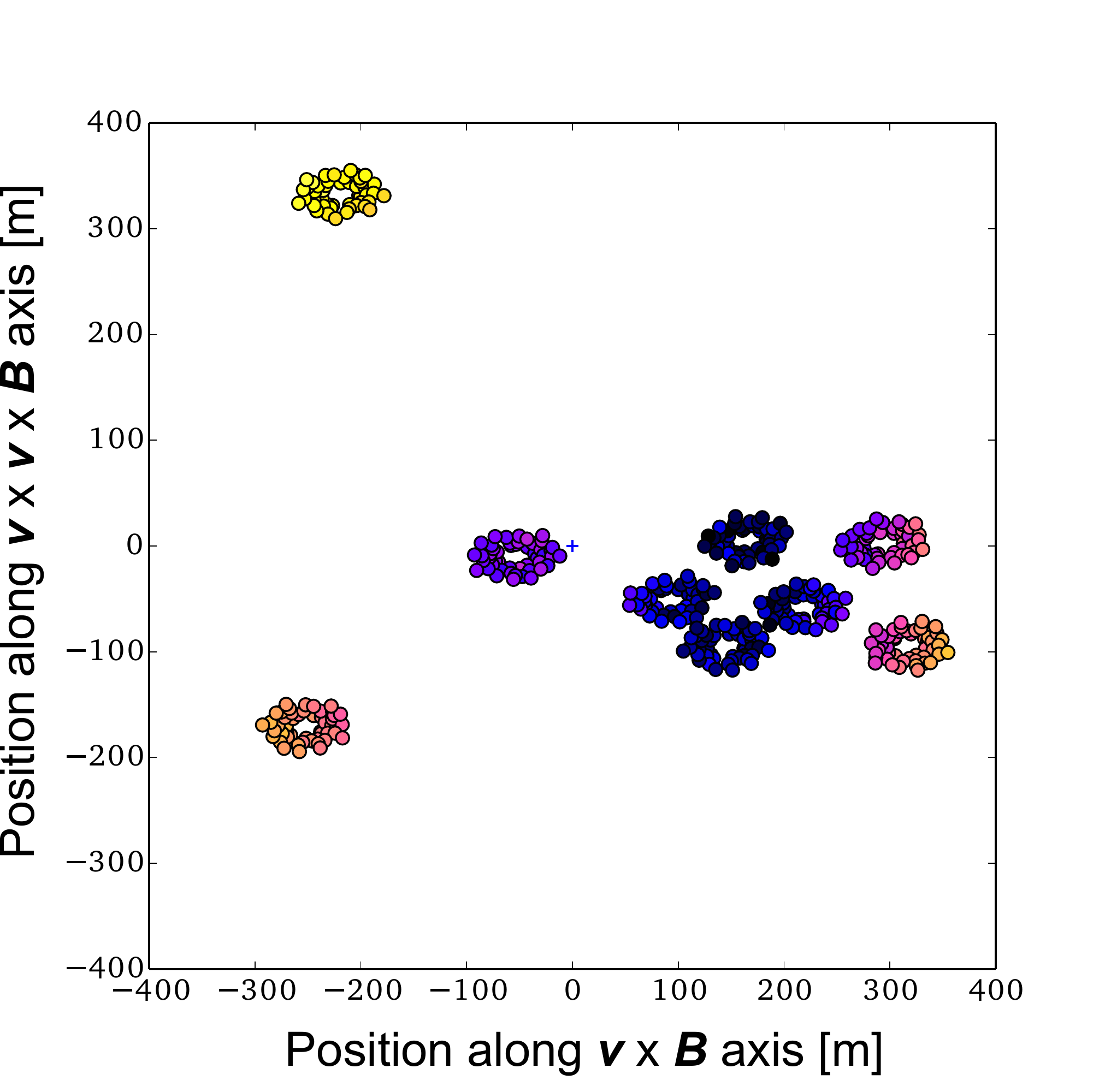}
  \includegraphics[clip=true,trim=50 0 45 0,width=0.35\textwidth]{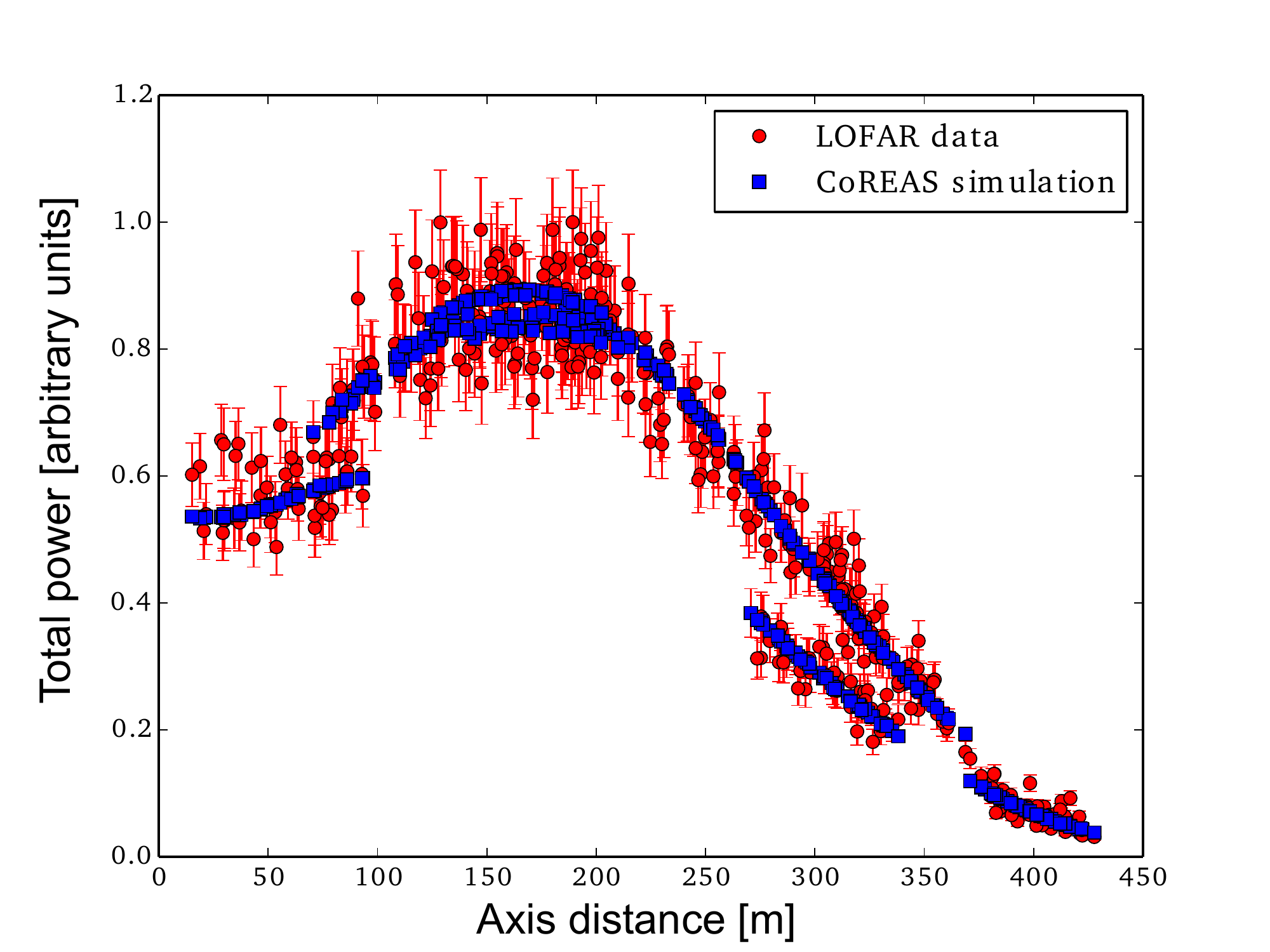}
  \includegraphics[width=0.31\textwidth]{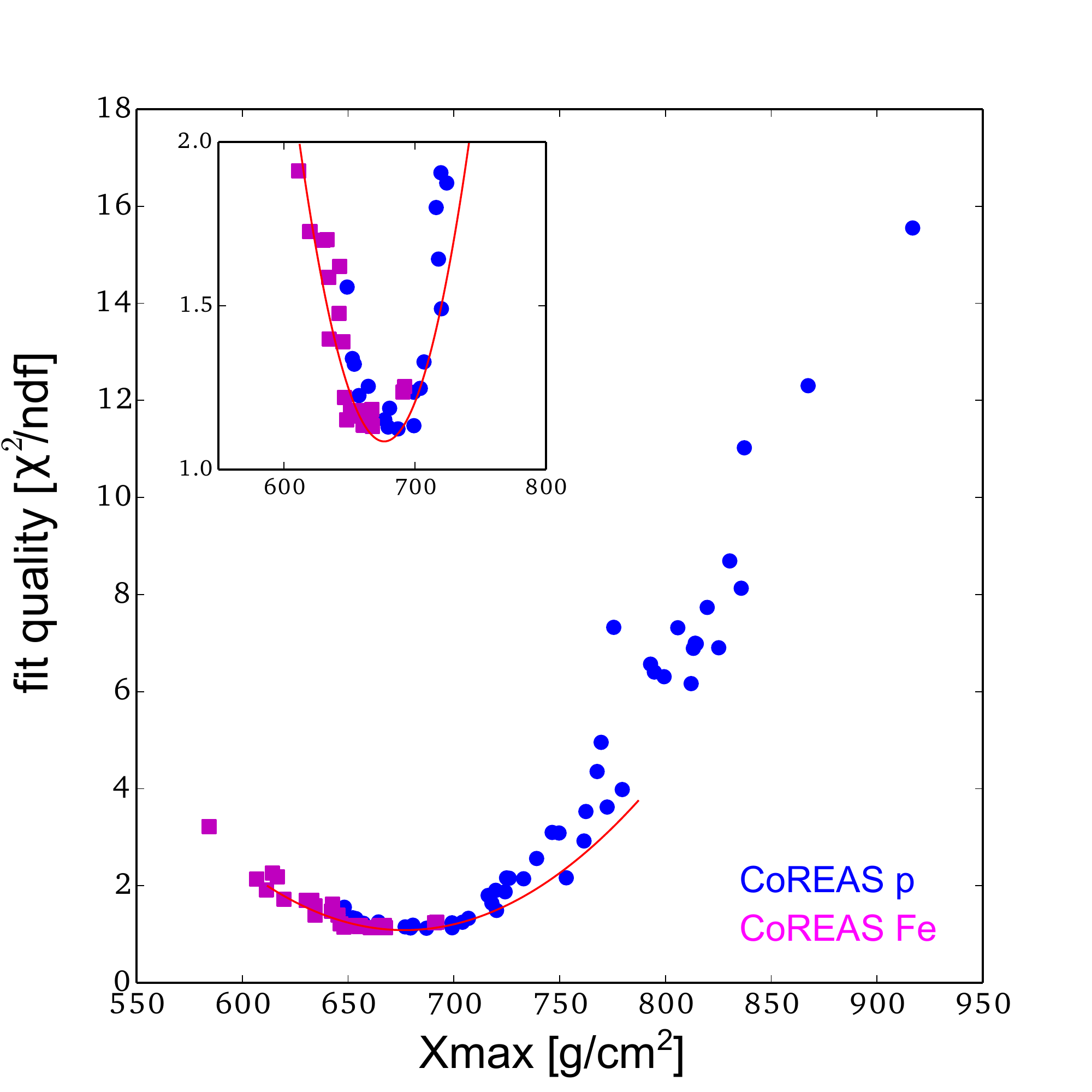}
  \caption{Left: Total power per area received at individual LOFAR 
  antennas (colored circles) compared with the two-dimensional 
  lateral signal distribution of the best-fitting  
  CoREAS simulations (background-color), for a specific air-shower 
  event. Depicted is the shower plane, defined 
  by the axes both along and also perpendicular to the direction of the Lorentz force.
Middle: One-dimensional projection of the 
  two-dimensional lateral signal distribution. Right: Quality of the 
  agreement between the total power distribution measured with LOFAR 
  and that predicted by an ensemble of CoREAS simulations of that air 
  shower event. The value of \xmax clearly governs the quality of the fit.
  All diagrams adapted from \citep{LOFARXmaxMethod2014}, reprinted from \citep{HuegePLREP}.}\label{fig:xmaxlofar}
\end{figure}

\begin{figure}[!htb]
\centering
\includegraphics[width=0.7\textwidth,clip=true,trim=0cm 0cm 0cm 16cm]{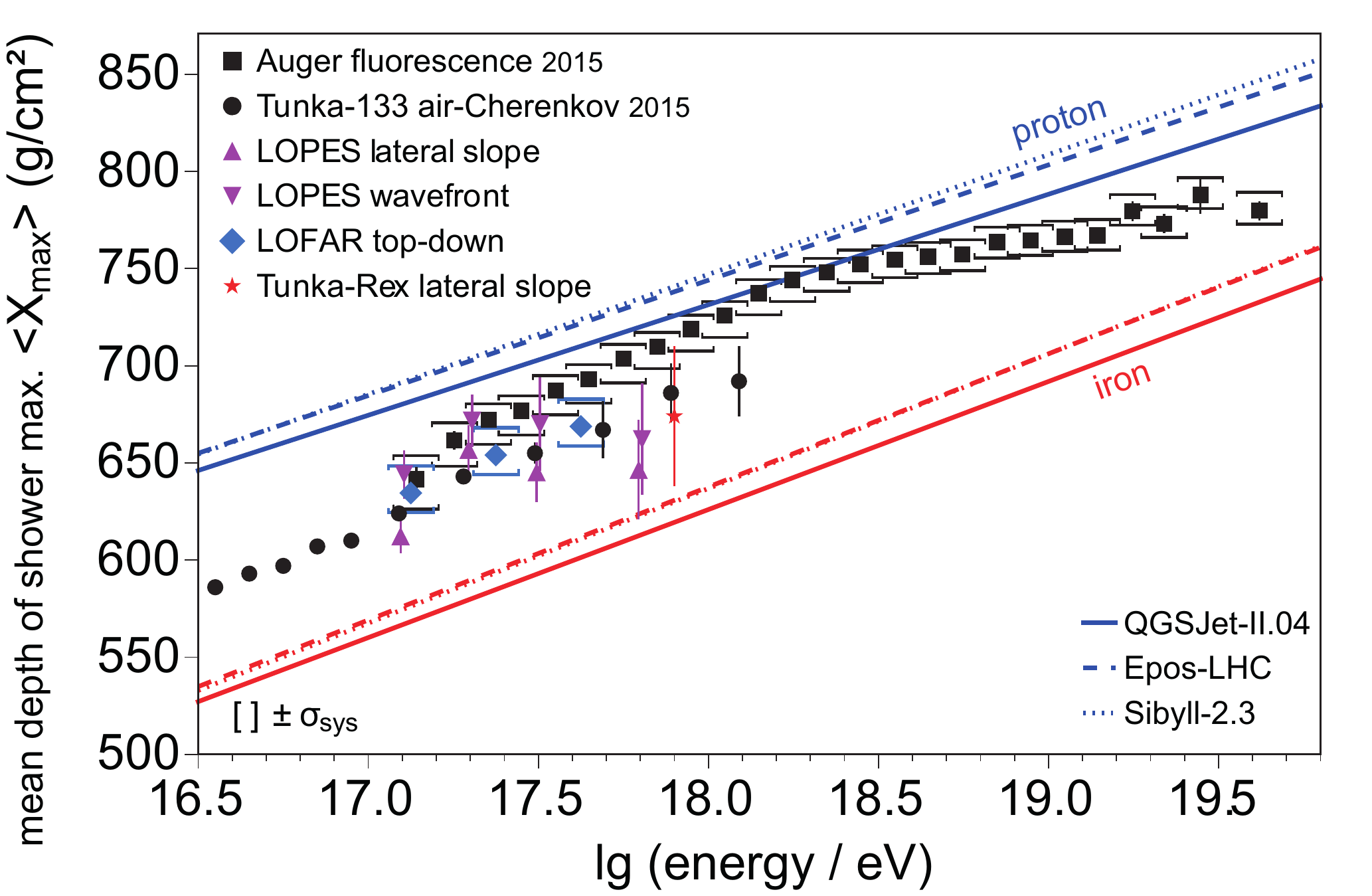}
\caption{Compilation of radio-based mean \xmax measurements as a function of cosmic-ray energy.
Please note that in some of these results, significant systematic uncertainties exist.
Reprinted from \citep{SchroederReview}.\label{fig:xmaxresults}}
\end{figure}

The most impressive \xmax reconstruction from radio data so far has been achieved with the very detailed measurements of individual air shower radio footprints by LOFAR.
A top-down simulation approach is used to determine the \xmax value for each individual air shower.
Dozens of CoREAS simulations of both iron- and proton-primaries are prepared for each measured event.
These are then shifted around in the shower plane to find the best-fitting core position.
The quantity that is compared between simulations and data is the total power measured in individual LOFAR antennas, as is shown in the left and middle panels of Fig.\ \ref{fig:xmaxlofar}.
The reduced $\chi^2$ value of the different simulations is then plotted versus their respective \xmax values, as shown in the right panel of Fig.\ \ref{fig:xmaxlofar}.
It turns out that the single quantity governing the agreement between simulations and data is \xmax, which can thus be read off the distribution with great precision.
The average \xmax precision achieved by LOFAR with this approach is $\approx 17$~g/cm$^2$.
Events where the non-uniform LOFAR antenna array samples the two-dimensional lateral signal distribution especially favorably can be reconstructed with an \xmax uncertainty of less than 10~g/cm$^2$.
Consequently, the upcoming SKA (Square Kilometer Array) with its much more uniform antenna density and wider observation band of 50-350~MHz should be able to reconstruct \xmax with a resolution well below 10~g/cm$^2$.
Further improvements may be realized by exploiting additional signal properties such as polarization, signal arrival time or pulse shape.
In Fig.\ \ref{fig:xmaxresults}, an elongation-rate plot based on radio measurements shows that the current \xmax reconstruction results are in agreement with those of classical detection methods.

In summary, radio detection has now achieved competitive resolutions for the reconstruction of direction, energy and depth of shower maximum of individual air showers.
The technique thus has clearly evolved from a pioneering phase to a solid detection technique that can be routinely applied in combination with other detection systems and provides valuable additional information in hybrid observation modes.

\subsubsection{Higher-frequency radio measurements}

The main focus of air-shower radio detection is in the band from approximately 30 to 80 MHz.
At these frequencies, the signals are generally coherent and the background is dominated by Galactic noise.
Below 30~MHz, short-wave band emissions and atmospheric noise make measurements generally difficult.
Between 80~MHz and 110~MHz, FM radio transmissions become problematic.
Above 110~MHz, Galactic noise decreases significantly, and in principle radio detection of air showers is feasible.
However, the radio signal at high frequencies is generally only coherent for observer positions on the Cherenkov ring.
The geometry for successful detection thus becomes rather specific, requiring dense antenna arrays for a high detection efficiency.
Nevertheless, successful detection at high frequencies has been made by several experiments.

LOFAR has measured radio emission from extensive air showers in the 110-200~MHz band \citep{Nelles201511}.
As expected, the radio-emission footprint has a clear Cherenkov ring, the diameter of which provides information on \xmax.

\begin{figure}[!htb]
\centering
\includegraphics[width=0.47\textwidth,clip=true,trim=0cm 8cm 0cm 5cm]{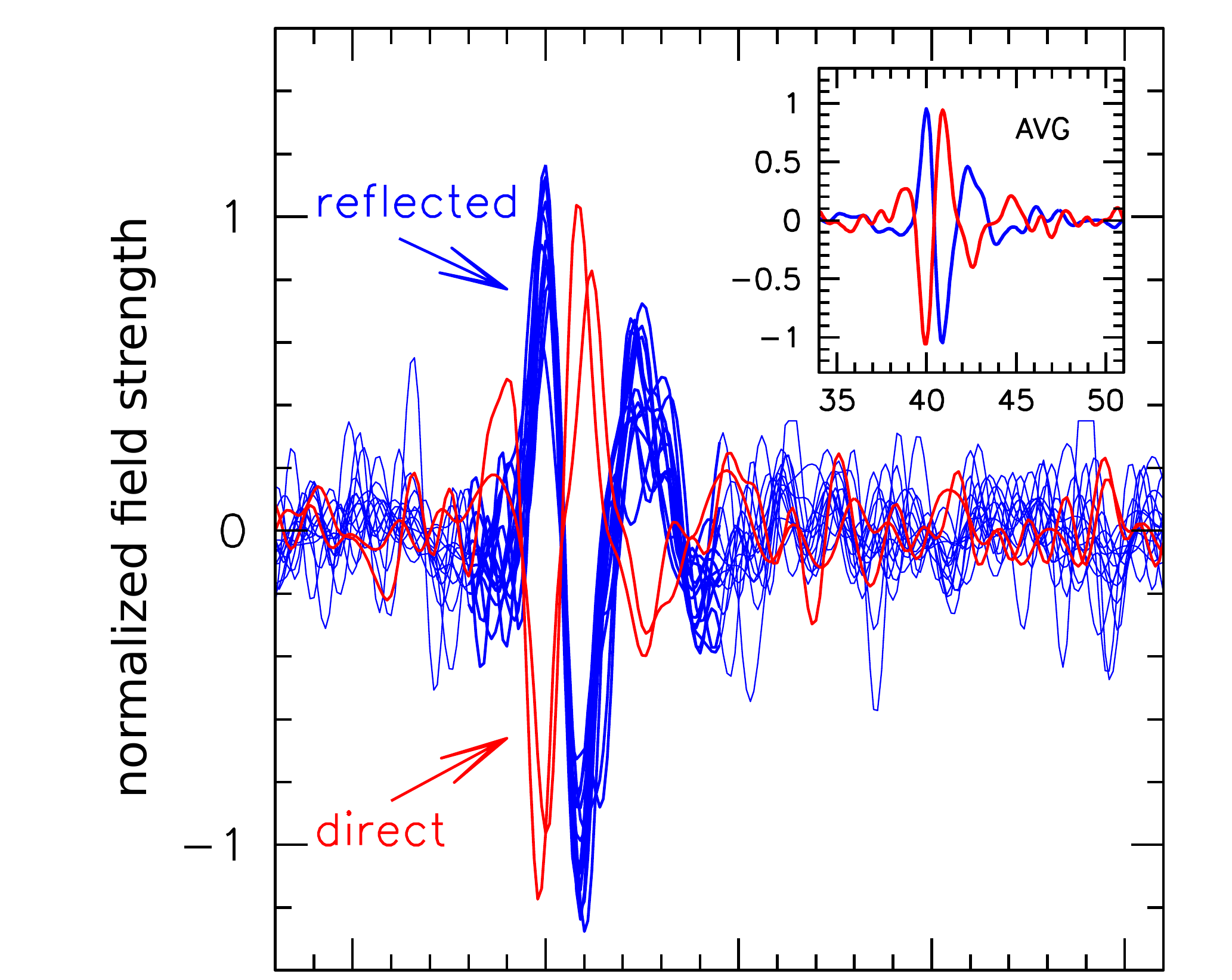}
\includegraphics[width=0.52\textwidth]{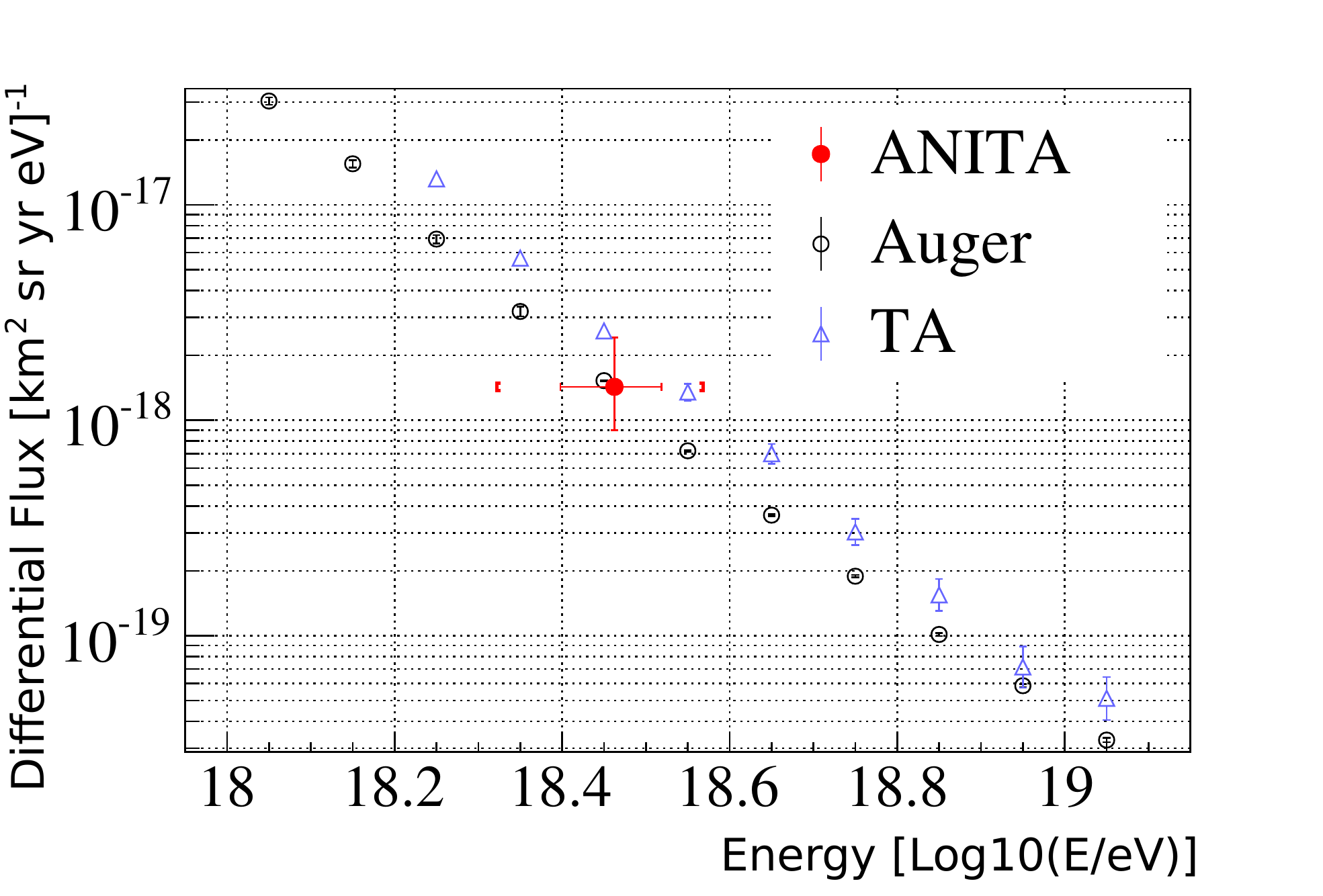}
\caption{Left: Radio pulses from 16 extensive air showers measured during the 
ANITA-I flight. 14 signals have been reflected off the ice surface of Antarctica;
2 signals result from Earth-skimming air showers. Adapted from 
\citep{HooverNamGorham2010}, reprinted from \citep{HuegePLREP}.
Right: Cosmic-ray flux as determined from the 14 events measured 
with the ANITA-I balloon flight. Adapted from \citep{ANITAEnergy}, reprinted from \citep{HuegePLREP}.\label{fig:anitaflux}}
\end{figure}

ANITA, originally intended to measure Askaryan emission from in-ice neutrino-induced cascades, has observed pulsed radio signals at 300-1200~MHz that are by now understood as air-shower signals reflected off the ice or from Earth-skimming events.
Such events are only detectable if the geometry is favorable, i.e., the ANITA antennas happen to lie on the (reflected) Cherenkov cone.
The emission can indeed be explained consistently by the same microscopic full Monte-Carlo simulations that were originally developed for frequencies below 100~MHz.
On the basis of such simulations, the energy of the individual events (left panel of Fig.\ \ref{fig:anitaflux}) could be determined and even a cosmic-ray flux could be calculated (right panel of Fig. \ref{fig:anitaflux}).

\begin{figure*}[ht!]
\centering
\includegraphics[width=0.8\textwidth]{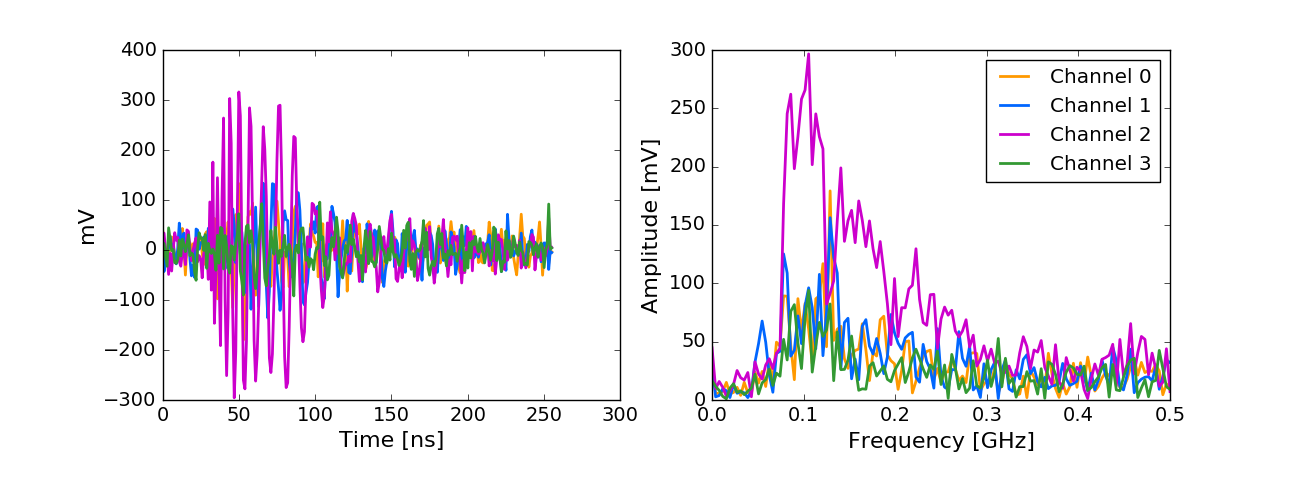}
\caption{Four-channel waveform recorded by ARIANNA Station 32, comprised of 4 upward-facing log-periodic dipole antennas (LPDA). Signal pattern and frequency content are consistent with downward-coming RF signal due to air showers  \citep{2016arXiv160907193N}.}
\label{fig:ARIANNA_cr}
\end{figure*}

\begin{figure}[htpb]
\centering
\includegraphics[width=0.7\textwidth]{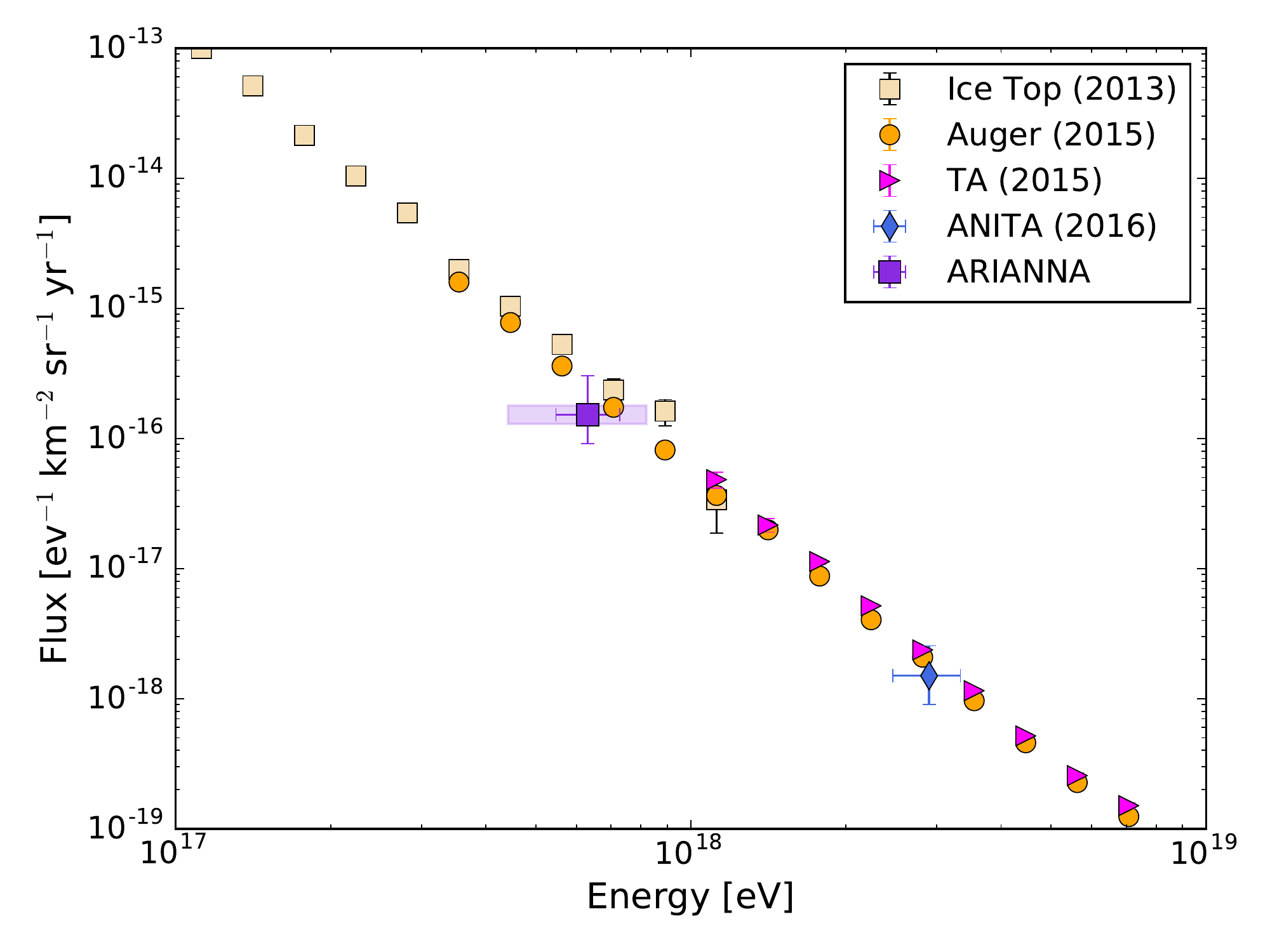}
\caption{ARIANNA cosmic ray flux vs. energy, compared with other experimental efforts \citep{2016arXiv160907193N}.} \label{fig:ARIANNAflux}
\end{figure}

The ARIANNA experiment \citep{Barwick:2014rca,Barwick:2014pca,Barwick:2014rga}, located on the Ross Ice Shelf of Antarctica, and, like ANITA, initially purposed for detection of cosmic neutrinos, has very recently also reported detection of radio emissions from approximately 40 cosmic ray-initiated air showers \citep{2016arXiv160907193N}. Figure \ref{fig:ARIANNA_cr} shows the four-channel waveform of a cosmic ray candidate detected in their station 32, deployed with upwards-facing log-periodic dipole array (LPDA) antennas, sensitive at 110-900 MHz frequencies.
The implied cosmic ray flux is commensurate with measurements by dedicated experiments such as Auger (Figure \ref{fig:ARIANNAflux}). 
If the full 1296-station (8 antennas per station spread over a 36 km x 36 km grid) ARIANNA would be realized in the future, projected event rates are of order 15,000 cosmic rays detected per year.

Finally, even at frequencies from 3.4-4.2~GHz the CROME experiment could measure air-shower radio emission which is strongly beamed along a Cherenkov cone and whose polarization characteristics are in agreement with predictions from CoREAS \citep{CROMEPRL}.
Isotropic ``Molecular Bremsstrahlung'' that would be a promising target for fluorescence-detection-like radio detectors was, however, not found by any experiment after the initial claim of its measurement at SLAC \citep{GorhamMBR}.
Neither air shower experiments such as CROME \citep{CROMEPRL}, MIDAS \citep{MIDAS}, EASIER and AMBER \citep{GHzAtAuger} nor the accelerator-based experiments AMY \citep{AMY}, MAYBE \citep{MAYBE} nor the Telescope Array Electron Light Source experiment \citep{ELSMBRIcrc2015} were able to find a signal.
Refined simulations of the signal strength confirm that the signal, if present, must be much weaker than originally expected (see ref.\ \citep{AlSamarai201526} and references therein).
Prospects for the use of ``radio-fluorescence'' techniques are thus very bleak.
A report of a moderately forward-beamed radio emission at 11~GHz from a 95~keV electron beam in air \citep{ContiSartori} should, however, be investigated further.

%----------------------------------------------------------------

\subsection{Future prospects}

With the successes of the past decade, the capabilities but also limitations of radio detection of cosmic rays have become increasingly clear.
The original hope that radio detectors could fully replace fluorescence detectors for ultra-high-energy cosmic rays, providing the same information with a ten-fold duty cycle, no longer seems realistic.
The main reason is the intrinsic beaming of the emission, which leads to fairly limited extent of the illuminated area, and thus requires rather dense antenna arrays.
With grid spacings of 200-300~m, radio detection is perfectly feasible at cosmic ray 
energies of 10$^{17}$ up to perhaps 10$^{19}$~eV.
Extension to the enormous areas needed at the highest energies, however, requires serious rethinking of current detection concepts with the goal to strongly decrease the cost per detector and make the detectors extremely easy to deploy and basically maintenance-free.

The situation is much more favorable if inclined air showers with zenith angles beyond 60$^{\circ}$ are targeted.
In that case, detector spacings of a km or more suffice \citep{KambeitzARENA2016} and very large areas can be covered.
If radio antennas are coupled with particle detectors, most of the infrastructure (communications, power supply, ...) can be shared and radio detection could be a very economical add-on.
This would allow clean separation of the muonic component (measured with particle detectors) and the electromagnetic component (measured with radio antennas), a powerful approach for composition measurements as well as studies of air-shower physics.
As of today, no other detection technique seems equally powerful for the measurement of the electromagnetic cascade in inclined air showers.

The pure sensitivity of radio detectors to the well-understood electromagnetic component of the air shower also suggests using radio detectors for an absolute calibration of the energy scale of cosmic-ray detectors.
This is advantageous from two standpoints: first, the emission can be predicted from first principles, and second, it does not suffer any relevant absorption or scattering in the atmosphere.
In addition to determining the energy scale of a cosmic-ray detector from calculations, radio detection also offers a way to cross-calibrate cosmic-ray detectors efficiently and accurately, especially when relying on the measured radiation energy \citep{AERAEnergyPRL}.

At cosmic-ray energies from $10^{17}$~eV to several $10^{18}$~eV, where radio detection is being successfully applied with today's detection concepts, radio detection is a valuable addition to any cosmic-ray experiment as it provides very useful additional information when applied in hybrid mode.
Already today, direction (better than 0.5$^{\circ}$) and cosmic-ray energy (15-20\%) can be reconstructed from radio data with competitive resolution.
The determination of \xmax with sparse radio arrays (grid spacings of a few hundred meters) has achieved resolutions of approximately 40~g/cm$^{2}$ and is thus not yet competitive with the established techniques which reach resolutions of approximately 20~g/cm$^{2}$.
However, analysis strategies are still under development and the resolution is expected to improve further.

\begin{figure}
\centering
\includegraphics[width=0.48\textwidth]{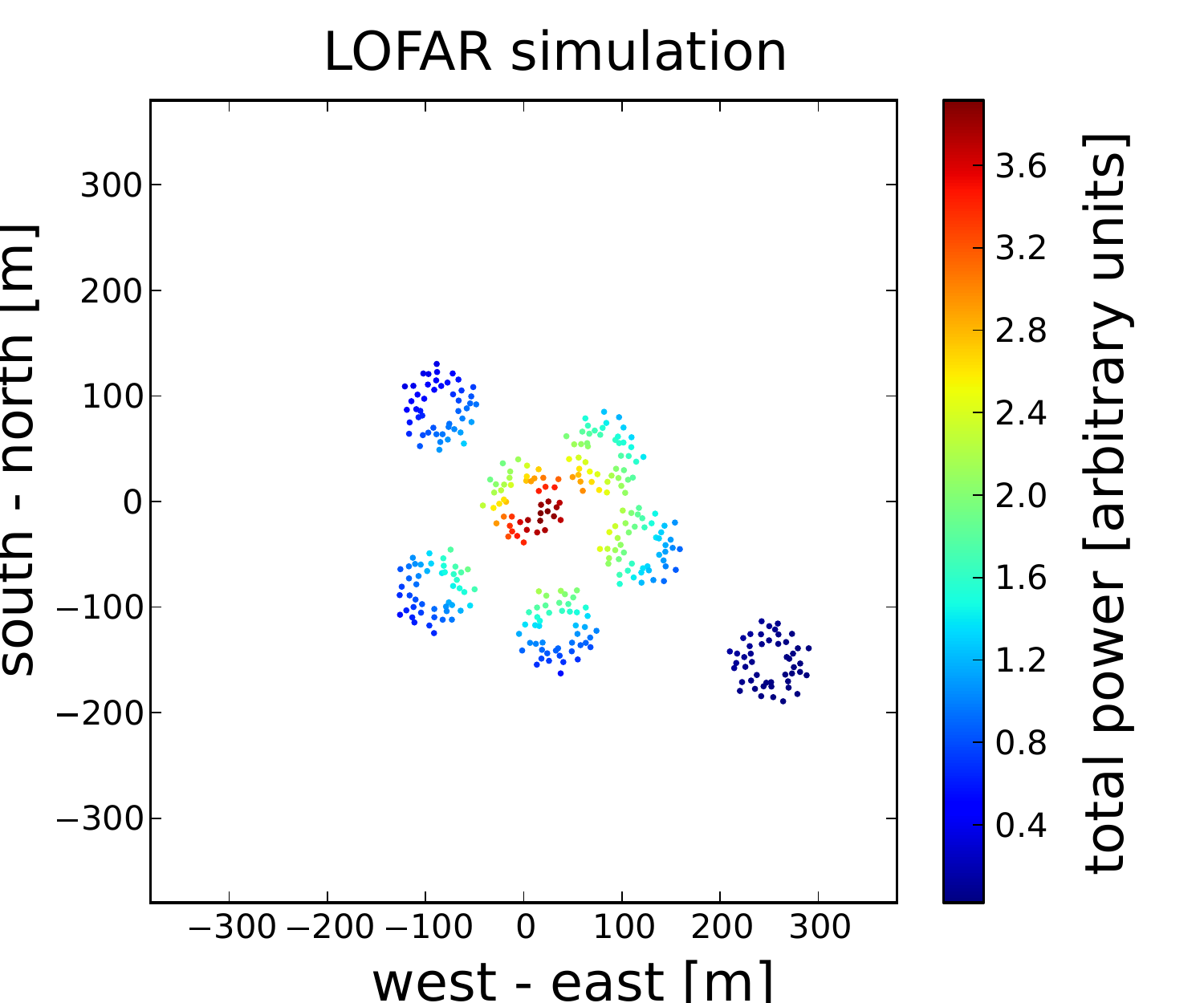}
\includegraphics[width=0.48\textwidth]{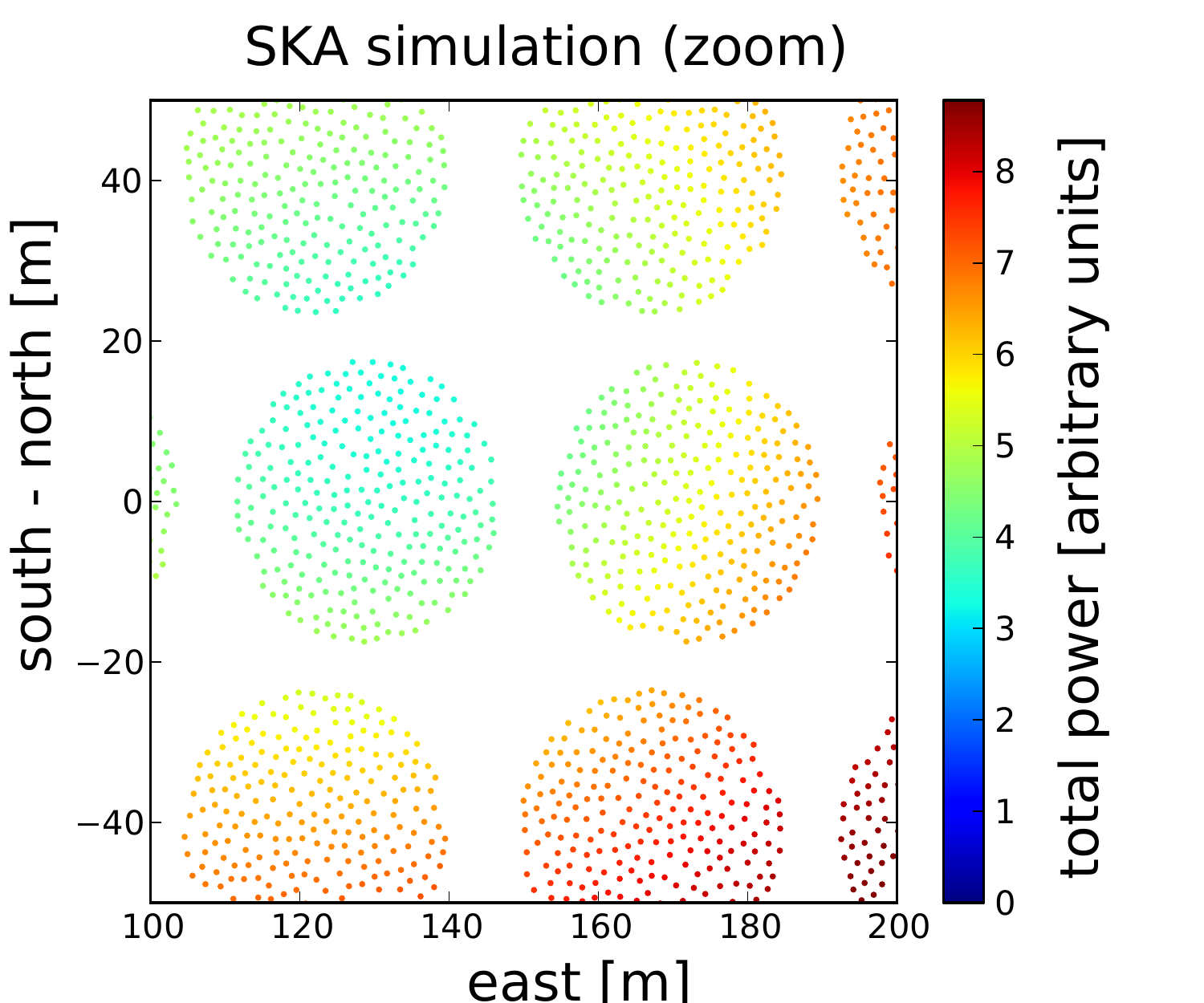}
\includegraphics[width=1.00\textwidth]{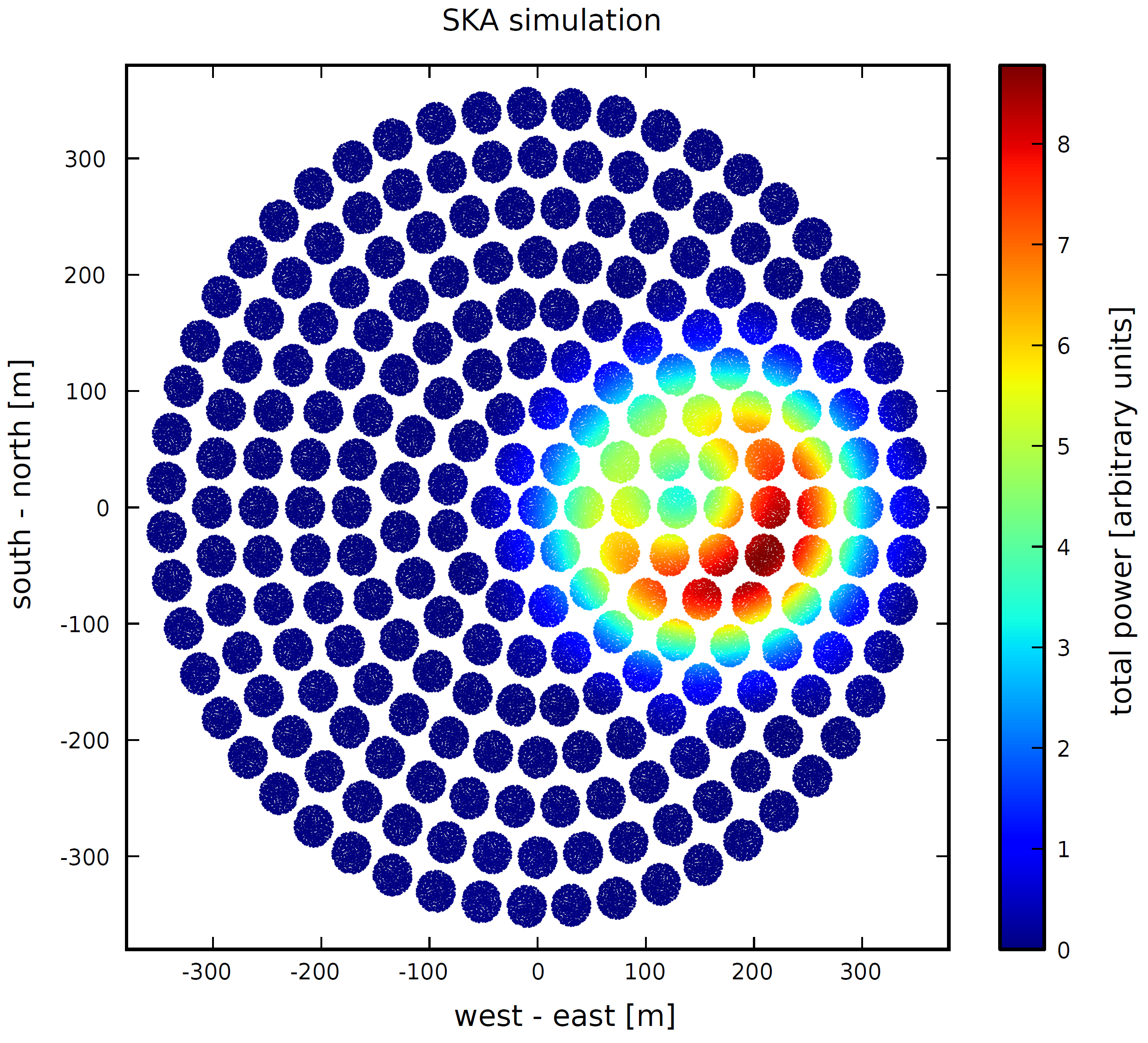}
\caption{CoREAS simulation of the radio emission from an air shower 
with 30$^{\circ}$ zenith angle and an energy of $10^{18}$~eV as sampled 
with LOFAR (top-left) and SKA-low (bottom and zoom-in at top-right). Each point
denotes a measurement with an individual dual-polarized antenna. SKA-low would
sample the air-shower radio signal extremely homogeneously, leading to a
very high quality measurement. The appearance of a Cherenkov ring in the SKA-low measurement is 
due to the measurement of higher-frequency components up to 
350~MHz. Adapted from \citep{HuegeSKAIcrc2015}, reprinted from \citep{HuegePLREP}. \label{fig:lofarvsska}}
\end{figure}

Finally, dense radio detection arrays such as LOFAR have demonstrated that, in principle, the radio signal carries enough information to reconstruct \xmax even more precisely than established techniques.
LOFAR in particular has already achieved an average \xmax resolution of 17~g/cm$^{2}$, without fully exploiting polarization or timing information of the radio signals.
Events that are sampled favorably by the non-uniform antenna array can be reconstructed by LOFAR with an \xmax resolution of better than 10~g/cm$^{2}$.
Even more dense and uniform antenna arrays, in particular the upcoming Square Kilometre Array (see Fig.\ \ref{fig:lofarvsska}), are thus expected to measure \xmax per event with a resolution better than 10~g/cm$^{2}$.
Such dense radio detectors could thus be a very valuable approach to do precision measurements in the energy region around 10$^{17}$~eV, where the transition from Galactic to extragalactic sources could well be taking place.

%%%%%%%%%%%%%%%%%%%%%%%%%%%%%%%%%%%%%%%%%%%%%%%%

\subsection{Radar Detection of Cosmic Ray Air Showers}
The concept of radar detection of air showers was introduced in the early 1940s by Blackett and Lovell \citep{blackett1941radio}
and has been revisited over the years \citep{gorham2001possibility}. The idea is simple: incoming cosmic rays with energy exceeding
100 PeV produce large primary ionization densities that should scatter electromagnetic waves of
order 10-100 MHz, permitting their detection via bistatic radar.
The bistatic radar technique, as applied to cosmic ray detection, is quite similar to the radar detection
of meteors, and one can use this fact as a starting point in understanding the radar
response of cosmic ray air showers, although ionization produced by cosmic rays occurs at
much lower altitudes ($<$10 km) than that produced by meteors ($>$80 km). 

The TARA (Telescope Array Radar Apparatus) experiment attempts to achieve large aperture for EAS detection by bi-static observation of radar reflections from the core of an EAS at radio frequencies \citep{TARA2014NIM,kunwar2015design}. 
Straightforward geometric considerations lead to the expectation that, since the total pathlength from 54.1 MHz transmitter to the position of the shower core and then to the receiver typically decreases with time for a down-moving shower, the received radar echo Doppler-redshifts by approximately 50\% over a period of several microseconds, resulting in a radio-frequency ``chirp'' with slope of order several MHz per microsecond. The chirp rate, in terms of the distance from shower to receiver, and also the relative inclination angle of the shower is a strong function of geometry, but has a typical value of --2 MHz/$\mu$s.

\begin{figure}
\centering
\includegraphics[width=0.65\textwidth]{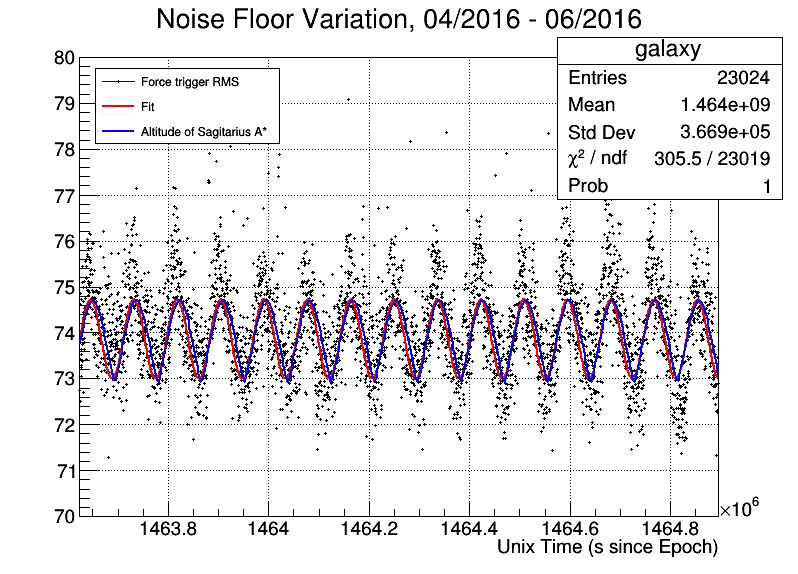} \caption{TARA remote station sensitivity check.} \label{fig:RSgalaxy} 
\end{figure}

Three months of data accumulated with the TARA ``remote stations'' verifies that the expected background is indeed galactic, as illustrated in Figure \ref{fig:RSgalaxy}, and attesting to the relative radio-quietness of the selected experimental site.
However, thus far, TARA has failed to observe any definitive radar echoes from air showers, either in coincidence with Telescope Array fluorescence detector triggers, or self-triggers using the TARA remote stations. Given the known flux of ultra-high energy cosmic rays, coupled with particle-level simulations of the expected radar reflected signal strength, one can estimate the effective `radar cross-section' presented to an incident transmitted radar signal. Based on the non-observation of signals, an upper limit of order 1 cm$^2$ has been set on this cross-section, considerably smaller than the expected dimensions of the plasma core of a descending air shower, expected to be of order 200 cm$\times$2 cm (based on CORSIKA simulations). Possible reasons for the non-observation are that either the electron charge constituting the plasma quickly recombines with the positive ion from which it was initially separated, and/or that the plasma lifetime may be substantially compromised by attachment to atmospheric diatomic $N_2$ and $O_2$ molecules; in fact, some calculations \citep{StasielakIcrc2015} give an unobservably small radar cross-section.

Following TARA's work, it has been suggested that reflections off in-ice shower targets may be a more advantageous approach, in so far as it avoids the aforementioned potential deficiencies of in-air targets \citep{deVries:2013qwa}. A beam test in winter 2015 using an ice block target and utilizing the Telescope Array Electron Linac Source has thus far yielded interesting, albeit inconclusive results \citep{de2015feasibility}. A follow-up run is planned for the winter of 2016-17. Alternately, it has been suggested that the 1 MW SuperDARN transmitter at the South Pole might be used as a radar source in conjunction with ARA-based receivers, however, the SuperDARN broadcast frequency of 10 MHz and the small duty cycle of that transmitter ($\sim$3\%) are not ideally matched to either ARA or the expected power spectrum of the plasma-reflected signal.

\section{Dense Media Radio Detection} \label{sec:askaryanDM}

The complement to atmospheric detection of radio signals due to cosmic-ray initiated air showers is (of course)
detection of showers in dense media. In this case, there is no geomagnetic-induced signal, but 
the coherent (Askaryan) signal outlined previously is enhanced \citep{Askaryan1962a,Askaryan1962b,Askaryan1965}.
The cut-off maximum frequency for radio coherence is set by the lateral scale of the shower. Since showers in-media are typically much
more compact than air, e.g., with a Moli\`ere radius $R_{\mathrm{Moliere}}\sim$10 cm for ice,
the coherent frequency limit for an in-ice shower extends approximately a 
factor of 20 higher than for in-air showers. Moreover, as the interaction length for hadrons is considerably 
shorter than that of neutrinos, such that UHECR will interact in the atmosphere, as described
above, detection of cosmic ray interactions in dense media centers on
ultra-high energy neutrino measurements. Nevertheless, there are backgrounds to neutrino searches 
due to UHECR cores
impacting the surface which must be calculated for experiments such as ANITA and ARIANNA.

Calculations of the signal-generation mechanism resulting from neutrino interactions 
are, in principle, straightforward, and 
analogous to the calculation of the radio-frequency Askaryan signal generated by in-air 
showers outlined previously.
For definiteness, consider a $\nu_e$ undergoing a 
charged current interaction, $\nu_e+N\to e+N'$. The primary UHE 
neutrino transfers most of its 
energy to the electron, which quickly builds an 
exponentially increasing shower of $e^+e^-$ pairs.  The number of pairs 
$N_e$ scales with the primary energy.  In the most populated region of the 
shower, at the ``bottom" of its energy range, a charge imbalance 
develops as positrons drop out due to annihiliation with pre-existing in-medium electrons 
and atomic electrons scatter in due to Compton scattering.  
The original detailed Monte Carlo calculations by 
Zas, Halzen and Stanev (ZHS) \citep{ZHS} 
and confirmed by later GEANT simulations \citep{RazzaqueSeunarineBesson2002,RazzaqueSeunarineChambers2004}
find that the net charge of the 
shower is about 20\% of $N_e$ for shower energies exceeding 1 PeV.  
The electric field produced by this 
relativistic net charge, evolving with the particle shower evolution, results in significant pulsed radiation for
wavelengths in the radio frequency region.
For wavelengths large comparable to the transverse size of the
shower, the relativistic
pancake can be treated as a single, extended, radiating charge.
(Clearly, in the limit $\lambda\to\infty$, the radiating region approaches a
point charge.). 

Simulations have continually evolved, and been refined, 
over time; the Santiago group \citep{AlvarezMunizLDFScheme,AlvarezMunizEngelGaisser2004,AlvarezMunizVazquezZas2000,AlvarezMunizRomeroWolfZas2010,AlvarezMunizAskaryan,AlvarezMunizANITASims,AlvarezMunizCarvalhoZas2012} have now developed a powerful 
software library capable of predicting the Askaryan signal for a wide range of 
materials (ice, salt, lunar rock)
and over a broad frequency interval.  
Of the possible neutrino targets,
cold polar ice offers several advantages 
which make it an extremely
attractive candidate.
In addition to comprising a huge, 
uniform terrestrial target, 
measurements to date 
indicate that cold
ice ($\le -50^\circ$C) has extremely long field attenuation 
lengths, of order 
$\ge$1-3 km for 100 MHz - 1 GHz radio
signals \citep{Bogo85,barwick2005south}, whereas
optical photons in ice have
absorption and scattering lengths which are typically an order of magnitude
smaller.

To set the scale, at the Cherenkov angle $\theta_C$, for which all frequency components
are in-phase and synchronous,
the Signal-to-Noise Ratio (SNR) for a
1 GHz bandwidth, thermal-noise limited 
receiver viewing a 1 PeV electron-induced shower at a distance 1 km 
is of order 1:1 \citep{Frichter:1995cn}. 
However, the solid polar angle $\sin\theta \mathrm{d}\theta$ at the Cherenkov angle is obviously 
vanishingly small as $\mathrm{d}\theta\to$0. As expected, the signal decoheres as one deviates
from the Cherenkov angle, with the highest frequency components diminishing earliest.
 Folding in system noise and realistic antenna response, 
the practical requirement of triggering 
at 5-6$\sigma_{kT+\mathrm{system~noise}}$ at one Cherenkov cone-width (of order 1 degree) off
$\theta_C$ pushes the practical threshold for an in-ice
detector to $\sim$50 PeV.
At such radio-frequencies, the power emission will be 
proportional to the shower energy squared. 
This rapid increase in power with energy substantially compensates for the decreasing 
neutrino flux, with an integral spectrum falling roughly as $1/E^2$. 

\subsection{Sources}
Sources of ultra-high energy cosmic rays, of all types, 
are described in detail elsewhere in this volume. 
Dense media radiowave neutrino detectors were originally proffered as a method to detect the
`guaranteed' cosmogenic neutrino flux, resulting from interactions of the known UHECR hadronic flux
with the CMB: $p_{\mathrm{UHECR}}+\gamma_{\mathrm{CMB}}\to\Delta^+\to p\pi^+$; $\pi^+\to\mu{\overline \nu_\mu}$; $\mu\to\nu_\mu{\overline \nu_e}e$. The resonant enhancement of 
this cross-section at the center-of-mass energy equivalent to the
$\Delta^+$ leads to the expectation that the UHECR proton 
flux will be severely attenuated by neutrino-generating
interactions at typical energies
$\sqrt{2E_p^{\mathrm{cutoff}}E_\gamma}\sim$1232 MeV 
implying a cut-off (the famed ``GZK'' limit) at 
approximately 100 EeV \citep{Greisen1966,ZatsepinKuzmin1966}.

Unlike UHECR below 10 EeV, ultra-high energy (UHE) neutrinos 
point back to sources of high energy cosmic rays, 
providing an identification of the source as well as a powerful test of 
models for the acceleration mechanism. Detection of UHE
neutrino fluxes simultaneous with gamma-ray bursts
would provide essential information on the nature of 
these extraordinarily luminous sources \citep{Asano:2016scy,Fargion:2016bsd}.
At even higher energies, ``GZK'' neutrinos may help
identify more exotic sources such as topological defects \citep{Berezinsky:1996ga}.
In the realm of particle physics, detection of UHE neutrinos 
from cosmological distances, if 
accompanied by flavor identification, may permit measurement 
of neutrino oscillation parameters 
over an unprecedented range of 
$\Delta m^2$ \citep{Halzen:1998be}, or
detect $\nu_\tau$ via ``double-bang'' signatures \citep{Shoemaker:2015qul,Fu:2014isa,Halzen:1998be,Moura:2007sr,Guzzo:2005fi,Beacom:2003nh}.
The angular distribution of upward-coming neutrino events
could be used to measure weak 
cross-sections at energies unreachable by man-made accelerators \citep{Klein:2013xoa}.
Alternately, if the high energy weak cross-sections are known,
they can be used to test Earth composition models
along an arbitrary cross-section (so-called
`neutrino tomography') \citep{Winter:2015zwx,Smirnov:2004zv}.

Most recently, as the statistics on the 
IceCube measurements of the extra-terrestrial atmospheric flux have been improved and the 
extrapolation of that flux into the $>$10 PeV energy range become more reliable, attention
has shifted towards the ``\"uber-guaranteed''\footnote{In the Joe Namath, rather than the Cam Newton sense.}
flux in the 10-100 PeV energy regime and efforts are increasingly focused on pushing down the
minimum radio-detection energy detection threshold towards $10^{16}$ eV.

\subsection{History}
Thus far, experimental efforts seeking detection of neutrinos via coherent radio frequency
signals have proposed or utilized two primary targets -
cold polar ice (either terrestrial or one of the ice-crusted moons of Jupiter or
Saturn) and the lunar regolith. Earlier interest in salt dissipated following 
measurements of disappointingly small RF attenuation lengths \citep{Connolly:2008rn}.

Unlike radio frequency detection of air showers, however, radio-based experiments have thus far detected zero
in-medium neutrino interactions and are, therefore, at a 
comparatively less developed stage. Much of the focus has therefore
been on a concerted study of the radio frequency properties of
candidate experimental sites and detector optimization, as we outline below.

\subsubsection{Initial Antarctic Work}
Twenty years after Askaryan's predictions, researchers 
at the Institute of Nuclear Research (INR, RAS) in Moscow
first proposed coherent radio as an experimental cosmic ray detection technique in 1986. INR-based
Russian collaborators performed the first
experimental studies and laid the initial
groundwork for development of a radiowave-based neutrino detection
experiment \citep{Boldyrev:1991fj,Markov:1986dx,Gusev:1985rx}.
That work includes the
definitive measurements at the Soviet station Vostok 
of the temperature profile of the ice down to 2~km (1.5 km above bedrock),
as well as measurement of the frequency and temperature dependence of the ice
absorption of radio waves.\footnote{We note that, from a radioglaciological standpoint,
Vostok is perhaps the optimal site for an ice-based radio detector. The relatively
thin firn layer (90 m thick rather than 160 m at the South Pole, leading to less ray tracing confusion),
the thicker, colder ice (typically 5 K colder than at South Pole, and approximately 0.5 km thicker) 
and the availability of a well-studied and complete ice core, permitting detailed characterization of
radio-frequency response) make this site decidedly preferred relative to South Pole. 
Additionally, the amount of anthropogenic noise at Vostok should be considerably smaller than
South Pole, given the smaller infrastructure.} 
A pilot experiment (``RAMAND'', for 
Radio wave Antarctic Muon And 
Neutrino Detector \citep{bogomolov1987background}) tested many aspects of
the neutrino radio detection idea at Vostok 
between 1985-1990, including critical
 initial measurements in Antarctica of natural and man-made radio impulse 
backgrounds.
Unfortunately, that nascent effort was abruptly terminated in 1991,
when the Soviet political infrastructure collapsed. 

\begin{figure}
\centering
\includegraphics[width=0.5\textwidth]{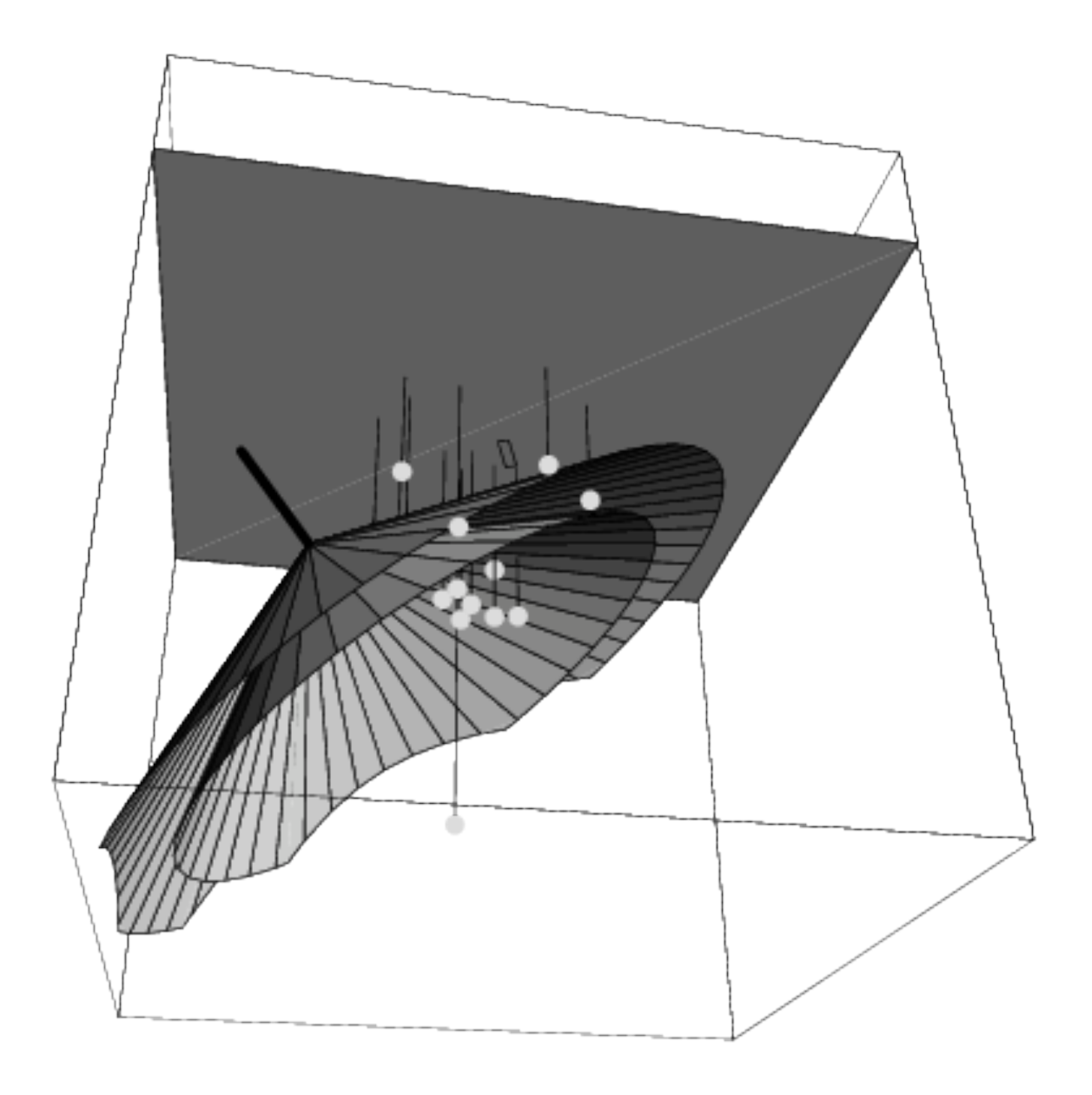}
\caption{RICE experimental schematic. \citep{Kravchenko:2002mm}}\label{fig:evdisp2}
\end{figure}

The RICE experiment \citep{Kravchenko:2001id,Kravchenko:2002mm,Kravchenko:2003tc,Kravchenko:2003gj,Kravchenko:2006qc,Kravchenko:2011im,Hogan:2008sx}, which first deployed in-ice antennas in 1995 
in conjunction with the AMANDA experiment and continued data-taking through 2011, 
represented a prototype of the in-ice neutrino detection technique and can perhaps be 
credited with first demonstrating the operational feasibility of such an effort. 
Figure \ref{fig:evdisp2}
shows the concept of the RICE in-ice array, illuminated by a characteristic Cherenkov cone produced by an ultra-high-energy cosmic neutrino at 10-100 PeV energies.  Neutrinos undergoing charged current reactions in the ice produce electromagnetic/hadronic showers, which generate radio waves coherently in the regime up to GHz frequencies. Owing to the LPM \citep{LandauPomeranchuk1953a,LandauPomeranchuk1953b} effect, at high energies, the target medium is increasingly length-contracted. As a result, shower electrons are unable to discriminate individual target positive (nuclear) or negative (electron) charges, resulting in a reduction in the effective electromagnetic interaction cross-section and an increase in the mean-free-path. Correspondingly, electromagnetic showers above 1 EeV are considerably elongated and the net radio signal strength consequently diluted; the primary experimental sensitivity above 10 EeV is therefore to hadronic showers. Timing coincidences of signals in radio receivers (``Rx'') are used to resolve the three-dimensional location, energy, and cone geometry to fix the parameters of individual neutrino sources. In the Figure, an incident $\nu_e$ interacting above the detector array initiates a shower which produces the characteristic Askaryan radiation with a pronounced Cherenkov ring. The $\pm$3 dB points of the emitted power (at 500 MHz) are shown as the nested cones. In the Figure, the RICE antenna array is drawn roughly to scale; the volume drawn corresponds to 500 m $\times$ 500 m $\times$ 500 m.

\subsection{Radio Frequency properties of ice} 

Unlike atmospheric RF-based detectors, which, aside from calculable elevation-dependent differences in atmospheric density, otherwise deal with a more-or-less monolithic target at radio frequencies, the RF properties of ice vary considerably across the Earth's ice caps, and the location-specific RF characteristics largely determine the sensitivity of a given experiment. Correspondingly, we therefore now discuss the current state of radioglaciological studies in both Greenland and Antarctica, in the effort to locate an ideal neutrino detection site.

\subsubsection{Dielectric Permittivity of Bulk Ice}
Perhaps the most important 
feature of any candidate neutrino telescope is the so-called effective volume ($V_{\mathrm{eff}}$), which
defines the equivalent ice volume, as a function of incident neutrino energy, over which interactions can be
measured. Equivalently, some experiments prefer to quote an effective area $A_{\mathrm{eff}}$, corresponding to the
two-dimensional projection of $V_{\mathrm{eff}}$. The effective volume is directly determined by the
radio-frequency ice response.
Ultimately, one seeks complete parameterization of the complex ice dielectric permittivity 
${\vec \epsilon}=\epsilon'(\omega)+\epsilon''(\omega)$ at 10 MHz--1000 MHz frequencies. 
The real component $\epsilon'$ 
determines limitations on $V_{\mathrm{eff}}$ 
from ray-tracing, while the
imaginary component limits $V_{\mathrm{eff}}$ due to absorption. Both of these are important
considerations for detection of neutrinos at very high energies ($E>$1 EeV), although
at the low end of the
sensitive energy regime (10 PeV), the effective volume is limited not by
attenuation or ray tracing, but by the RF Signal-To-Noise (SNR) triggering threshold.

In general, the ``loss tangent'', defined as the ratio of the imaginary to the real
components of the permittivity ($\tan{\delta}=\epsilon''/\epsilon'$, related to the attenuation
coefficient $\alpha$ by $\alpha=8.686(2\pi f/c)\sqrt{\epsilon'}\tan\delta$ and the field attenuation 
length $L_\alpha=1/\log\left(10^{\alpha/20}\right)$) \citep{Bogo85},
of both the upper-ice ``firn'' and the deeper ice (in the Ih phase) should be
fully mapped from HF (3--30 MHz) to UHF (300--3000 MHz).  
The radio regime of hundreds of MHz, between the two Debye resonances
at kHz frequencies and the infra-red peak,
and where the ice response is relatively flat (the loss tangent is a minimum at $\sim$ 300 GHz \citep{Warren:1984,Warren:2008}), is 
a therefore suitable frequency region for neutrino detection experiments.

To date, much of our information about the interior of the polar ice sheets has been derived from extensive compilations of radar survey data such as those of BEDMAP \citep{BEDMAP} or CReSIS \citep{CReSIS}, in which a transmitter (Tx) beams radio-frequency signal towards the surface from a height of order 1 km, and the returns from the ice below are then recorded. These data have been used to map out ice sheet internal layers and bedrock topography, 
which can then inform models of ice flow and mass balance. Some of this information, in conjunction with ice core data, has been sensational -- the characteristics of some of the returns are consistent with acid layers deposited following volcanic eruptions. 
Among the most comprehensive studies of the polar ice sheets in the wavelength interval relevant to Askaryan signals is the nearly 20-year data sample accumulated in Greenland and Antarctica, by the Center for Remote Sensing of Ice Sheets \citep{CReSIS} group, based at the University of Kansas (KU). This allows measurements of the ice dielectric permittivity using 150--195 MHz radar depth sounding (RDS) data accumulated by CReSIS. 
These measurements have been complemented by site-specific bistatic radar probes at a variety of locales across the ice sheets. 

Of course, to understand the internal structure of the ice sheet, it would be preferable to directly extract ice cores, which provide {\it prima facie} ice chemistry data, however, such an undertaking is expensive and logistics-intensive. 
In addition to the
SPICE core recently extracted at South Pole, multi-km core data taken from Law Dome (1200 m),
Vostok (3623 m), 
Siple Dome (1003 m),
Dome C (3270 m),
Dome Fuji (3035 m), 
Wais Divide (3435 m),
and Byrd Station (2164 m) are all available for analysis from the US 
National Ice Core Lab (NICL; www.icecores.org), with the latter the only Antarctic core reaching bedrock. 

Much of the difficulty in defining how appropriate a site may be for neutrino detection, particularly
for aerial-based survey experiments such as ANITA, is due to the fact that
the radio-frequency response depends on the specifics of the ice crystal alignment within the ice sheet at that particular locale. In the context of a simplified coupled-oscillator model of electromagnetic propagation through ice, the transmission of electric field through the ice sheet is different for the case where, e.g., the ice crystal plane is oriented vertically vs. horizontally, resulting in differences in wavespeed. Typically, the wavespeeds have orientation-dependent extrema, defined by a minimum wavespeed (the so-called `extraordinary' axis direction) and a maximum wavespeed (the so-called `ordinary' axis direction, which is generally, although not always, perpendicular to the extraordinary axis direction, depending on whether the dielectric tensor has large off-diagonal elements).

Radar (from airborne surveys or site-specific) and optical (from ice cores) analysis of glacial ice can be used to reveal the crystal orientation (dependent on the ice thermal history as well as local stress/strain) fabric (COF) and also the so-called ``grain'' (defined as the local region over which alignment is uniform) structure within the ice sheet. For example, crystallization of ice occurring at the time of the last glacial maximum `encodes' the thermal history of the sheet at that time.
The amount of alignment is depth-dependent: as the ice overburden increases with depth and in the absence of any stress tensor, one expects the c-axis (defined as the normal to the plane of the hexagonal 
Ih crystal) to increasingly orient vertically, i.e., perpendicular to the `glide plane' defined by the lateral bulk ice flow. However, if the free-energy is sufficiently high and the stress tensor is non-zero, the c-axis of newly formed crystals will often align at 45 degrees relative to the vertical in order to best relieve the local shear stress -- to the extent that the stress tensor has large off-diagonal elements (the case where there is no preferred local horizontal bulk ice flow), there is rotational symmetry of the stress tensor about the vertical and a so-called `girdle' develops, with the c-axis vector tracing an ellipse (or, for a perfectly symmetric stress tensor, circle) in the horizontal plane, and centered on the vertical. Subsequent bulk flow can effectively rotate those crystals back into the vertical plane. In general, ice with strong fabrics will behave anisotropically under stress and will exhibit different flow compared to weaker fabrics; most importantly for
neutrino-oriented experiments, they will display anisotropic response to RF propagation. By contrast, ice with a weak fabric will not show a bulk directional asymmetry in wavespeed. 

\subsubsection{Birefringence} Askaryan signals are polarized perpendicular to the Cherenkov cone. Unfortunately, in general,
the propagation velocity of radio waves is polarization-dependent, according to the degree of bulk COF alignment.
The response of ice as a function of polarization (``birefringence''), as outlined above,
is characterized
by differences in either wavespeed and absorption along linear
(generally orthogonal) axes, or alternately, in terms of a 
left-handed circular polarization (LCP) vs. right-handed circular polarization (RCP) basis. 
Formally, the two asymmetries in the dielectric constant ($\delta_{\epsilon'}$, 
real and $\delta_{\epsilon''}$, imaginary) are linked by the Kramers-Kr\"onig dispersion 
relation
-- if one is non-zero the other must be non-zero, as well. 
The importance to radio experiments is obvious -- a radio signal traveling through a dense medium
will resolve itself along the basis axes defined by the `ordinary' and `extra-ordinary'
birefringent axes, with a time delay comparable in magnitude to the signal duration between the 
arrival time of the two components.

Birefringence is a well-known property of standard terrestrial ice Ih, as was first described by 
Brewster two centuries ago \citep{Brewster}.
At optical frequencies, Ih ice has an extraordinary index of 1.313 and an ordinary index of 1.309, corresponding to a wavespeed asymmetry of 0.3\% between the fast (ordinary) and slower (extraordinary) components.\footnote{In principle, IceCube likely has sufficient statistics to observe this directional dependence as a systematic offset in the PMT arrival times of photons coming from different directions,} % \citep{OpticalBiref}. 
For an experiment detecting neutrino interactions 1 km distant, this corresponds to a time lag of
about 15 ns in the arrival time of the two components, assuming maximal asymmetry.

\begin{figure}[htpb]
\begin{minipage}{18pc}
\centerline{\includegraphics[width=1.0\textwidth]{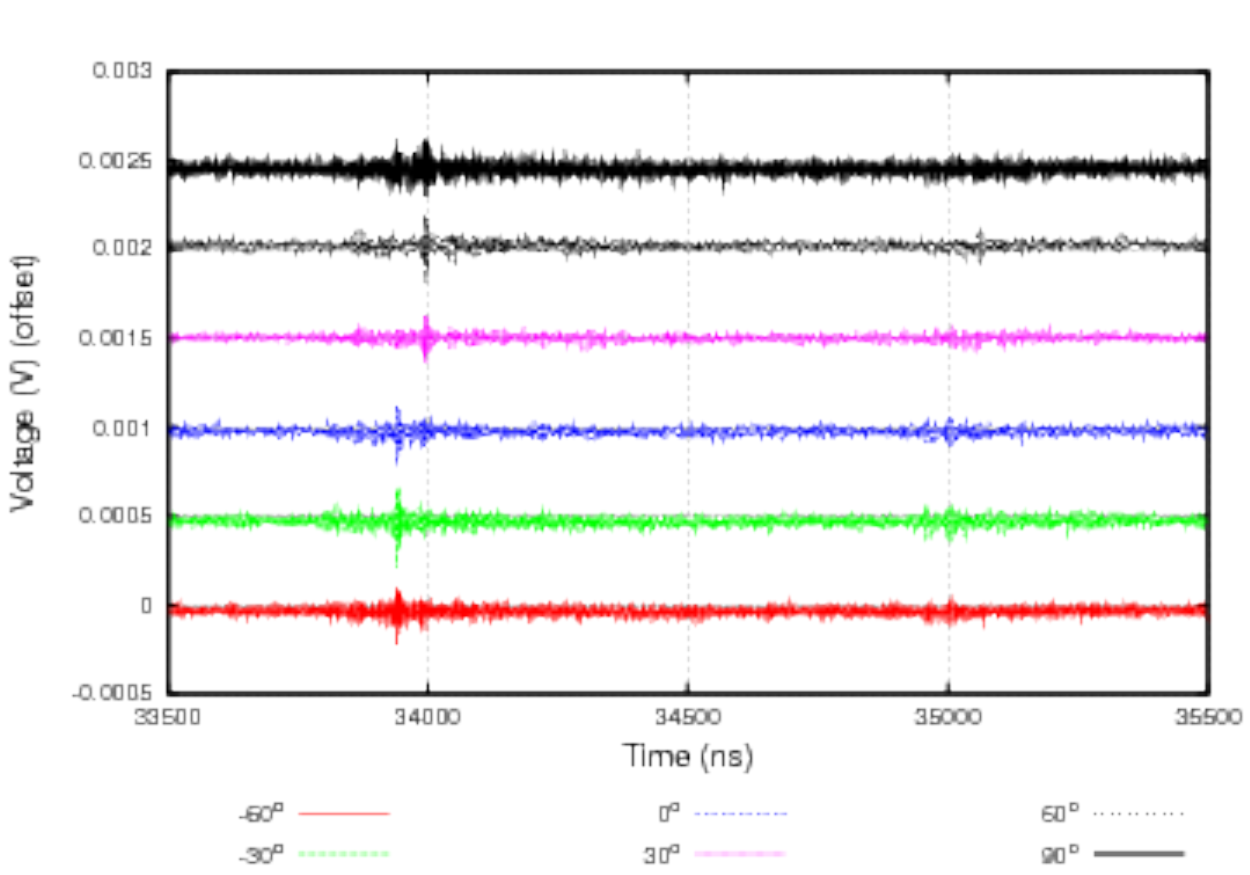}}\caption{Voltage measured, as a function of time, for bedrock reflections, after subtraction of rms noise, as a function of azimuthal electric field polarization vector (in the horizontal plane) \citep{Besson:2010ww}.}
\label{fig:RT09_V1V2}
\end{minipage}
\hspace{2mm}
\begin{minipage}{18pc}
\centerline{\includegraphics[width=1.0\textwidth]{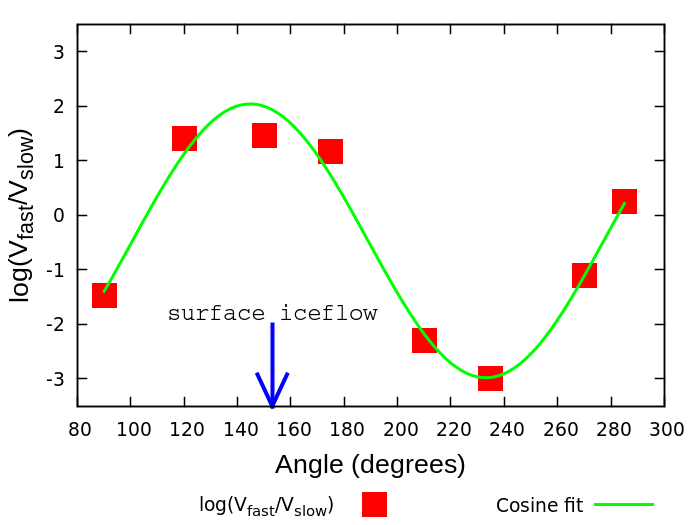}}\caption{Ratio of amplitudes for ``fast'' vs. ``slow'' reflection signals, as a function of azimuthal orientation of co-polarized surface horns.  Blue arrow indicates azimuth corresponding to local horizontal ice flow direction which is expected to play a role in crystal orientation fabric alignment\citep{Besson:2010ww}.}\label{fig:RT09_V1V2_a}
\end{minipage}
\end{figure}

Site-specific bistatic radar operating on the surface of the ice sheet 
is useful in revealing the birefringent asymmetry 
dependence on depth into the ice sheet, by comparing the synchronicity of echoes registered at various
azimuthal angles from internal layer reflections vs. bedrock reflections.
Previous papers \citep{Besson:2007jja,Besson:2010ww} investigated the basal echo times 
at both Taylor Dome and also South Pole as a function of polarization. Those locale-specific
studies demonstrated that reflections through the full ice sheet, off the bedrock,
do exhibit the 
azimuthal dependence of echo times and voltages characteristic of birefringence. Specifically,
for polarizations aligned parallel to the putative ordinary axis, only one return is observed, with a voltage $V_{\mathrm{fast}}^{\mathrm{max}}$ and a negative
echo time offset
relative to that observed for polarizations aligned with the putative extraordinary axis,
presumably rotated by $\pi$/2 radians. 
The South Pole study explicitly demonstrated the expected dependence of
detected amplitude on polarization (Figure \ref{fig:RT09_V1V2}). Designating the voltage observed
for alignments parallel to the ordinary axis as $V_{\mathrm{fast}}^{\mathrm{max}}$, then, for polarizations at $\pi$/4 radians relative to each
axis, two returns should be observed, each with amplitude $1/\sqrt{2}$ as large as the
co-aligned case; i.e., $V_{\mathrm{fast}}(\pi/4)/V_{\mathrm{slow}}(\pi/4)=1$. Additionally, for this case,
$V_{\mathrm{fast}}/V_{\mathrm{slow}}=V_{\mathrm{fast}}^{\mathrm{max}}/\sqrt{2}$.
Fig. \ref{fig:RT09_V1V2_a} \citep{Besson:2010ww}
shows measurements consistent with these expectations, and thus,
a correlation consistent with ice flow direction.
However, reflections from internal layers are synchronous to within 1 ns down to a depth of 1500 m, indicating that
the birefringent asymmetry is generated in the deepest half of the ice sheet, where the 
ice flow velocity gradient is greatest, and, presumably, the ice strain. 

\paragraph{Summary of Birefringent Results}
Table \ref{tab:birefsum} summarizes some recent birefringence measurements. From the Table, it 
seems clear that birefringence is a `generic' property of ice sheets -- depending on signal 
orientation, there may therefore be a delay between the arrival time for signals propagating along the ordinary vs. extraordinary bases. There is one important caveat here -- thus far, all {\it in situ} measurements of birefringence have been conducted along the vertical, whereas 
signals from incoming neutrinos will be largely oblique. It is therefore important that
complementary measurements at inclined angles be made in the future, such as those measurements
planning to broadcast 1 km from the South Pole SPICE core hole horizontally to the embedded ARA radio array at South Pole.

\begin{table}[htpb]
\caption{Summary of recent birefringence measurements. \label{tab:birefsum}}
\begin{center}
\begin{tabular}{c|c|c|c} 
Publication & Locale & $\delta_{\epsilon'}$ Result & Comment \\ \hline
\citep{Besson:2007jja} & Taylor Dome & 0.12\% & time-domain \\ 
\citep{Besson:2010ww} & South Pole & 0.3\% & time-domain \\ \hline
\citep{hargreaves1977polarization} & Greenland & 0.024--0.031\% & \\
\citep{doake2002polarization} & Brunt Ice Shelf & $>$0.14--0.47\% & \\
\citep{doake2003applications} & George VI Ice Shelf & $>$0.05--0.15\% & \\
\citep{matsuoka1997precise} & Lab Ice & $\sim$3.4\% & 1 MHz -- 39 GHz \\
\citep{fujita1993measurement} & Lab Ice & (3.7$\pm$0.6)\% & 9.7 GHz \\ 
\citep{woodruff1979depolarization} & Bach Ice Shelf & 0.52\% & \\
\citep{fujita2003scattering} & Mizuho Station & measurable & \\ 
\citep{fujita2006radio} & Mizuho & 1.5\%-3.5\% & frequency-domain \\ \hline
\citep{stockham2016radio} & Thwaites Glacier & 0.03\% & CReSIS \\ 
\citep{stockham2016radio} & Eastern Greenland & 0.17\% & CReSIS \\ \hline
\end{tabular}
\end{center}
\end{table}

\subsubsection{Extraction of Average Attenuation Length from Surface and Bedrock Echoes in Aerial Radar Surveys} 
In addition to the site-specific measurements,
the CReSIS data permit an estimate of the average attenuation length between the surface and the bedrock directly from the relative strengths of those two echoes, after applying a geometric correction, dependent on the altitude of the plane when the data were collected and the depth of the bedrock. This allows a comparison of any putative terrestrial
neutrino detection experiment, based in either Greenland or Antarctica.
Systematic errors in this measurement include:
\begin{figure}
\centering
\includegraphics[width=0.43\textwidth]{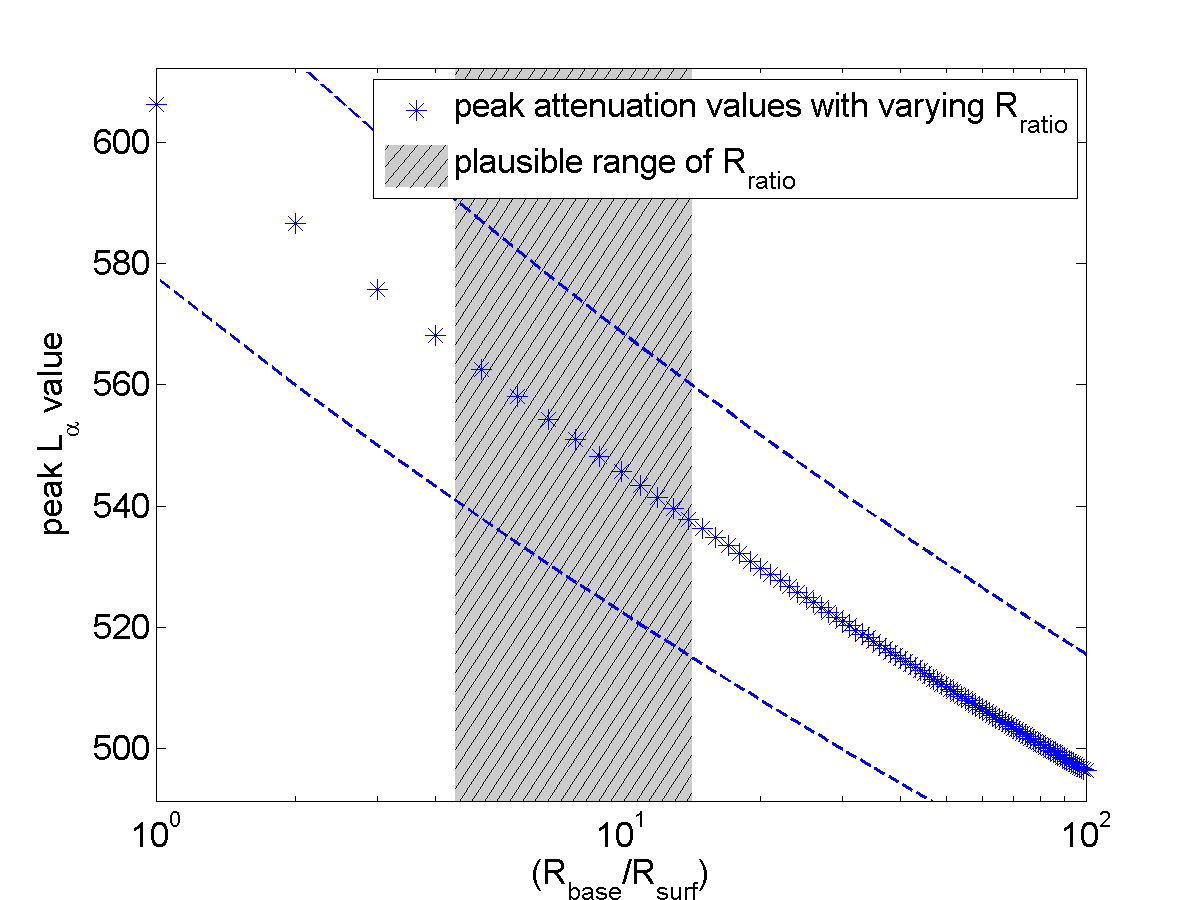}
\caption{Calculated value of GRIP site (Greenland) field attenuation length as a function of assumed relative bedrock:surface reflectivity. The dashed lines represent the estimated uncertainty in the attenuation length. \citep{stockham2016radio}} \label{fig:LattenVRefl}
\end{figure}
\begin{itemize}

\item The several dB uncertainty in the gain of the CReSIS receiver system at the time the two returns are recorded. In particular, the gain of the system at the time the surface return is recorded is smaller than the system gain at the time the bedrock return is recorded, to avoid saturation of the DAQ dynamic range by the bright surface return.
\item Uncertainty in the reflection coefficient of the surface and also bedrock returns due to uncertainty in the dielectric contrast \citep{paren1981reflection}. 
For the surface return, the reflected power at normal incidence plausibly varies from 1.9$\to$2.5\%, corresponding to surface index-of-refraction
values of 1.32$\to$1.38. For bedrock, assuming a dry 
rock-ice interface, the permittivity may range from
4--5 ($2<n<2.23$) for sedimentary rock vs. 6--10 ($2.45<n<3.16$) 
for massive rock, resulting in a possible variation in the 
reflected power from 11.1$\to$26.9\%. If there is a water layer
between bedrock and the bottom of the ice sheet, then the reflectivity
increases to $\sim$65\%, although this should be clearly evident as a marked discontinuity in 
the calculated attenuation length, as a function of position, which is not readily observed.

Typically, one assumes a ratio of 10:1 basal power reflection coefficient:surface power reflection coefficient; results are fairly insensitive to reasonable variations in this assumption. Figure \ref{fig:LattenVRefl} shows the dependence of the extracted attenuation length at the GRIP site on the assumed value of the relative bedrock:surface reflectivity, with the default value (10) and the estimated error bars indicated.
\item Uncertainty in the `smoothness' of the surface and bedrock 
reflectors. Here, we assume that the scale of surface inhomogeneities is
small relative to one wavelength, corresponding to 2 meters in air and roughly
1.3 meters in ice. Under that assumption, scattering is considered to be 
specular, corresponding to $1/r^2$ power diminution rather than the
$1/r^4$ that would be characteristic of diffuse, incoherent scattering. If there
were, in fact, total decoherence at the bedrock, the scattered signal would be unobservably small.
\end{itemize}
The width of the observed variation in the measured surface return strength over a Greenland
GRIP echogram is of order 2 dB, which should bracket the possible effects of surface roughness. The derived attenuation lengths for Greenland and their associated errors, as well as for Antarctica, are shown in Figure 26 
\citep{stockham2016radio}. 
We note that ANITA-2 mapped the Antarctic Solar radio-frequency signal (integrated from 200 MHz -- 1200 MHz), as observed in it's surface reflection \citep{Besson:2013jm}. By comparing the strength of the surface reflection with the direct solar radio-frequency signal strength, the reflectivity of the surface was derived. That study found good agreement with the Fresnel coefficients expected at the snow-air interface, assuming specular scattering.
\begin{figure}
\subfloat[Summary of Greenland attenuation lengths derived from direct ratio of measured surface return strength to bedrock return strength.]{\includegraphics[width=0.48\textwidth]{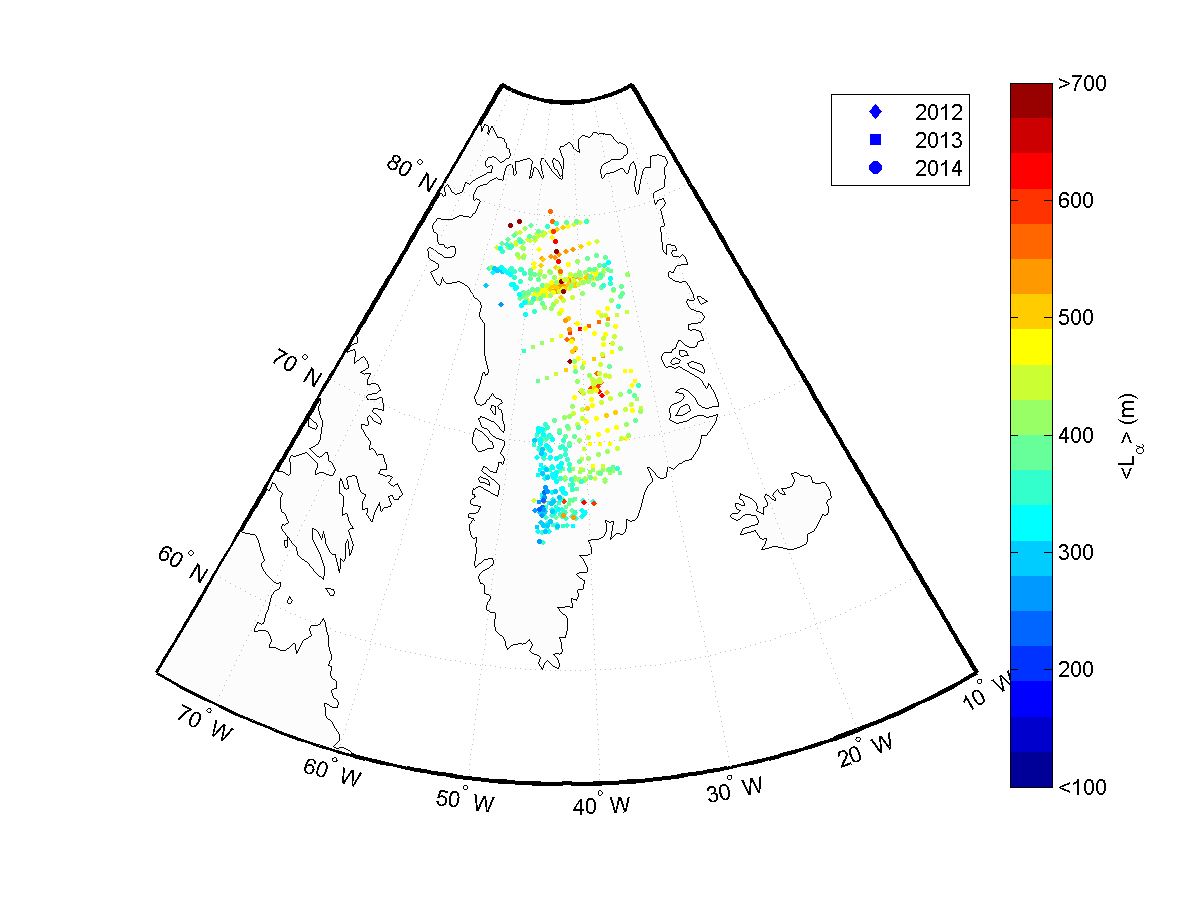}}
\subfloat[Estimated error in Greenland attenuation lengths.]{\includegraphics[width=0.48\textwidth]{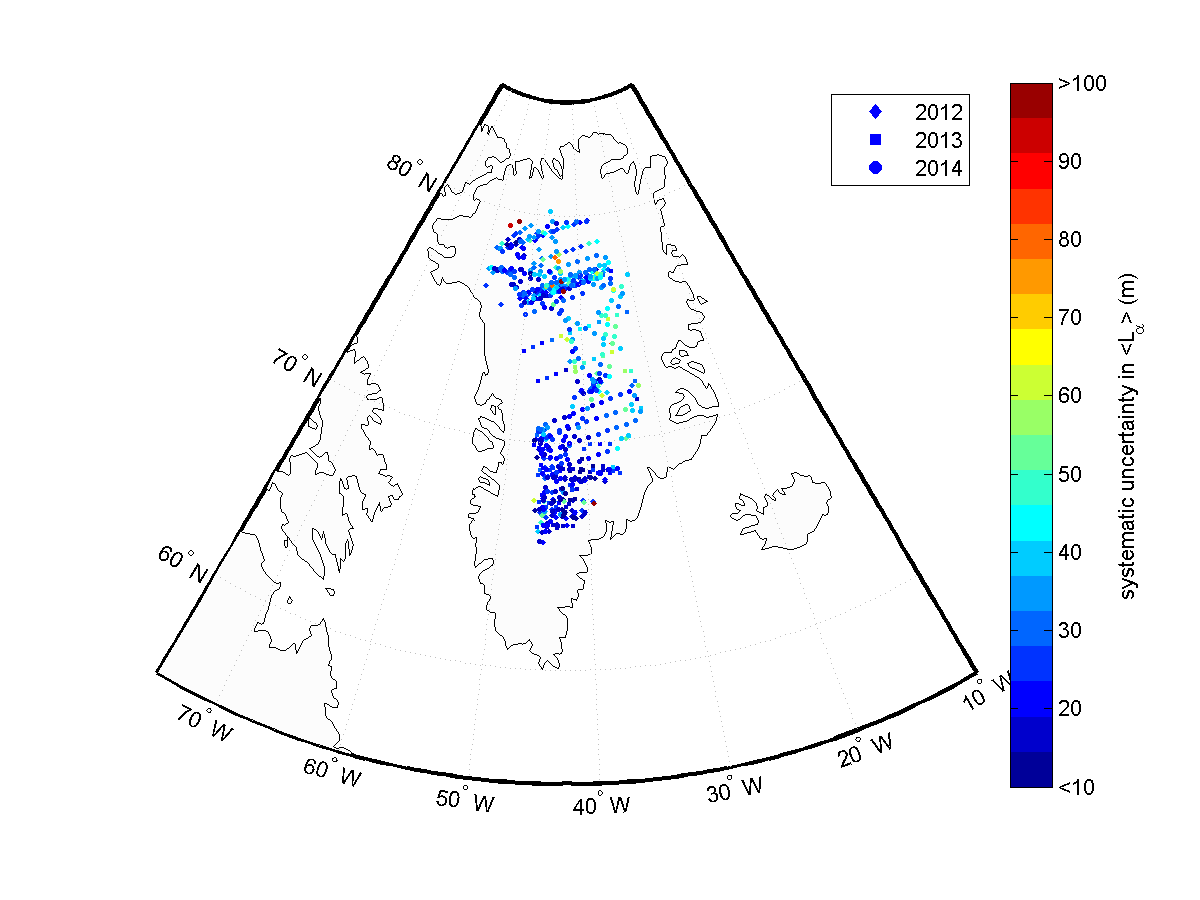}} \\
\subfloat[Summary of Antarctic attenuation lengths derived from direct ratio of measured surface return strength to bedrock return strength.]{\includegraphics[width=0.48\textwidth]{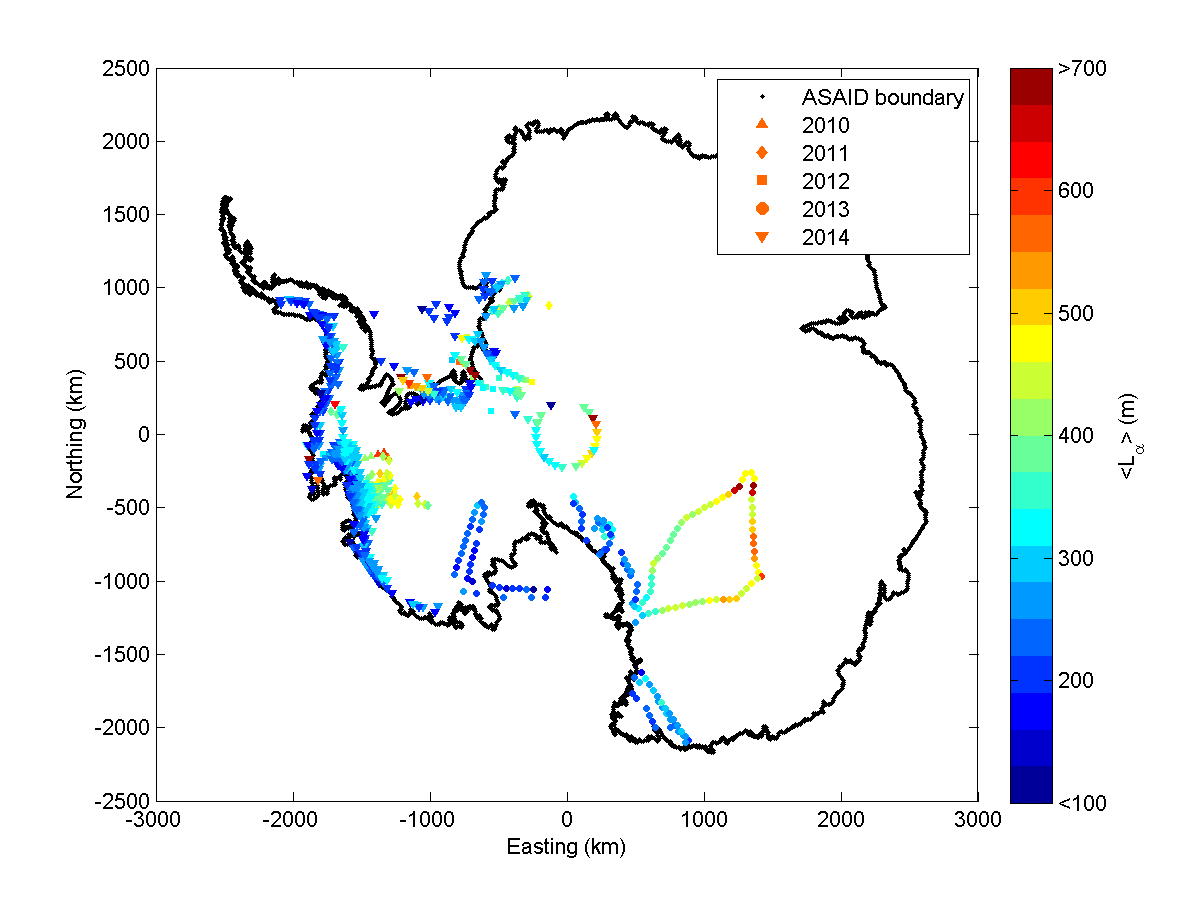}}
\subfloat[Estimated error in Antarctic attenuation lengths.]{\includegraphics[width=0.48\textwidth]{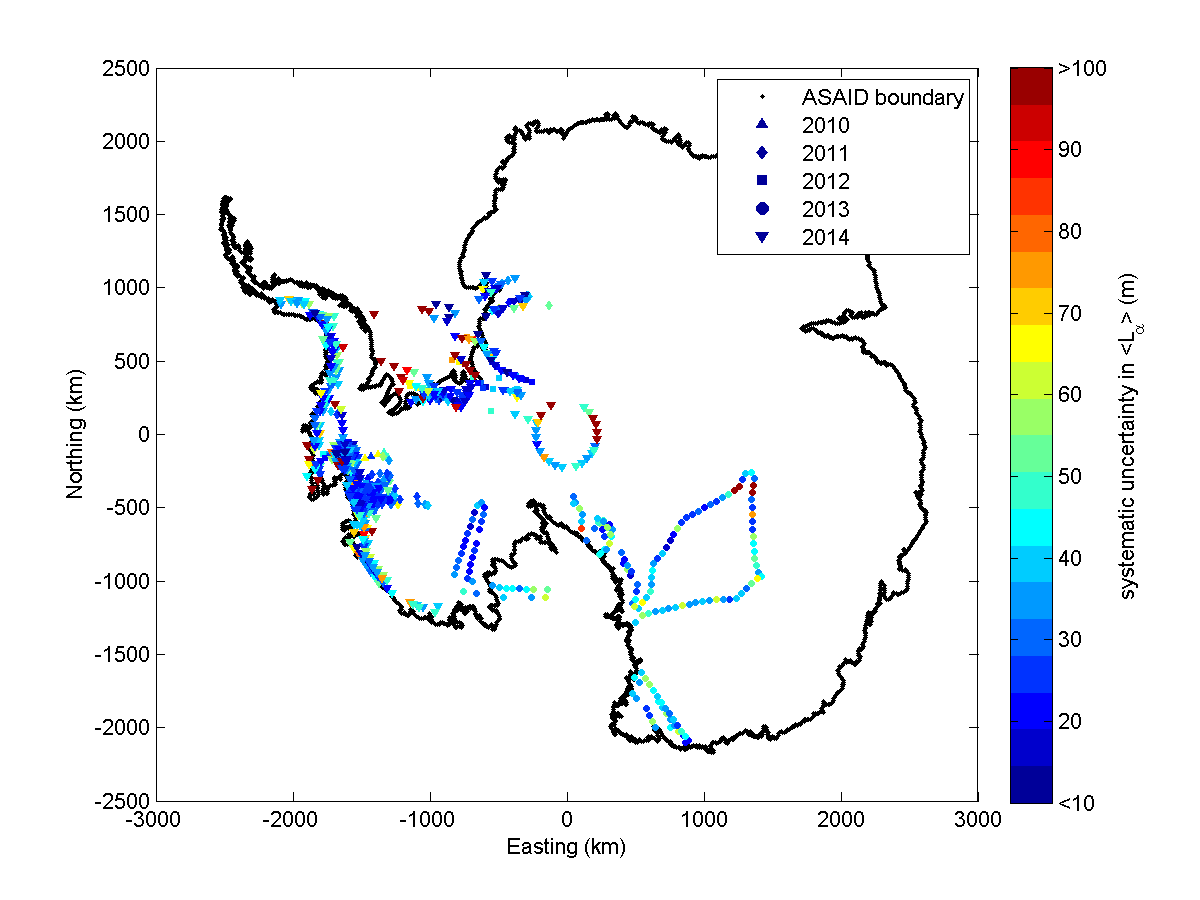}}
\label{fig:Lattens}
\caption{Compilation of attenuation lengths in Antarctica and Greenland, with their associated errors.}
\end{figure}

In general, the Antarctic attenuation lengths are somewhat longer than those for Greenland, again consistent with the expected colder Antarctic ice.

\subsubsection{Dispersion}
By performing a Fourier transform on the waveforms captured in bottom bounce
studies at South Pole, one can investigate
the dispersive characteristics of ice over the frequency range 200 MHz -- 900 MHz. 
Synchronicity of received power is observed to within 4 ns for all frequencies considered, implying a variation in the real part of the dielectric constant less than 0.012\% over the range of frequencies relevant for neutrino detection. 
This asymmetry is consistent with the
naive expectation that the variation in dielectric constant with frequency should be relatively small at radio frequencies, far from Debye resonances for ice. By contrast, in their extensive review of radio ice sheet sounding, Dowdeswell and Evans \citep{dowdeswell2004investigations} suggest a variation in the dielectric constant of approximately $\Delta\epsilon'\approx0.04$ over the interval 1--100 MHz, corresponding to a wave speed variation of about 1.25\%.

\subsubsection{Possible Effects of Scattering Layers on Signal Reconstruction}If there are many scattering layers, with a high dielectric contrast relative to the surrounding ice, then it is, in principle, possible for rays to refract and scatter at grazing incidence angles, and incur some (perhaps considerable) cumulative loss of amplitude. Measurements of attenuation length at South Pole using transmitters deployed at $\sim$2 km depths \citep{Allison:2011wk} indicate that this effect is likely not significant.
We note that in-ice radar reflecting layers, observed by the RICE experiment in 2007, have recently been correlated with ash layers detected by the South Pole Ice Core Experiment (SPICE) -- in particular, one of the two most prominent ash layers observed thus far by SPICE in 2016, at a depth of 1180 m, was predicted by RICE data taken nearly a decade earlier \citep{Besson:2010ww}.

We again emphasize here that usual measurements of ice properties made by neutrino-search
experiments are based on vertically-propagating
`bottom bounce' measurements, in which $L_{atten}$ is derived from the magnitude of the return power, and therefore generally suffer from four principal differences relative to radio signals emanating from an in-ice neutrino interaction, which will generally not be vertical:
\begin{enumerate}
\item Because the index-of-refraction increases with depth, rays are focused down to the normal as they emerge from the surface. There is a defocusing that occurs on the return path which, however, does not compensate the focusing; the magnitude of this flux focusing, and therefore the boost in received power, is approximately equal to $(n_{bottom}/n_{top})^2$, with $n_{bottom}\sim$1.78 and $n_{top}\sim$1.0 for surface transmitter antennas in-air. Not including this effect results in an over-estimate of the radio attenuation length $L_{atten}^\nu$ appropriate for neutrino detection.
\item Depending on the bedrock roughness, there may be significant cross-polarized power emerging after the bedrock reflection; in the extreme case, the cross-pol may comprise up to half the initially beamed power. Although the bed at South Pole (site of the ARA experiment) is likely to be `flat', not including this effect and only measuring received co-polarized power results in an under-estimate of $L_{atten}^\nu$.
\item Birefringence will tend to result in a splitting of signals along two axes -- in the worst case, the received power is reduced by a factor of two. Not including this effect results in an overestimate of the effective volume for neutrino detection.
\item As mentioned above, the cumulative effect of multiple impurity layers, at glancing angles, may be significant, although ARA's observation of distant in-ice transmitters puts a limit on how large this effect might be\citep{ARA}. 
\end{enumerate}

\subsubsection{Surface Ice Properties}Properties of the ice surface are important, particularly for the ANITA experiment, in their inference of both the energies of neutrinos, based on the transmission of signal across the air-ice interface, and up to the ANITA gondola, as well as their inference of the energies of cosmic rays based on their observation of radio reflections, as mentioned previously.
In general,
accurate inference of the energies of
these cosmic rays requires understanding the transmission/reflection of radio wave
signals across the ice-air boundary.
Satellite-based measurements of Antarctic surface reflectivity, using a co-located
transmitter and receiver, have been performed more-or-less continuously for the last few 
decades. Our comparison of four
different reflectivity surveys, at frequencies ranging from 2--18 GeV
and at near-normal incidence,
yield generally consistent maps of high vs. low reflectivity, as a function of 
location, across the continent. 
Using the 
Sun as an RF source, and the ANITA-3 balloon borne radio-frequency 
antenna array as the
RF receiver, the surface reflectivity can also be measured
at elevation angles of 12-30 degrees. Such measurements yield
good agreement with the expected reflectivity as prescribed by the Fresnel equations, with
no obvious dependence on location \citep{Besson:2013jm}. 

To measure surface reflection at near-glancing elevation angles ($<5^\circ$) 
inaccessible to the Antarctic Solar technique and unprobed by previous aerial
surveys, a trailer balloon transmitting a fast, high-amplitude signal using
inexpensive piezo-electric technology can be used (``HiCal''). The HiCal-1 mission in
January, 2015 trailed ANITA-3 by distances of 650-800 km, and registered
approximately 100 `golden' doublets, in which the HiCal signal was observed by
ANITA both in it's direct transmission as well as it's surface-reflection, arriving
some seven microseconds later.
Unlike previous satellite-based measurements,
HiCal-ANITA constitute a 
transmitter-receiver pair separated horizontally by hundreds of kilometers.
Data taken with HiCal, between 200--600 MHz show 
a significant departure from the Fresnel equations, with the 
deviation increasing with obliquity of incidence, which can perhaps be attributed to the
combined effects of possible 
surface roughness, radar clutter and/or shadowing of the
reflection zone due to Earth curvature effects.

\begin{figure}[htpb]
\centerline{\includegraphics[width=1.0\textwidth]{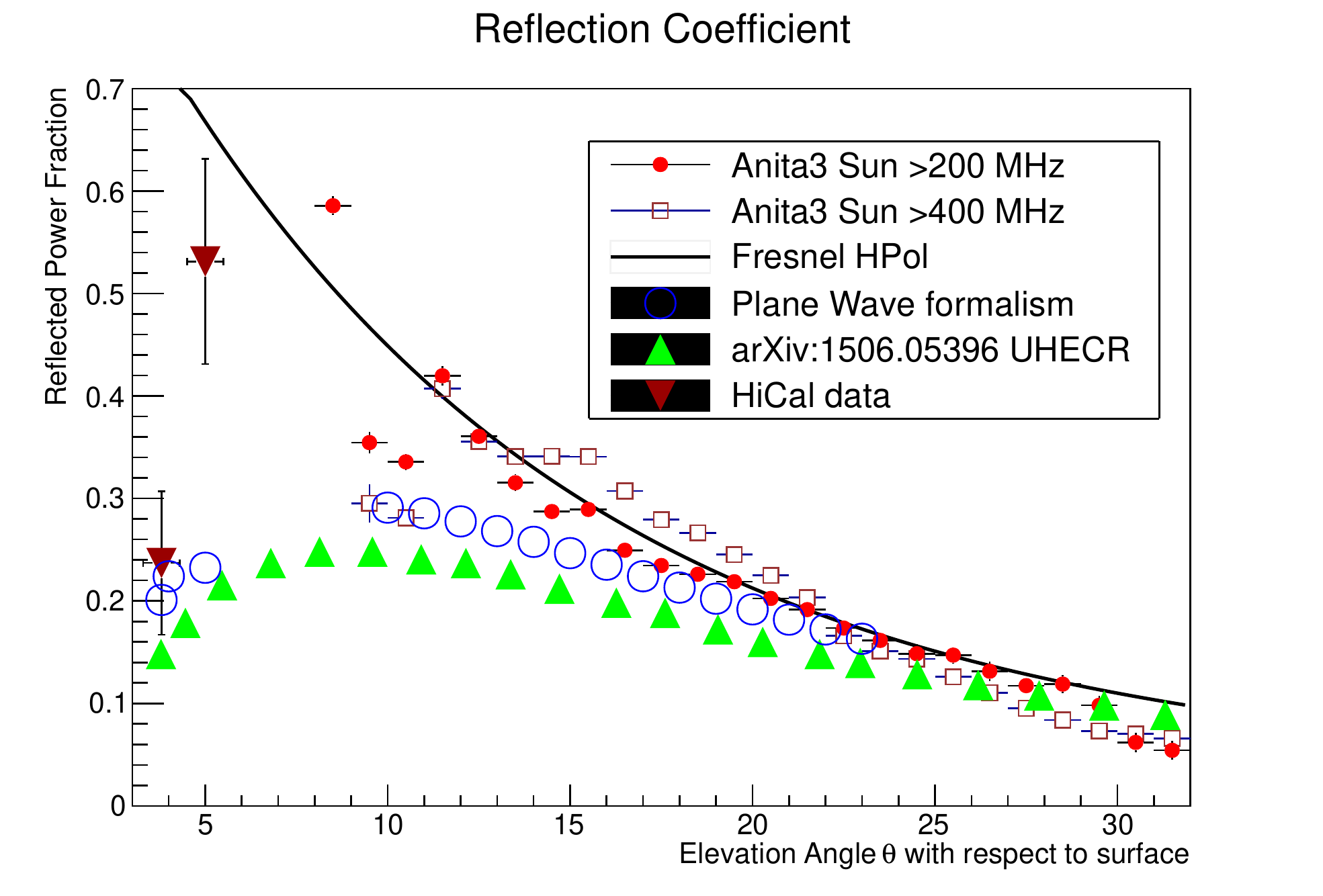}} 
\caption{Summary of ANITA-3 Antarctic radio frequency surface reflectivity measurements. Curve shows HPol Fresnel power reflection coefficient, assuming surface index-of-refraction of 1.35. Filled red circles are ANITA-3 Solar observations over full band 200-1000 MHz; open brown squares show Solar reflectivity result, after high-pass filtering above 400 MHz. Inverted brown triangles show results obtained using HiCal-1 triggers observed by ANITA-3. Filled green triangles show currently applied corrections to UHECR energys; open blue circles show results of similar calculation using plane-wave formalism.}\label{fig:overlay.pdf}
\end{figure}

Figure \ref{fig:overlay.pdf} shows the result of surface reflection studies done for the ANITA-3 experiment,
including both Solar reflection measurements as well as the data from the HiCal-1 experiment.
We note that the calculations done thus far for estimating the energies of UHECR recorded by ANITA-1 have used
a lower estimate of the reflection coefficient than indicated by the current data, suggesting an overestimate of
the quoted corrected UHECR energies by approximately 10--20\% \citep{Schoorlemmer:2015afa}.

\subsubsection{Possible Surface Wave RF Propagation}
It has been suggested that radio flux emanating from an 
in-ice shower may be ``trapped'' on the 2-d surface,
in which case the dimunition of signal with distance will be softer (1/r) than for the case where flux propagates in three dimensions ($1/r^2$)\citep{Ralston}. If so, this represents a potentially extremely lucrative technique for pushing down the lower-energy threshold of the neutrino detection experiments.
The notion of a surface propagating wave has been understood 
as a special solution to 
Maxwell's Equations since the 1800's, but was first definitively articulated by
Jonathan Zenneck over one century ago \citep{Zenneck1907}, and also famously
considered as a promising avenue for the then-burgeoning wireless industry by
Nikolai Tesla\footnote{See also Jack and Meg White's discussion of this phenomena in
Jim Jarmusch's ``Coffee and Cigarettes'', also in http://www.youtube.com/watch?v=sL9bq3YmHJo}. 
Zenneck obtained the dispersion relation for propagation along a boundary with
permittivity $\epsilon$ in terms of the free-space permittivity $\epsilon_0$ as: $k^2=k_0^2\epsilon_0\epsilon/(\epsilon_0^2+\epsilon^2)$, with
the standard definition $k=\omega/c$, leading to the expectation that, in so far as 
dielectrics are characterized by $\epsilon>\epsilon_0$, surface waves should propagate
with phase velocities exceeding the vacuum velocity of light $c_0$. We note that
this is distinct
from superluminal phase velocities propagating along the conducting surface of a waveguide \citep{SLP}, or group velocities propagating at $v>c_0$ in regions of anomalous dispersion.
Crudely, such waves
represent the solution to Maxwell's Equations in the limit where the incident
angle is equal to the critical angle $\theta_{crit}$ -- intuitively, surface flux trapping must be
the case here since there is neither a transmitted nor a reflected wave. In addition
to 2-dimensional rather than 3-dimensional flux spreading, another prediction in this limit is the expectation of waves following the curvature of the surface
(or the Earth, as a whole), contrary to the usual expectation that, e.g.,
radio waves must follow line-of-sight. Upon learning of observations 
that low-frequency
radio waves were detected well beyond simple line-of-sight,
Sommerfeld (1909) followed with
an expansion of Zenneck's investigations \citep{Sommerfeld1909}, suspecting that those
far-propagating radio waves were carried along the Earth-atmosphere boundary (later it was
realized that this phenomena was primarily due to ionospheric reflections), with particular attention
given to the vertical vs. horizontally polarized components (``VPol'' vs. ``HPol'') of 
surface waves. Sommerfeld concluded that, of the two, the transverse component
is most efficiently transported at the interface. 

More recently, the behavior of waves propagating along a boundary of two dielectrics (characterized by permittivities $\epsilon_0$ and $\epsilon_1$) at near-glancing incidence has been re-visited \citep{Ralston} in the context of neutrino detection, and the qualitative conclusion that RF waves may couple to the interface and travel exclusively along the surface, experiencing flux spreading considerably smaller than in the bulk dielectric, re-derived. That first-principles calculation predicts, specifically, that:
\begin{itemize}
\item Unlike the well-known evanescent solutions to Maxwell's equations corresponding to ``total internal reflection'' incidence angles, surface waves are superluminal, with phase velocity $v_p=\sqrt{{\epsilon_0+\epsilon_1}\over\epsilon_1}$, corresponding to $v_p\sim 1.34c_0$ at the ice-air boundary typical of Antarctica.
\item The amplitude attenuation length for a surface wave is intrinsically a factor of $2\sqrt{2}$ times longer than the attenuation length in the bulk. This effect is distinct, and augments, the effect of two-dimensional rather than three-dimensional flux spreading, corresponding to power reduction with distance-from-source as $1/r$ vs. $1/r^2$, respectively. In fact, for so-called neutrino-induced ``Askaryan'' signals, one expects an overall ratio of $E_{surface}/E_{bulk}\sim\sqrt{\omega r/c}$; taking typical values of $\omega\sim$250 MHz and r=2800 m \citep{Ralston}, this results in an enhancement of nearly 300 in the predicted surface electric field amplitude.
\item Surface coupling is expected when the incidence angle grazes to within one degree of the surface.
\end{itemize}

Probing surface waves represents a considerable experimental challenge.
Systematics are minimized using a propagation medium which is at least several
wavelengths thick, and, ideally, semi-infinite in lateral extent. These requirements are not
easily satisfied given the typical confines of a standard laboratory, unless the wavelength can
be limited to ${\cal O}$(1--10 cm). Previous experimental probes of such effects are, correspondingly, not readily found in the literature. 

Hansen \citep{Hansen} 
attempted to quantify the effects of rough ocean surfaces by propagating electromagnetic waves between 4 and 32 MHz, over a 235 km path length. In particular, he sought to verify calculations by Barrick \citep{Barrick} 
that prescribed the numerical loss of signal strength as a function of wave height, wind and sea state. He observed remarkably good agreement with Barrick's predictions, up to a frequency of 18 MHz, beyond which his data showed a linearly increasing excess of measured power relative to expectation, rising to a value of 15 dB higher measured power than expected at 30 MHz. Hansen was aware of surface wave propagation and cites work by Wait \citep{Wait},
however, that work was not explicitly connected to Zenneck waves, nor is there explicit measurement of wave velocity.

The first unequivocal claim of surface Zenneck waves was made by Soviet groups \citep{SU1,SU2,SU3}
in a series of experiments beginning in 1980.
They measure group velocities exceeding $c_0$ for electromagnetic waves traveling on the surface
of high-salinity (35\%) water, over the frequency regime from 700-6000 MHz. Moreover, they also
claim observation of the expected $1/\sqrt{r}$ diminution of amplitude with distance from the
source, as well as the transformation from a surface wave to a bulk wave in the case where the 
propagating surface wave encounters a surface discontinuity. 

Mugnai {\it et al} \citep{Mugnai} claimed free space
superluminal (by several per cent) wave propagation using a pulsed microwave beam at 8.6 GHz, by measuring the
leading edge arrival time between a transmitter and receiver over distances ranging from
30 cm to 130 cm, although the statistical significance of this result has been contested \citep{BigelowHagen2001},
and a re-analysis of the original data finds inconsistency with $v=c_0$ at less than 2$\sigma$ significance \citep{RingerMacherMead}.
Ranfagni {\it et al} \citep{Ranfagni}, within the last decade, 
claimed observation for surface wave excitations at $\sim$10 GHz frequencies, using custom microwave
horns and microwave absorbers to minimize multi-path effects.
Their measurements focused on wavespeeds over the frequency range 7.5--10 GHz using a transmitter system stepping through this band at 10 MHz increments. Although their received signals were often evidently contaminated by secondary, downstream interference effects, the `leading edge' of their received signals often indicated phase velocities exceeding $c_0$, although this was not systematically quantified. No direct study of amplitude was reported in that experiment.

\paragraph{Investigation of Flux Spreading using Variable Transmitter Depth}
To the extent that flux is trapped on the surface, of course, it can therefore no longer be measured within the bulk ice. I.e., the near-constancy of received signal by the englacial RICE receiver array, comparing the case where the transmitter is near-surface, and therefore more likely to excite surface coupling, to the case where the transmitter is fully submerged, similarly implies a near-constant signal flux spreading and inconsistency with surface flux trapping. 
This was the motivation behind an experiment aimed at measuring surface waves in 2008.
As a transmitter antenna is lowered into a borehole, one can measure the
\begin{figure}
\centering
\includegraphics[width=0.48\textwidth]{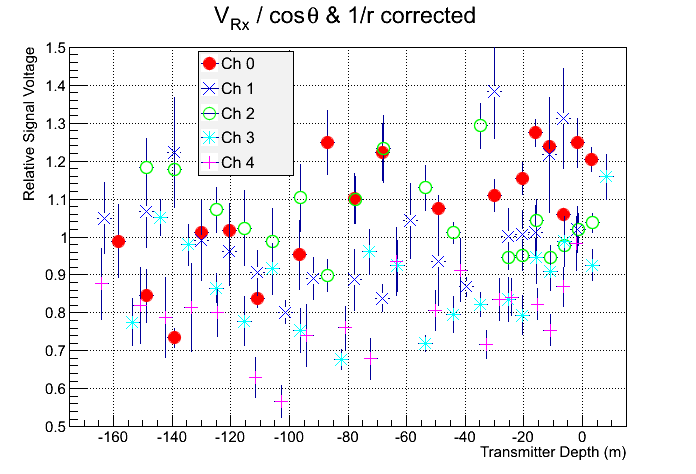}
	\caption{Comparison of received, corrected signal amplitudes in RICE receiver array, as a function of transmitter depth being lowered into ice borehole. Received signal strength is observed to be approximately independent of transmitter depth. \citep{alvarez2015experimental}} \label{fig:IOtimes}
\end{figure}
peak amplitude of the signal received in several of the RICE channels 0--4, themselves varying between 110 m and 200 m depths. Knowing the geometry of the transmitter and receiver, correcting for the dipole beam pattern of the dipole antennas ($G(\theta)\propto \cos\theta$), and assuming bulk electric field spreading varying as $1/r$, we observe reasonable agreement (Fig. \ref{fig:IOtimes}), down to the bottom of the transmitter borehole, between the signal received at the surface compared to the signal received at depth, consistent with the expected volumetric flux dimunition for all tested depths, including with the transmitter at the surface, and inconsistent with significant flux-trapping for near-surface transmitter depths. 
\subsection{Current Status of In-Ice Experiments}

Prospects for detection of ultra-high energy neutrinos have been brightened with the recent observation of ultra-high energy astrophysical neutrino candidates by the IceCube experiment \citep{icecube2013evidence}; the current number of detected neutrinos of non-atmospheric origin is now of order 50 \citep{aartsen2016search}. Although their origin has yet to be established, this observation nevertheless has, at long last, ushered in the era of the `neutrino telescope'. 

There are currently several Antarctic initiatives which seek to take advantage of the excellent radio-frequency clarity of cold ice to achieve first-ever detection of the cosmogenic neutrino flux. 
Each has its own particular strengths and weaknesses, which we attempt to highlight below. The possibility of a neutrino detector sited
near Summit, Greenland (the Greenland Neutrino Observatory \citep{wissel2015site}) has also been explored and an acceptably long attenuation
length measured at 75 MHz \citep{avva2015situ}, although a mature prototype has not yet been 
deployed.

\subsubsection{The ANITA (ANtarctic Impulsive Transient Antenna) Experiment}
Initiated in 2003,
the Antarctic Impulsive Transient Antenna (ANITA) experiment is a balloon-borne antenna array primarily designed to detect radio wave pulses caused by neutrino interactions with matter, specifically ice. The basic instrument consists of a suite of 40 quad-ridged horn antennas, optimized
over the frequency range 200-1200 MHz, with separate outputs for
vertically vs. horizontally incident radio frequency signals, mounted to a high-altitude balloon.
From an elevation of
$\sim$38 km, the balloon scans the
Antarctic continent in a circumpolar trajectory.
Details on the ANITA hardware, as well as the triggering scheme crucial to the analysis described herein, are provided elsewhere \citep{GorhamAllisonBarwick2009}.

\begin{figure*}[ht!]
\begin{minipage}{18pc}
\includegraphics[width=1.0\textwidth]{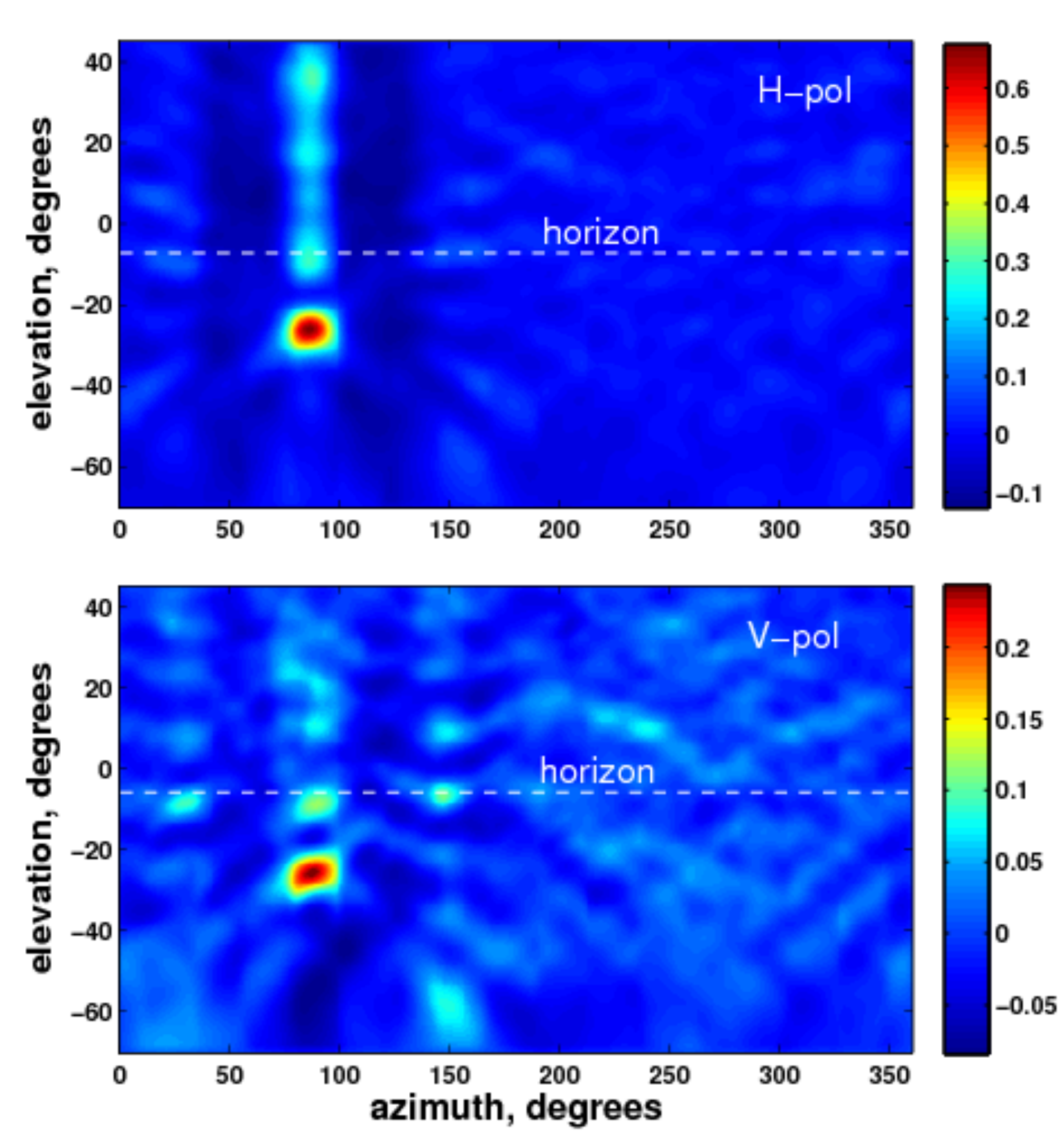} \caption{ANITA interferogram from one of their UHECR candidates. \citep{Hoover:2010qt}.} \label{fig:ANITA_UHECR_IFG} \end{minipage}
\hspace{1mm}
\begin{minipage}{18pc}
\includegraphics[width=1.0\textwidth]{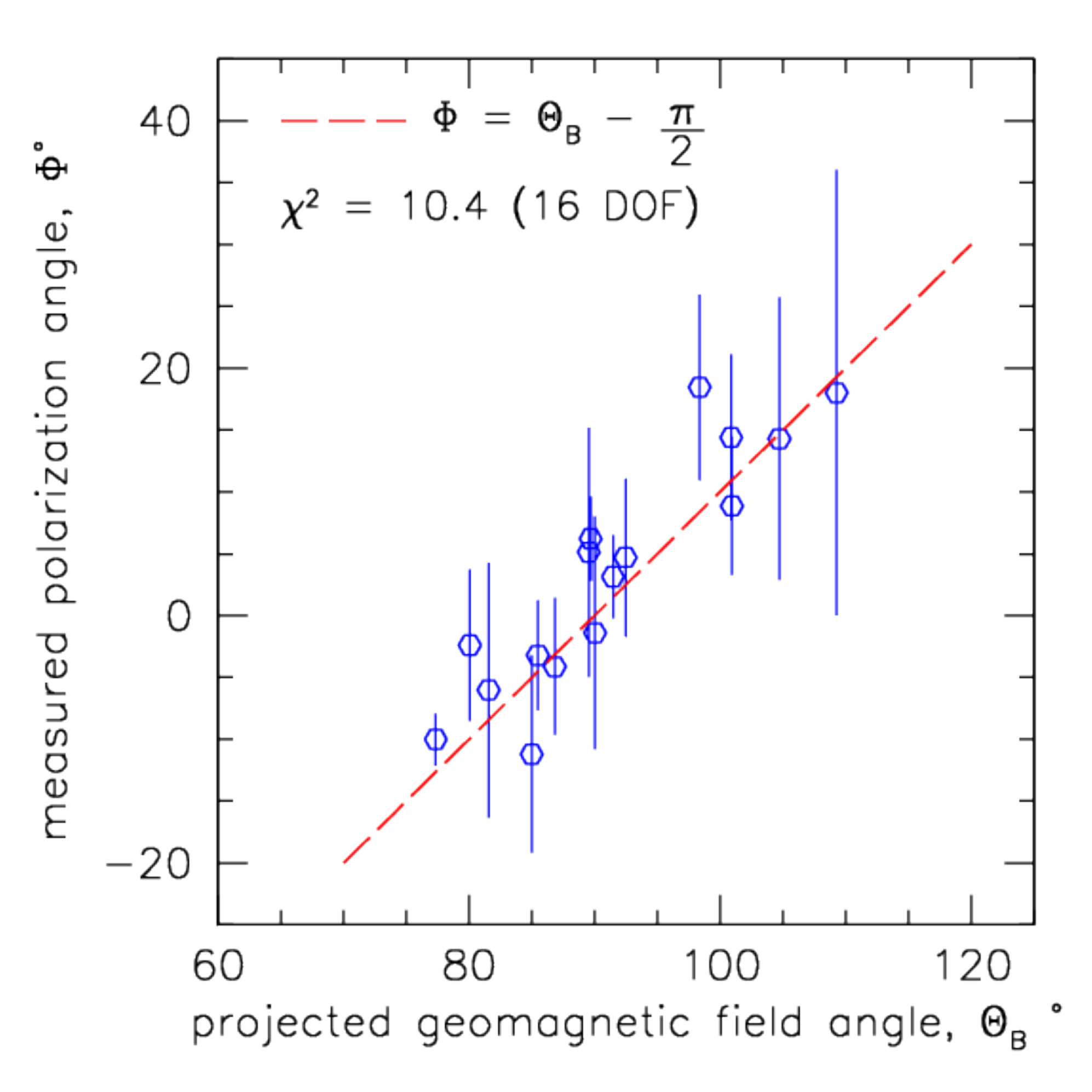} \caption{Ratio of VPol:HPol signal strength for ANITA UHECR candidates, compared with expectation knowing the UHECR event geometry and the local Antarctic geomagnetic field orientation. \citep{Hoover:2010qt}} \label{ANITA_polangle} 
\end{minipage}
\end{figure*}

Two one-month long missions (ANITA-1; Dec. 2006-Jan. 2007 and ANITA-2; Dec. 2008-Jan. 2009) have yielded world's-best limits on the UHE neutrino flux in the energy range to which
ANITA is sensitive \citep{HooverNamGorham2010,gorham2010observational}. Analysis of data taken during the 23-day
ANITA-3 flight (2014-15) is currently underway. As discussed previously,
initial analysis of the ANITA-1 data sample 
also provided a statistically large (16 events) sample of radio frequency signals attributed to
radio-frequency radiation associated with cosmic-ray
induced extensive air showers
(EAS) \citep{HooverNamGorham2010}. One such event, reconstructed interferometrically, 
is shown in Figure \ref{fig:ANITA_UHECR_IFG}; the high signal-to-noise ratio of these events is evident from the
interferogram. Critical in convincing the 
community that these events were indeed cosmic rays was the excellent agreement between the relative 
VPol:HPol signal strengths observed
in these events compared to expectation, knowing the local geomagnetic field direction (shown in Fig.\ \ref{ANITA_polangle}).

The heirarchical trigger and small occupancy allows ANITA to collect data very close to the intrinsic thermal noise floor, resulting in excellent sensitivity. Although no neutrinos have been found thus far, ANITA has, as mentioned previously, nevertheless demonstrated the ability to self-trigger on radio emissions from air showers.
ANITA-IV is scheduled to launch in December, 2016 and may be the last flight in the ANITA series. 

Among the most interesting recent developments from ANITA has been the re-analysis of their 2006-07 ANITA-1 data, including closer scrutiny given to an `upcoming' (i.e., emanating from below the horizon and apparently moving upwards from a point on the Antarctic surface) triggered event \citep{gorham2016characteristics}, which was not included in the original ANITA report of cosmic ray observations. This event is particularly interesting since it does not appear to exhibit the expected signal phase inversion which would otherwise mark it as a surface radio reflection, and is characteristic of the other surface-reflected cosmic ray candidates reported in the original ANITA-1 publication. However, the steep angle at which this event is observed indicates that, if it were a neutrino, it would have had to traverse a much larger chord of the Earth than should have been allowed by Standard Model neutrino cross-sections. Neglecting that, the topology is consistent with a $\nu_\tau\to\tau$ which then escapes the ice and decays hadronically in-air.

There are two primary strengths of the ANITA experiment. The first is the enormous amount of ice which can be 
scanned by the gondola at any given time, and corresponding to a disk of size $\pi r^2t$, where $r$ is
the distance to the observable horizon from a balloon
altitude $h$ ($\sqrt{2hR_{Earth}}\sim$650 km), and $t$ is the thickness of ice from
which neutrinos can be observed. In principle, of course, $t$ is the entire ice sheet thickness at any
given point, however, given the expected
attenuation length profiles as a function of depth, which are expected to diminish as
the ice warms closer to the bedrock, the upper half of the Antarctic ice sheet, corresponding to
$t\sim$1.5 km, is most favorable as a neutrino target.

Second, in those regions of the continent where there are no Antarctic bases, the environment is
remarkably low-noise. Indeed, it is this property of Antarctica that allowed ANITA-3 to register
the radio-frequency pulses from an inexpensive piezo-electric signal
generator mounted on a trailer balloon (``HiCal''), emitted from a distance in
excess of 700 km.

One drawback of ANITA is the fact that its livetime is limited by the caprices of the
circumpolar vortex. Although the first two ANITA flights, in 2006-07 and 2008-09, respectively,
lasted through over 3 revolutions around the continent, spanning a period in excess of one month,
the ANITA-3 mission of 2014-15 was cut down after 1.5 orbits owing to concerns that the balloon might
drift off the continent, in which case an ocean recovery of the data would have been nearly impossible.

Note that NASA is planning for an ultra-long duration balloon program (ULDB), which would
target loft times exceeding 100 days, at constant altitude. This compares with the current
LDB balloons, using flexible mylar, and which ascend and descend
in the stratosphere, depending on the solar elevation in the sky, 
and consequently the degree of
solar heating.
This would be ideally matched to the goals of the EVA
(ExaVolt Antenna) project \citep{EVABalloon}, which is an ambitious proposal to compensate for the large inherent 
synoptic distance-to-vertex with antenna gain. Specifically, by instrumenting the equatorial plane of
the balloon itself with a set of focusing antennas, 
with the primary receiver in the interior of the balloon
itself, one can achieve directional factors of order $10^2$ compared to the current
$\sim 10^1$, with a corresponding reduction in the experimental energy threshold down to the 
exavolt level. 
This is no small engineering feat, however, as the antenna array must `unfurl' as the balloon inflates.
Nevertheless, such a project, incorporated into a 100-day ULDB mission, flying over some combination
of Antarctic ice (optimized for neutrino detection) plus seawater (with a reflection coefficient likely
twice as large as that for Antarctic ice) 
should accumulate unprecedented self-triggered cosmic ray statistics.

\subsubsection{The ARA Experiment at the South Pole}
As a complement to ANITA, the Askaryan Radio Array (ARA) employs a buried antenna approach, with 37 proposed 16-antenna `stations' spread, at 2-km spacing, over a hexagonal grid at South Pole \citep{allison2012design}.
Since the expected `cosmogenic neutrino' flux is expected to peak at energies of approximately 300 PeV, and since the Askaryan signal power is distributed within a `fat' Cherenkov cone of frequency-dependent width $\Gamma(f)\propto 1/f$, with signal power at any frequency rising roughly proportional to $\sqrt{f}$, the most efficient embedded neutrino detection scheme involves multiple radio-frequency detectors (i.e., antennas) located in a close-packed geometry, with typical separations not more than 10 m, and thus offering the ability to image neutrino interactions over the entire solid angle comprised by the target ice volume. Achieving low energy thresholds also requires that the detectors themselves be close to the neutrino interaction point (i.e., `embedded') to mitigate 1/R signal amplitude losses. Each `station' therefore comprises an individual neutrino detector, with stations spaced at distances of order one radio-frequency attenuation length. Note that this geometry does, in fact, imply a high threshold ($>$10 EeV) for illumination of multiple stations -- correspondingly, the scientific overlap with the co-located IceCube experiment is modest, at best.

Thus far, in addition to the original pilot `testbed', three additional stations have been deployed (ARA-1, ARA-2, and ARA-3), with three more stations (ARA-4 -- ARA-6) planned for 2017-18.
In its final form, ARA proposes 37 stations spaced over a surface area of order 100 square kilometers, at a total cost of approximately \$10 M.   

The drilling of holes for ARA builds on experience gained with the Enhanced Hot Water Drill (EHWD) used for IceCube operation, although the smaller and shallower dry holes required for ARA necessitated design of a dedicated drill {\it ab initio}. The ARA hot-water drill used for ARA stations 2 and 3 in December 2012 drilled 200-meter holes in approximately 12 hours. Unlike the EHWD, the ARA drill includes separate reels for simultaneously sending, as well as pumping out the hole water -- at the conclusion of drilling operations, the resulting hole is therefore 'dry'. (Indeed, the first version of this drill, used for ARA01 operation was based on sequential drilling, followed by pumping of the melted ice water. An underestimate of the hole refreezing rate, however, resulted in ARA01 holes drilled to only 40\% of the desired 200 m. depth.)

\begin{figure*}[ht!]
\begin{minipage}{18pc}
\includegraphics[width=1.0\textwidth]{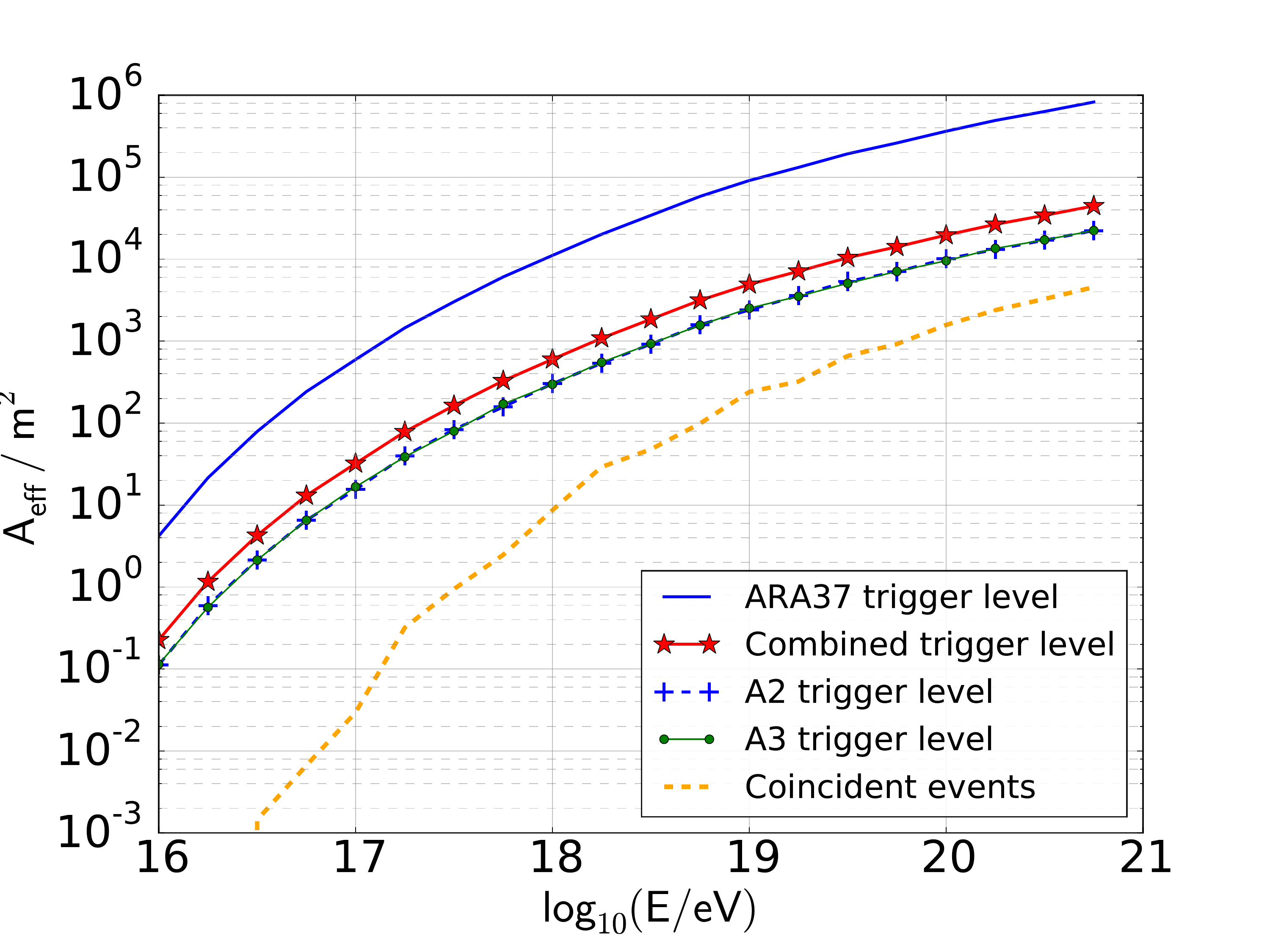} \caption{Effective area of current, and projected full ARA-37 array. \citep{allison2016performance}.} \label{fig:ARAAeff} 
\end{minipage}
\hspace{1mm}
\begin{minipage}{18pc}
\includegraphics[width=1.0\textwidth]{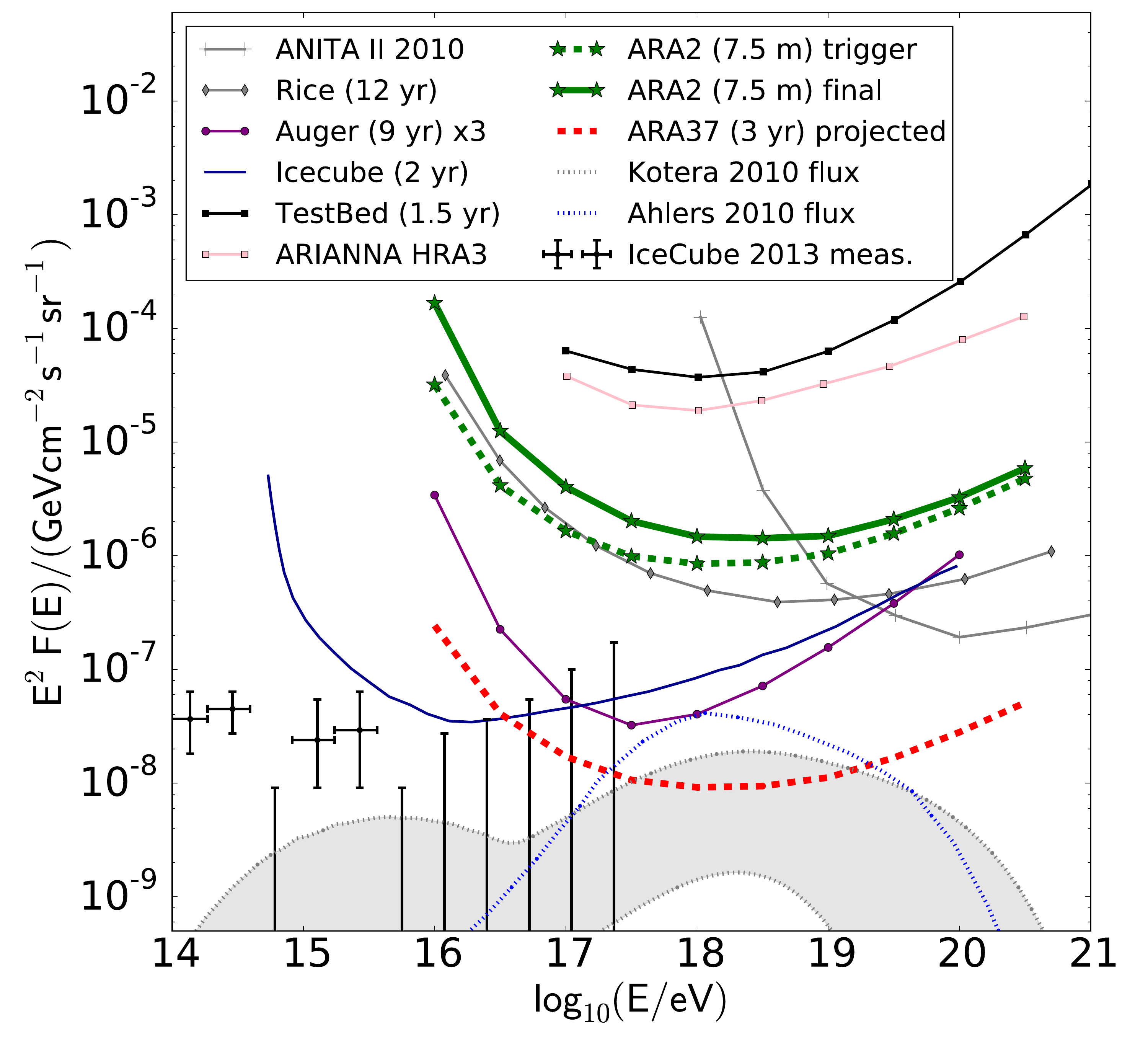} \caption{Neutrino senstivity of current, and projected full ARA-37 array \citep{allison2016performance}.} \label{fig:ARAUL} 
\end{minipage}
\end{figure*}

Neutrino reconstruction for ARA, as with ARIANNA and also ANITA consists of a series of initial event selection to suppress events with non-neutrino-like waveforms, followed by source localization using signal arrival times in the antennas comprising the array and, in the final step, testing waveforms against a neutrino `template' waveform. The so-called effective area $A_{\mathrm{eff}}$ for the most recent ARA neutrino search analysis, based on data from their ARA-2 and ARA-3 stations, and representative of the equivalent cross-sectional area over which intercepted neutrinos should be detected, is shown in Figure \ref{fig:ARAAeff}, with the limit implied by their non-observation of neutrino signal candidates in Figure \ref{fig:ARAUL}.

Thus far, the ARA experiment has extracted neutrino flux limits from their shallow `testbed', as well as flux
limits from their first two `deep' stations, with 16 antennas deployed to 180 meters \citep{allison2015first,allison2016performance}.
In the near future, ARA plans to investigate two initiatives: a) a phased-array approach that will allow them to reduce their experimental trigger threshold by an expected factor of $\sim$4\citep{vieregg2016technique}, and b) the possibility of drilling with a very fast, lower-overhead drill to a depth of 70 meters, with narrower holes. This would reduce the drilling overhead, and also reduce the overall cost of the experiment by anywhere from 25-40\%, with some reduction in the effective volume at very high neutrino energies due to ray tracing effects.

\subsubsection{The ARIANNA Experiment}
Whereas ARA is sited at South Pole, and draws power off the South Polar grid, ARIANNA is sited at Minnabluff, just below the snow surface, on the Ross Ice Shelf, in a region shielded by local mountains from the radio background presented by McMurdo Station. As a result, the ARIANNA hardware design targets minimal power consumption, with none of the narrow-band notch filtering required by ARA to suppress South Pole station backgrounds.

Each of the ARIANNA stations \citep{Gerhardt:2010js,Barwick:2014boa,Barwick:2014rca} 
consists of 8 downward-facing commercially-available
log-periodic dipole antennas (LPDA) arranged octagonally at a depth of $\sim$2 m at close spacing ($\sim$2 m) relative to a 
central Data Acquisition box. Contrary to ARA, and partially due to the comparatively shallow depth of the Ice 
Shelf at the selected site, the ARIANNA detection scheme relies, at least in part, on the fidelity of 
the neutrino-generated RF signal being preserved upon reflections at both the underlying shelf-sea water as well
as surface shelf ice-air interfaces.
Since the radio frequency attenuation length is 
\begin{figure}
\centering
\includegraphics[width=0.5\textwidth]{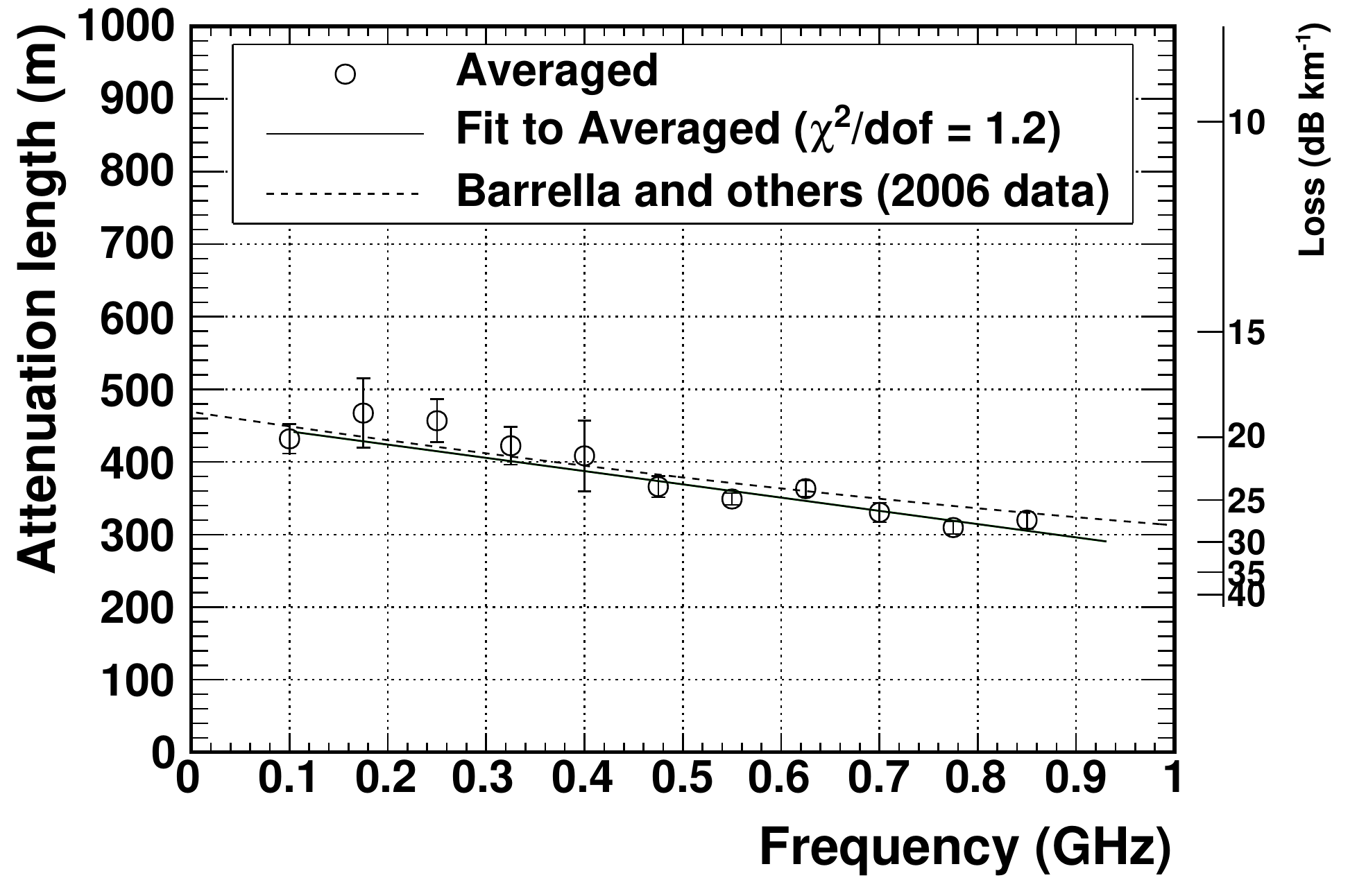} \caption{ARIANNA attenuation length as a function of frequency, measured at the experimental site \citep{hansJGlac}.} \label{fig:ARIANNA_att} \end{figure}
approximately a factor of three lower than the
RF attenuation length in the upper 1.5 km at the South Pole (Fig. \ref{fig:ARIANNA_att}), the so-called effective
volume $V_{\mathrm{eff}}$ per station is smaller for ARIANNA than ARA, although at somewhat reduced deployment
logistical overhead, as ARIANNA foregoes the need for
drilling. Moreover, since the neutrino flux is falling geometrically, by focusing detection on 
neutrinos at energies an order-of-magnitude lower than ARA, ARIANNA searches in an energy
regime not limited by attenuation effects.

\begin{figure*}[ht!] \begin{center}\includegraphics[width=1.0\textwidth]{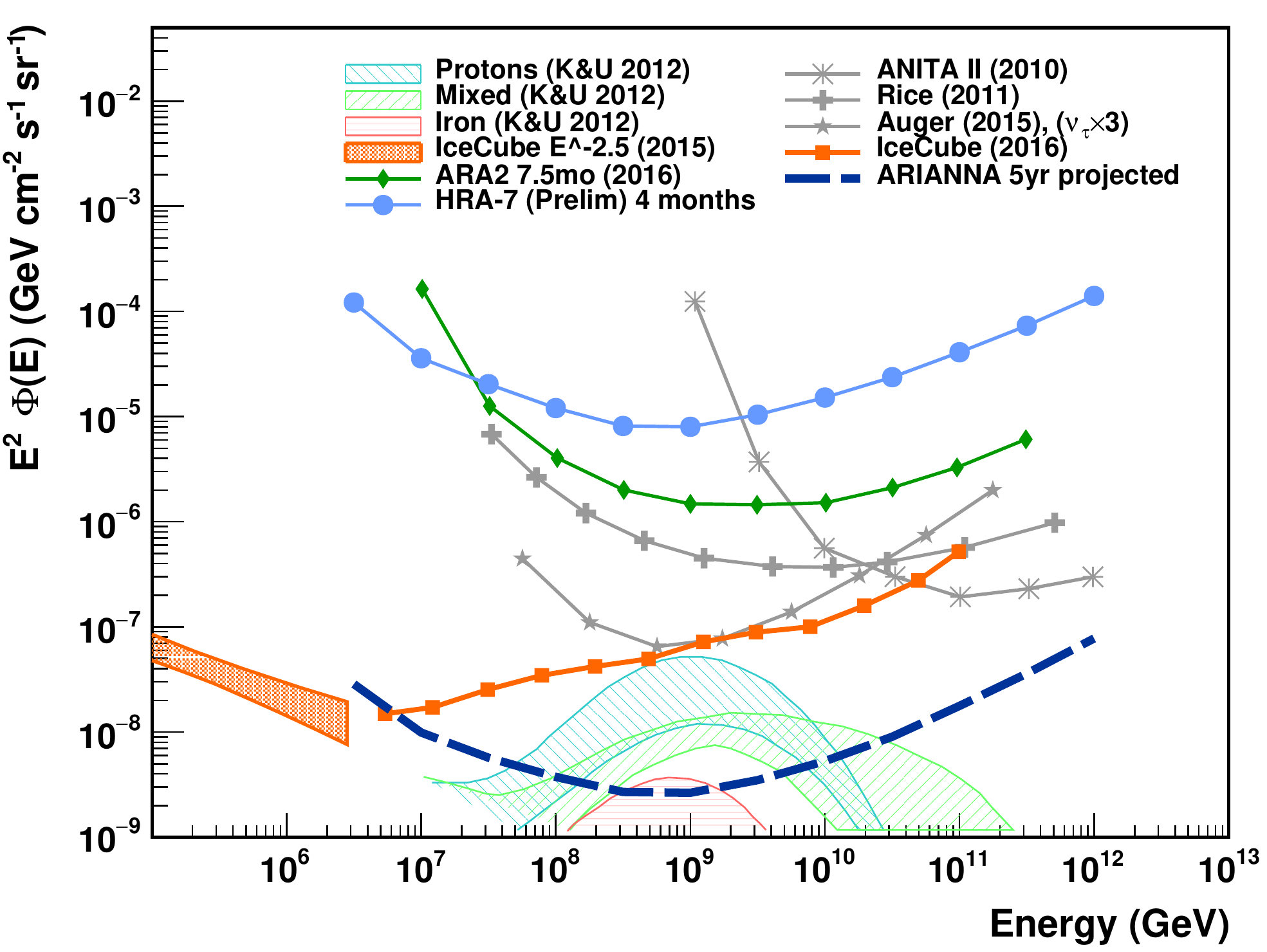} \caption{Projected neutrino senstivity of the ARIANNA-1296 array \citep{Barwick:2014pca}.} \label{fig:ARIAsensitivity} \end{center} \end{figure*}

ARIANNA have, thus far, performed extensive characterization of the Ross Ice Shelf neutrino target \citep{Barwick:2014rga} and 
produced first limits on the neutrino flux \citep{Barwick:2014pca}.
Currently funded for commissioning of the first seven ARIANNA stations by a US Major Research Instrumentation grant (``High Resolution Array'', or HRA), the full array is envisioned to fill a 36-array x 36-array station grid, 
with spacing between stations of order 1 km and frequency reach to 50 MHz (lower than ARA and ANITA). 
Once fully realized, 
Monte Carlo simulations indicate that the full ARIANNA-1296 outperforms ARA-37, albeit with a higher pricetag. 
The ultimate projected science performance of ARIANNA is summarized in Figure \ref{fig:ARIAsensitivity}.

\subsubsection{Direct comparison of Radio Experiments}

\begin{figure}[htpb]\centerline{\includegraphics[width=1.0\textwidth]{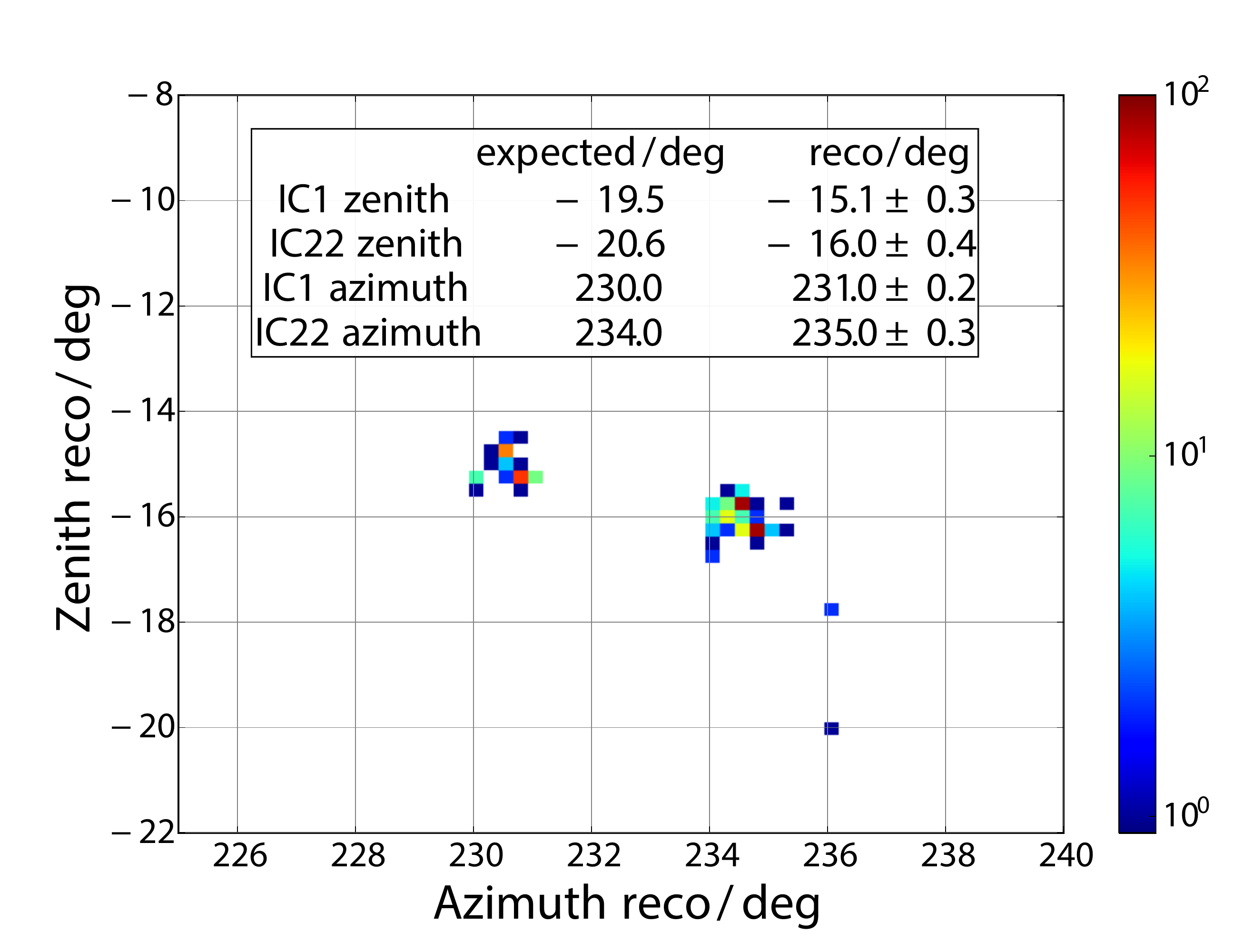}} \caption{ARA reconstruction of deep transmitters in IceCube holes 1 and 22 \citep{allison2016performance}.} \label{fig:A3Txrecon} \end{figure}

\paragraph{Calibration}
Both ARA and ARIANNA rely on pre-existing, deployed {\it in situ} transmitters to monitor their sensitivity, accumulating
hundreds of such calibration pulser events per day, whereas ANITA relies on either ground station transmitters or
in-air trailing balloon-borne transmitters (as in the HiCal concept) for calibration. However, the ANITA experiment
has been most successful thus far in using the Sun as a constant calibration source; ARIANNA has also clearly tracked
the Sgr A* galactic center radio source. ARA has similarly observed the sidereal galactic signal strength variation
in their surface antennas, but, thus far, has not shown a galactic signal from the in-ice array.
ARA is unique in their access to transmitters embedded deep into the ice sheet, in holes originally drilled for the
IceCube experiment. As shown in Figure \ref{fig:A3Txrecon}, ARA can reconstruct in-ice sources, down to 2 km source depths, with sub-degree precision in both elevation and azimuth, although ray tracing effects can systematically bias the elevation reconstruction by up-to several degrees. 
This is, thus far, the only demonstrated reconstruction of a deep source, and more typical of a neutrino geometry, from any of the competing neutrino detection projects.

\paragraph{Backgrounds}
South Pole Station, supporting a population of anywhere from 100-200 people during the
austral summer can be fairly noisy in the RF-band. Most of this background can be suppressed using the excellent
directional reconstruction of ARA, however, the possiblity of mis-reconstructions due to sub-threshold backgrounds
superimposed on noise requires higher experimental thresholds between 
October and station close (ca. Feb. 15 of each year). 
Proximity to the main station
also requires notch filtering (particularly to suppress the local Land Mobile Radio 
carrier at 450 MHz), which is present during the entire year.
ARIANNA, at the remote MinnaBluff site, is insulated from backgrounds due to McMurdo Station and runs at a relatively
constant threshold during the eight months when the sun is far enough above the horizon to power the array (ARIANNA
is considering wind power to improve their livetime to year-round operation). The ANITA flights have had to contend
with on-ice anthropogenic backgrounds, which reduced their effective area by approximately 45\% for, e.g., the ANITA-2
neutrino search, as well as satellites, which produce large backgrounds in the interval from 
260 to 380 MHz, resulting in compromised 
trigger sensitivity to sources at latitudes north of the payload. Once triggers have been taken, however,
those backgrounds can be easily mitigated in offline analysis. ANITA-IV will include pre-triggering
adaptive digital filtering of such backgrounds.

In addition to anthropogenic backgrounds (largely non-existent for ARIANNA, and suppressed with source directional reconstruction for both ARA and ANITA), calculations of the transition radiation signal resulting from air showers impacting the Antarctic surface indicate that such backgrounds may be comparable in both rate as well as signal strength to the sought-after neutrino-induced coherent Cherenkov signal\citep{deVries:2015oda}.

\paragraph{Radioglaciological Considerations}
From the radioglaciological standpoint, ARA is best situated 
among the existing radiowave cosmic ray experiments.
The Antarctic ice at South Pole is perhaps the best-characterized polar ice on the planet, with extensive
measurements of the complex permittivity. The 2.85 km thick ice at South Pole, thanks to studies conducted by
the IceCube experiment, 
has a well-measured temperature profile, which can be immediately translated into an attenuation length profile as a function of depth, using measurements of integrated two-way attenuation. 
Moreover, the South Polar Ice Core Extraction (SPICE) project presents the opportunity for additional
studies down to 1.5 km depth, and also offers considerable impurity characterization. 
The possibility of drilling below the firn, and also the potential scientific overlap with the IceCube experiment
further buttress the case for neutrino physics with ARA.

\paragraph{Science reach comparison}
ANITA takes in a huge field-of-view, encompassing approximately $2\times 10^6$ cubic km of ice, although the typical
distance-to-vertex, of order 100 km, implies an effective neutrino detection threshold of order $10^{19}$ eV. 
The detection of radio emissions from air showers by ANITA and ARIANNA is a testament to the generally low-background
environments in which these experiments operate. ARA has deployed surface antennas with the intention of observing
such air showers, however, their current surface antenna trigger is inoperable. In principle, there should be sensitivity
to down-coming signals using the in-ice `deep antenna' ARA
trigger, although those signals must also be separated from 
possible station RFI.

Table \ref{tab:compare} compares the three extant experiments. We note that all three experiments achieve sub-degree source reconstruction angular resolution in both azimuth and polar angle, and are architecturally very similar.
\begin{table}[htpb]
\begin{tabular}{cccc} \\ \hline
                  & ARA & ARIANNA & ANITA \\ \hline
$<L_{atten}(1500m)>$  & 1500 m  &  500 m & 500-1500 m \\
antennas & Fat dipoles  &  LPDA  & quad-ridge \\
         &  Slot-Cylinder     &    & dual-pol horns \\
Frequency band  & 150--1000 MHz & 110--1100 MHz & 200--1200 MHz       \\
LiveTime (days) & 270/yr   &  240/yr    &  30/flight   \\
Trigger Rate    & 5--10 Hz  & 0.1--1 Hz    & 50 Hz \\ 
Trigger SNR threshold &  4.5  & 4   &  4 \\
impulsive/CW RFI & SPS (summer) & Tribo-electric & Comms satellites \\
$E_\nu^{min}$ & 100 PeV    & 30 PeV    & 10 EeV \\ 
CR reconstruction? & No  & Yes  & Yes \\
Future Plans & ARA-4/-5/-6  & ARIANNA-1296 & ANITA-IV (2016-17) \\
             & (2017-18)    &  (proposed)              &  ANITA-V, EVA \\
\end{tabular}
\caption{Comparison of competing extant neutrino experiments}
\label{tab:compare}
\end{table}

\subsubsection{Future of in-ice efforts}
Given that the National Science Foundation (NSF) is unlikely to financially
support
multiple in-ice neutrino detection experiments, future progress will likely
require coalescence of the relevant stakeholders. As the IceCube statistics
on extra-terrestrial neutrinos grows, interest similarly grows in using
radio, at the lower end of the sensitive energy range, to match to the
upper end of the extrapolated IceCube neutrino energy spectrum. Currently,
both ARIANNA and ARA are approximately five years into their R\&D program, with comparable
hardware and performance. Which, if either of the two sites is selected for future
expansion depends, ultimately, which can deliver the most science per dollar. Each
experiment currently presents substantial logistical challenges -- for ARIANNA, delivering
hardware to
the remote 30 km $\times$ 30 km MinnaBluff site will require considerable infrastructural support
from NSF and its logistical subcontractor (Antarctic Support Corporation). For ARA,
the primary logistical challenges are (currently) drilling holes and also the
construction of roads and laying of cable to meet ARA's ultimate power requirements.
The most economical scheme is likely the solar-powered, and therefore (in principle)
limitless scheme proposed by ARIANNA, but with antennas deployed in shallow holes or
trenches. Note that the effective volume gain achieved by deploying below
the firn can be offset, to a large extent, by using high-gain antennas such as those
employed by ARIANNA, on the surface. An additional advantage of the ARIANNA scheme is that
such antennas can be made arbitrarily large, and therefore take advantage of the larger 
solid angle subtended by the lower-frequency components (below 100 MHz) of the Cherenkov signal.
Moreover, downward-facing high-gain antennas are
less susceptible to the transition radiation signal caused by air shower cores
impacting the Antarctic surface which, it has been argued, may be as copious as the
sought-after cosmogenic signal \citep{deVries:2013qwa}. By contrast, deep deployment
would require separation of a possible TR signal from true radio emissions from air
showers, which otherwise can serve as an extremely valuable calibration tool.
Ice purity and transparency arguments favor deployment at South Pole.

From the standpoint of science, the IceCube flux measurements can, with increasing reliability, now be extrapolated into the 10--100 PeV science threshold for the in-ice radio experiments. Unlike the cosmogenic flux, with large variations in the predicted levels, this provides a reliable ``handle'' on which the radio sensitivity can be pinned. Correspondingly, both ARIANNA and ARA are now developing strategies for pushing down their experimental sensitivity.
\subsection{Lunar and Salt targets}
\subsubsection{Earth Moon target}
The original two papers authored by Gurgen Askaryan recognized the science reach of the effect which now
bears his name \citep{Askaryan1962a,Askaryan1962b}, 
and also suggested the lunar regolith as an appropriate neutrino target. 
At the time,
there had been 15 years of studies of cm-meter wavelength lunar surface reflections, although more
quantitative understanding of the RF properties of lunar rock was not achieved until moon rocks
were returned by Apollo missions for terrestrial studies. Taking a value for the
measured loss tangent at 
2 GHz frequencies to be $\tan\delta\sim$0.003 (other experiments such as NuMoon \citep{NuMoonWesterbork} target lower frequencies for their neutrino searches) implies an RF attenuation length of
order 10 meters; however, the circumference of the lunar limb implies an enormously large potential
sensitive volume. This experimental scheme has been used many times thus 
far \citep{Scholten:2009ad,Jaeger:2009zb,James:2009qh,James:2008ff,Beresnyak:2005yh,Gorham:2003da}.
In the future, the Square Kilometer Array (SKA) is expected to have excellent sensitivity to 
both UHECR as well as neutrinos interacting in the regolith \citep{SKALunarARENA2016}.

The primary disadvantages of the lunar technique are: a) the 390,000 km distance from Earth to Moon
pushes the effective threshold up beyond 
$10^{20}$ eV, b) the observed signal should ideally be corrected for atmospheric dispersion effects, and c) the inability of experiments to easily distinguish
between neutrino-induced showers and hadronic-induced showers. The former shortcoming can be addressed, in part,
by using a radio receiver in orbit around the moon, as proposed for the LORD \citep{ryabov2009ultrahigh} experiment.
Alternately, one might hope to take advantage of the cold ice crusts on Jovian and Saturnian moons such as
Europa and Enceladus (and Ganymede and Callisto) \citep{gusev2010ice,shoji2012efficiency,miller2012pride} 
using a dedicated orbiter. As with LORD, one must cleverly
package a receiver sensitive to 100 MHz wavelengths in a small space, and with absolutely 
minimal power requirements. Moreover, for satellites in proximity to Jupiter, one must also contend
with the enormous Jovian decametric emissions. Beyond those obstacles, there are additional uncertainties
associated with the smoothness of, e.g., the Enceladus ice surface, as well as the actual purity of the
ice surface target, and the unknown effect of such things as liquid water inclusions.

\subsubsection{Salt Targets}
As an alternative to Antarctic ice, it was also suggested that large salt deposits may be appropriate
neutrino targets \citep{salt1,salt2}. However, measurements of RF propagation in typical rock salt at the
Hockley Salt Mine in Mississippi approximately 15 years ago 
yielded disappointing RF attenuation lengths, of order tens of meters. Salt domes,
having favorable volumes of order tens of cubic kilometers and often found near the Earth's surface
surface, also have
more favorable loss tangents than rock salt due to their formation process, which typically
purifies the sodium choride. Their proximity to oil fields has resulted in over 500 salt domes
being identified in the US, typically along the Gulf Coast. However, unlike Antarctica,
use of domes for scientific purposes is often obscured (at least in the US), 
and even overshadowed, by surface land use, as well as
deeper mineral rights and permits issues. At this time, there is no clear candidate for an appropriate 
salt-based neutrino detector.

\section{Conclusions}
The radio technique for measurement of cosmic particles is attractive for several reasons. First, in contrast to, e.g., photomultiplier tubes, the front-end component in the signal chain (namely, an antenna) can be modeled using freeware simulation codes and inexpensively constructed from easily accessible materials. Second, the duty cycle of radio instruments is, in principle, 100\%. 
Third, the ability to capture signals over a wide frequency band allows one to use sophisticated pattern and signal-recognition techniques for noise suppression. Finally, the meter-scale wavelengths are well-matched to the size of graduate students. Perhaps the most notable drawback of radio techniques is that radio-frequency backgrounds, particularly anthropogenic, are ubiquitous.

Following progress over the last decade, radio-frequency detection now rests on a solid theoretical and experimental basis. The former is provided by increasingly sophisticated and detailed simulation codes which are now able to absolutely predict expected signal strengths for both in-air and in-medium radio signals from basic electrodynamics and that do not involve any free parameters. The latter has been achieved by a series of testbeam experiments which, under controlled conditions, verify the calculations, as well as by air-shower measurements with several detector arrays which are well in agreement with signal predictions. While radio detection of neutrinos has yet to be demonstrated, cosmic-ray radio detection has by now become a routine business. In fact, for air-shower radio detection it has been shown that competitive accuracies in the reconstruction of air-shower parameters have already been achieved (better than $0.5^{\circ}$ in arrival direction, $\approx 15\%$ in energy, and $\approx 20$~g/cm$^{2}$ in depth of shower maximum).

Looking to the future, experiments such as the Square Kilometer Array will usher in the next generation of radio observation facilities. Although not purposed primarily for cosmic ray detection, the SKA can nevertheless offer powerful cosmic ray detection sensitivity. The broadband frequency response from 50 to 350~MHz and the very dense antenna spacing will allow extremely precise measurements of cosmic-ray air showers. For in-medium particle detection, there are several current efforts, although no single, funded, future project which will unambiguously establish the cosmogenic neutrino flux. Likely, some coalescence of current efforts will define the path forward.

%\bibliographystyle{ptephy}
%\bibliography{references}

\end{document}